\documentclass[12pt,letterpaper]{article}
\pdfoutput=1
\usepackage{jheppub_mod}
\usepackage{slashed}
\usepackage{bigints}

\newcommand{\bCentering}{\centering}
\newcommand{\bCaption}{\caption}

\def\Z{\mathbb{Z}}
\def\R{\mathbb{R}}

\def\Q{\mathbb{Q}}

\def\ov{\overline}

\def\Sym{\mathbf{Sym}}
\def\Anti{\mathbf{Anti}}

\def\Tr{\text{Tr}}
\def\IM{\text{Im}\,}
\def\RE{\text{Re}\,}
\def\ov{\overline}
\def\1{{\bf 1}}
\def\2{{\bf 2}}
\def\3{{\bf 3}}
\def\4{{\bf 4}}
\def\6{{\bf 6}}
\def\OR{\Omega\mathcal{R}}

\def\targ#1#2{\genfrac{[}{]}{0pt}{}{#1}{#2}}
\def\targ2#1#2{\genfrac{}{}{0pt}{}{#1}{#2}}

\definecolor{mygr}{rgb}{0,0.6,0}
\definecolor{mygrey}{rgb}{0,0.1,0.2}
\definecolor{myblue}{rgb}{0,0.5,0.9}
\definecolor{myblue2}{rgb}{0,0.5,0.5}
\definecolor{myorange}{rgb}{1,0.5,0}
\definecolor{mypurple}{rgb}{0.6,0,1}
\definecolor{mygolden}{rgb}{1,0.8,0.2}
\definecolor{myblue3}{rgb}{0,0.1,0.9}
\definecolor{myblue4}{rgb}{0,0.6,0.6}

\newcommand{\cG}{\mathcal{G}}

\newcommand{\ud}{\mathrm{d}}

\makeatletter
\newsavebox\myboxA
\newsavebox\myboxB
\newlength\mylenA

\def\mathtab#1#2#3{\begin{table}[th]\bCentering$#1$\bCaption{#3}\label{tab:#2}\end{table}}

\newcommand*\xoverline[2][0.75]{%
    \sbox{\myboxA}{$\m@th#2$}%
    \setbox\myboxB\null% Phantom box
    \ht\myboxB=\ht\myboxA%
    \dp\myboxB=\dp\myboxA%
    \wd\myboxB=#1\wd\myboxA% Scale phantom
    \sbox\myboxB{$\m@th\overline{\copy\myboxB}$}%  Overlined phantom
    \setlength\mylenA{\the\wd\myboxA}%   calc width diff
    \addtolength\mylenA{-\the\wd\myboxB}%
    \ifdim\wd\myboxB<\wd\myboxA%
       \rlap{\hskip 0.5\mylenA\usebox\myboxB}{\usebox\myboxA}%
    \else
        \hskip -0.5\mylenA\rlap{\usebox\myboxA}{\hskip 0.5\mylenA\usebox\myboxB}%
    \fi}
\makeatother

\title{Large Field Inflation from Axion Mixing}

\date{\today}

\author[a,b]{Gary Shiu,}
\author[c]{Wieland Staessens,}
\author[a,b]{and Fang Ye}

\affiliation[a]{Department of Physics, University of Wisconsin-Madison, Madison, Wisconsin, USA}
\affiliation[b]{Institute for Advanced Study, Hong Kong University of Science and Technology, Hong Kong}
\affiliation[c]{Instituto de F\'isica Te\'orica UAM-CSIC, Cantoblanco, 28049 Madrid, Spain}

\emailAdd{shiu@physics.wisc.edu}
\emailAdd{wieland.staessens@csic.es}
\emailAdd{fye6@wisc.edu}

\abstract{We study the general multi-axion systems, focusing on the possibility of large field inflation driven by axions. We find that through axion mixing from a non-diagonal metric on the moduli space and/or from St\"uckelberg coupling to a U(1) gauge field, an effectively super-Planckian decay constant can be generated without the need of
``alignment" in the axion decay constants. We also investigate the consistency conditions related to the gauge symmetries in the multi-axion systems, such as vanishing gauge anomalies and the potential presence of generalized Chern-Simons terms.
Our scenario applies generally to field theory models  whose axion periodicities are intrinsically sub-Planckian,
but it is most naturally realized in string theory.
 The types of axion mixings invoked in our scenario appear quite commonly in D-brane models, and we present its implementation in type II superstring theory. Explicit stringy models exhibiting all the characteristics of our ideas are constructed within the frameworks of Type IIA intersecting D6-brane models on $T^6/\OR$ and Type IIB intersecting D7-brane models on Swiss-Cheese Calabi-Yau orientifolds.
}

\preprint{MAD-TH-14-11\\ IFT-UAM/CSIC-15-006}

\begin{document}

\maketitle

%%%%%
%%%%%
\section{Introduction}
%%%%%
%%%%%

Observational results on the cosmic microwave background (CMB) and the large scale structure of our universe
continue to provide increasingly strong support for the inflationary paradigm. 
While the generic predictions  of inflation are in good agreement with data,  its theoretical underpinnings remain to be uncovered. 
An observable that plays a decisive role in discriminating classes of models are primordial gravitational waves, imprinted in B-mode polarization of the CMB.
Although a detectable level of primordial B-mode is not a must for inflation, 
such signal if observed would naturally point us to ``large field" inflationary models\footnote{Among the assumptions in \cite{Lyth:1996im} is that both the scalar and tensor perturbations are generated by vacuum fluctuations. Exceptions involving gravitational waves sourced by particle production during inflation can be found in \cite{Senatore:2011sp,Ozsoy:2014sba,Mirbabayi:2014jqa,Sorbo:2011rz,Cook:2011hg,Barnaby:2012xt,Mukohyama:2014gba,Ferreira:2014zia}.}.
Models in which the inflaton transverses super-Planckian distance in field space are sensitive to the ultraviolet completion of gravity. Thus, a proper formulation of large field models calls for inputs from quantum gravity.

In this regard, axions are a particularly well-motivated inflaton candidate.  Other than their abundance in string theory, the approximate shift symmetries that they enjoy serve to protect the inflaton potential over a large field range. Symmetry protection is what 
underlies the idea of natural inflation \cite{Freese:1990rb}. Non-perturbative effects breaking an otherwise exact shift symmetry generate a sinusoidal potential with the periodicity of the canonically normalized field set by
 the axion decay constant. However, detailed studies \cite{Banks:2003sx,Svrcek:2006yi} surveying different formulations of string theory have concluded that axions with super-Planckian decay constant do not seem to arise in controlled (i.e., weak coupling and large volume) regimes of 
string theory.

A way out of this conundrum is to break the periodicity of the axions perturbatively \cite{Silverstein:2008sg,McAllister:2008hb,Palti:2014kza,Marchesano:2014mla,Blumenhagen:2014gta,Hebecker:2014eua,Ibanez:2014kia,Arends:2014qca,McAllister:2014mpa,Franco:2014hsa,Blumenhagen:2014nba,Hebecker:2014kva,Ibanez:2014swa}. Common features in string compactifications such as fluxes, branes, and torsional cycles can provide sources of monodromies, leading to a change in
the axion potential upon transport around a (naive) cycle.
As pointed out in \cite{Marchesano:2014mla}, the monodromy inflation idea can be implemented in supersymmetric compactification (concrete realizations can be found in \cite{Marchesano:2014mla,Blumenhagen:2014gta,Hebecker:2014eua})
if the monodromy is induced by an F-term potential;
 the shift symmetry is spontaneously broken rather than explicitly broken,
and the F-term monodromy inflationary models 
have a direct connection with the 4d effective framework developed in \cite{Kaloper:2008fb,Kaloper:2011jz,Kaloper:2014zba}. Unlike natural inflation, the form of the inflaton potential is not universal. A variety of potentials have been found \cite{Marchesano:2014mla,McAllister:2014mpa} and thus the signatures of these large field models depend  on the sources of monodromy. 

Instead of breaking the axion periodicity, the inflaton field range can also be enhanced when one extends natural inflation to multiple axion fields.
Cumulative wisdom from earlier works \cite{Kim:2004rp,Dimopoulos:2005ac,Berg:2009tg} to recent investigations \cite{Choi:2014rja,Higaki:2014pja,Tye:2014tja,Kappl:2014lra,Bachlechner:2014hsa,Ben-Dayan:2014zsa,Long:2014dta,Higaki:2014mwa,Bachlechner:2014gfa,Burgess:2014oma,Gao:2014uha,Kenton:2014gma} has 
highlightened several mechanisms for field range enhancements, including kinetic alignment \cite{Dimopoulos:2005ac} from eigenvector delocalization \cite{Bachlechner:2014gfa} and axion decay constant alignment \cite{Kim:2004rp}.
A common feature shared by these multi-axion models is that the field range enhancement 
$f_{\rm eff}/f$ is tied to the number of low energy degrees of freedom 
 (including the axions and the rank of the non-Abelian groups which generate the non-perturbative instanton effects). Thus, the enhancement needed for super-Planckian field 
 excursion also takes away the elegance and simplicity of natural inflation.

 In this paper, we propose a new way to realize large field inflation
without breaking the axion periodicity or introducing large number of fields. 
In addition to kinetic mixings and mass mixings arising from the non-perturbative instanton potential,  there are in general other mixings in a multi-axion system. In the presence of St\"uckelberg $U(1)$'s, axion mixings are induced from their couplings to the Abelian gauge fields.
Each St\"uckelberg $U(1)$ gauge field gains a mass by eating a combination of axions.
As we will show, these St\"uckelberg couplings not only give a perturbative mass to the combination of axions that are eaten, but also extend the field range of the axions that survive.
The field range enhancement does not require a large number of fields. In one of our simple examples, the low energy degrees of freedom (below the St\"uckelberg $U(1)$ mass scale) involve only a single axion and some chiral fermions that are required in any case for anomaly cancellation. Our proposal is therefore a minimal realization of natural inflation in theories with sub-Planckian axion decay constants.

The axion couplings invoked in this work are rather generic. In fact, the St\"uckelberg mechanism lies at the heart of anomaly cancellation in string theory and arises frequently in D-brane constructions of particle physics\footnote{It was recently pointed out in~\cite{Shiu:2013wxa,Feng:2014eja,Feng:2014cla} that the mass mixings of St\"uckelberg $U(1)$'s provide an interesting and natural portal into dark sectors.}. The lagrangian for the multi-axion system considered here is more general and hence subsumes the considerations  of previous proposals. Our generalization thus provides an interesting starting point for further  studies of multi-axion inflation, and their statistical
analysis using random matrix theory.

This paper is organized as follows.  In section~\ref{sec:mixing} we examine kinetically mixed axions in three different scenarios and the possibility to generate a super-Planckian decay constant in each case. We also discuss the gauge-invariance problem induced by an axion eaten by the Abelian gauge field. In section~\ref{sec:string axions} we implement the axion mixing scenarios in string theory using Type II superstring compactifications with D-branes and provide some explicit examples in the frameworks of intersecting D6-brane models in Type IIA and of intersecting D7-brane models in Type IIB. The concluding remarks are given in section~\ref{sec:con}.

A summary of our conventions can be found in appendix~\ref{A:Conventions}. In appendix~\ref{A:ConsAxions} we offer a short discussion about the definition of the axion decay constant (in two different yet equivalent representation schemes) and how to read off the decay constant for kinetically mixing axions. Appendix~\ref{A:ChiralRotations} contains a brief review on chiral rotations and their relation to the scalar potential for axions. Appendix~\ref{A:Dualisation} provides technical details about the dualization procedure between two-forms and zero-forms in four dimensions. In appendix~\ref{A:Generalisation} we provide the full generalization of the system studied in section~\ref{sec:mixing}. And appendix~\ref{A:Fermion} discusses methods to find an explicit field theoretic model for the set-up analysed in~\ref{Ss:U1mixing}.

%%%%%%%%%%%%%%%%%%%%%%%%%%%%%%%%%%%%%%%%%%%%%%%%%%%%%%%%%%%%%%%%%%%%%%%%%%%%%%%%%%%%%%%%%%%%%%%%%%%%%%%%%%%%%%%%%%%%%%%%%%%%%%%%%%%%%%%%%%%%%%%%%%%%%%%%%%%%%%%%%%%%%%%%
\section{Mixing Axions in a Field Theory Setting} \label{sec:mixing}
Axions are CP-odd real scalars whose continuous shift symmetry can only be violated by nonperturbative effects such as gauge instantons, D-brane instantons, etc. However, the residual discrete shift symmetry still constrains how axions interact with other sectors and anticipating some of the considerations presented in section~\ref{sec:string axions}, the effective lagrangian for a system of $N$ axions $a^i$ with $i\in\{1,\ldots, N\}$ can be written as follows: 
\begin{eqnarray}\label{Eq:GeneralLagrangian}
{\cal S}_{axion}^{\rm eff} &= & \bigintsss \left[- \frac{1}{2}\, \sum_{i,j=1}^N \cG _{ij} (\ud a^{i}-k^{i}A) \wedge \star_4 (\ud a^{j}-k^{j}A) -\frac{1}{g_{1}^{2}} F \wedge \star_4 F  -\frac{1}{g_{2}^{2}} \Tr(G\wedge \star_4 G) \right. \notag \\
&& \left. \qquad  +\frac{1}{8\pi ^2}\left( \sum_{i=1}^N r_i a^i \right) \text{Tr} ( G \wedge G ) \right], 
\end{eqnarray}
where ${\cal G}_{ij}$ represents the metric on the axion moduli space. All axions are assumed to carry a charge $k^i$ under the single $U(1)$ gauge group (with potential $A$, field strength $F$ and gauge coupling $g_1$) and couple simultaneously to the topological density associated with a non-Abelian gauge theory (with field strength $G$ and gauge coupling $g_2$), and the coefficients $r_i$ correspond to model-dependent discrete parameters.\footnote{This set-up can be generalized straightforwardly to configurations of multiple axions carrying charges under multiple $U(1)$ gauge groups and coupling non-perturbatively to various non-Abelian gauge groups, see equation~(\ref{Eq:CompleteGeneralAxionGauge}) of appendix~\ref{A:Generalisation}. For simplicity, we will consider a minimal set-up with respect to the number of gauge groups to illustrate our scenario.}

To simplify the analysis, we choose a basis and a normalization such that all the matter fields carry integer charges under the $U(1)$ and $k^{i}$'s and $r_i$'s are integers\footnote{The integers $k^{i}$'s can be understood as ``axion charges". For closed string axions these charges are geometric in nature, as they depend on how the D-brane supporting the $U(1)$ gauge group wraps the  cycles along the internal space, as discussed in section~\ref{Ss:ClosedStringAxions}. 
}. The St\"uckelberg type couplings between the axions and the $U(1)$ gauge potential are invariant under the local transformation:  
\begin{equation}\label{Eq:tran}
\forall \, i: a^i \rightarrow a^i + k^i \eta, \qquad A\rightarrow A+\ud \eta. 
\end{equation}
By virtue of the St\"uckelberg mechanism the gauge boson acquires a mass when at least one of the $k^i\neq 0$. In case various $k^i$ are different from zero, the gauge boson eats a linear combination of the respective axions with $k^i\neq 0$. 

In our set-up, we consider an anomalous coupling of the axions to a strongly coupled non-Abelian gauge group, whose gauge instantons are considered to be the dominant non-perturbative effect in the action (\ref{Eq:GeneralLagrangian}), imposing a periodicity for the linear axion combination of the form:
\begin{equation}\label{Eq:CollectiveShiftSymmetry}
\sum_{i=1}^N r_i a^i \simeq \sum_{i=1}^N r_i a^i + 2 \pi. 
\end{equation}
The global continuous shift symmetry of the axions, manifestly preserved by the kinetic term for the axions, is therefore explicitly broken for this particular axionic direction. Independently, the axions $a^i$ can couple to other instanton effects (such as D-brane instantons), causing a periodicity of the form:\footnote{For axions charged under a $U(1)$ gauge symmetry, the field identification set by the $U(1)$ gauge symmetry reads:
\begin{eqnarray}
a^i\rightarrow a^i+2\pi k^i\,\nu^i.
\end{eqnarray}
In this respect, the axion periodicity (\ref{Eq:shift}) can be interpreted~\cite{BerasaluceGonzalez:2012vb} as a ``fractional" $1/k^i$ $U(1)$ gauge transformation, or equivalently to a transformation of the axion field under a discrete $\Z_k$ gauge symmetry. Such discrete symmetries remain present at low energies after the St\"uckelberg mechanism has taken place, given that they are also preserved by the non-perturbative corrections. Consequently, discrete $\Z_k$ symmetries can be used to constrain perturbative $n$-point couplings at energies much lower than the St\"uckelberg scale, see e.g.~\cite{Chen:2012jg,BerasaluceGonzalez:2011wy,Anastasopoulos:2012zu,Honecker:2013hda}.}
 
\begin{eqnarray}\label{Eq:shift}
a^i\rightarrow a^i+ 2 \pi \nu^i , \qquad \nu^ i \in \Z,
\end{eqnarray}
for each axion separately, yet which do no necessarily contribute effectively to the action (\ref{Eq:GeneralLagrangian}). Which non-perturbative effects contribute to the effective action, is in practice a model-dependent consideration. The most straightforward examples clarifying these statements can be found for closed string axions, which emerge from the dimensional reduction of the various differential $p$-forms along closed $p$-dimensional cycles. In case such a closed $p$-cycle can for instance be deformed due to the presence of massless deformation moduli, the Euclidean D-brane instanton supported by the $p$-cycle will most likely not contribute to the effective superpotential due to the presence of unsaturated fermionic deformation zero-modes. Nonetheless, the axion associated to the ``non-rigid" $p$-cycle is characterised by a periodicity set by the D-brane instanton. We will discuss the instanton contributions more explicitly in section~\ref{Ss:ClosedStringAxions} and the explicit examples considered in section \ref{subsec:ex} will allow us to clarify these statements even further.

In this section, we will investigate the physical effects of kinetic mixing among axions as reflected in the lagrangian (\ref{Eq:GeneralLagrangian}) and discuss configurations for which one of the axion decay constant can exceed the reduced Planck mass. To this end, we distinguish kinetic mixing among axions due to a non-diagonal metric ${\cal G}_{ij}$ on the moduli space (metric mixing) and kinetic mixing due to the St\"uckelberg couplings ($U(1)$ mixing). For simplicity, the number of axions is set to $N=2$, enabling us to highlight the differences between the two mixing scenarios as well. The formulae presented here can be generalized straightforwardly to set-ups with three or more axions, as we lay out in appendix~\ref{A:Generalisation}. Our analysis is divided into three parts: in a first phase purely metric mixing for axions will be considered, after which we continue the analysis with purely $U(1)$ mixing. As a last step we combine both mixing scenarios and discuss the most generic case.

%%%%%%%%%%%%%%%%%%%%%%%%%%%%%%%%%%%%%%%%%%%%%%%%%%%%%%%%%%%%%%%%%%%%%%%%%%%%%%%%%%%%%%%%%%%%%%%%%%%%%%%%%%%%%%%%%%%%%%%%%%%%%%%%%%%%%%%%%%%%%%%%%%%%%%%%%%%%%%%%%%%%%%%%
\subsection{Metric Kinetic Mixing}\label{Ss:MetricMixing}
In four dimensional supergravity theories and compactifications of string theories, it is customary for scalar fields to be characterised by non-canonical kinetic terms or a $\sigma$-model like action, which clarifies the presumed presence of the non-trivial metric ${\cal G}_{ij}$ in the lagrangian (\ref{Eq:GeneralLagrangian}). In order to expose the physical effects of this metric at fullest, we simplify the two-axion system by assuming that neither of them is charged under a $U(1)$ gauge field, i.e.~$k^1 = 0 = k^2$.
The kinetic terms for the axions then reduce to the following expression,
\begin{eqnarray}\label{Eq:KineticTermsOnlyMetricG}
{\cal S} ^{\rm kin}_{axion} = - \bigintssss  \frac{1}{2}\, \sum_{i,j=1}^{2}\mathcal{G}_{ij} (\sigma)\,  \ud a^i \wedge \star_4 \ud a^j,
\end{eqnarray}
where the metric ${\cal G}_{ij}$ on the axion moduli space depends on other (usually CP-even) moduli fields labeled collectively by $\sigma$ encoding geometric information about the internal manifold.\footnote{In general, the axion shift symmetry can take a much more intricate form than an affine realisation of a $U(1)$ symmetry. In that case, the Lie-derivative of the metric with respect to the Killing vector fields has to vanish, see e.g.~\cite{BerasaluceGonzalez:2012vb}, which constrains the dependence of the metric on the axions. In this paper, we will assume that the shift symmetries of the axions are affine realisations of $U(1)$ symmetries, in which case the metric ${\cal G}_{ij}$ does not depend on the axion fields $a_i$.} We will pay more attention to this point in section~\ref{sec:string axions} and assume for now that the moduli fields $\sigma$ are stabilised with non-vanishing vevs. With respect to the axion basis $(a^1, a^2)$ the symmetric metric $\cG _{ij}$ reads:
\begin{align}\label{Eq:MetricG}
\cG = \left(\begin{array}{cc} {\cal G}_{11} & {\cal G}_{12} \\
{\cal G}_{12} & {\cal G}_{22} \end{array}\right), \qquad \text{with } {\cal G}_{11}, {\cal G}_{12}, {\cal G}_{22} \in \R \backslash \{ 0\}.
\end{align}
The requirement that the metric is positive-definite boils down to the following two constraints by using Sylvester's criterion: 
\begin{eqnarray}\label{Eq:constraint0}
{\cal G}_{11} >0, \qquad {\cal G}_{11} {\cal G}_{22} -{\cal G}_{12}^2 >0.
\end{eqnarray}
The symmetric matrix ${\cal G}_{ij}$ can be diagonalized to a matrix with eigenvalues: 
\begin{equation}\label{Eq:MetricMixingMetricEigenvalues}
\lambda_\pm = \frac{1}{2}\left[ ({\cal G}_{11} + {\cal G}_{22}) \pm \sqrt{4 {\cal G}_{12}^2 + ({\cal G}_{11}-{\cal G}_{22})^2} \right],
\end{equation}
with associated normalized eigenvectors:
\begin{equation} 
\vec{u}_- = \left(\sin \frac{\theta}{2}, -\cos \frac{\theta}{2} \right), \qquad \vec{u}_+ =  \left(\cos \frac{\theta}{2}, \sin \frac{\theta}{2} \right),
\end{equation}
and where the parameter $\theta$ appears through the parametrization:
\begin{equation}\label{Eq:MetricMixingParametrisation}
\cos \theta = \frac{{\cal G}_{11}-{\cal G}_{22}}{\sqrt{4 {\cal G}_{12}^2 + ({\cal G}_{11}- {\cal G}_{22})^2}}, \qquad  \sin \theta=\frac{2{\cal G}_{12}}{\sqrt{4 {\cal G}_{12}^2 + ({\cal G}_{11}- {\cal G}_{22})^2}}, \qquad \text{with } \,0\leq \theta <2\pi.
\end{equation}
This parametrization enables us to expose the $SO(2)$ rotation used to diagonalize the metric $\cG _{ij}$. With these set of manipulations the kinetic action for the axions reduces to a diagonalized form:  
\begin{equation}\label{Eq:KineticTermsDiagonal}
{\cal S}^{\rm kin}_{axion} =  - \bigintssss \left[  \frac{1}{2} \lambda_-  \ud a^- \wedge \star_4 \ud a^- + \frac{1}{2} \lambda_+ \ud  a^+\wedge \star_4 \ud a^+  \right] , 
\end{equation}
where we introduced the new axion basis $(a^-, a^+)$:
\begin{equation}\label{Eq:OrthNonDiagKahler}
\left(\begin{array}{c} a^- \\  a^+ \end{array} \right)
 = \left(  \begin{array}{cc} \sin \frac{\theta}{2} & - \cos \frac{\theta}{2} \\  \cos \frac{\theta}{2} & \sin \frac{\theta}{2}   \end{array} \right)   \left(\begin{array}{c}  a^1 \\ a^2 \end{array} \right). 
\end{equation}
In order to correctly determine the effective axion decay constants for $a^-$ and $a^+$ respectively, we also have to apply the $SO(2)$ rotation on the anomalous coupling to $\Tr(G\wedge G)$: 
\begin{eqnarray}\label{Eq:PotKahlerMixingPhysBasis}
{\cal S}^{\rm anom}_{axion} &=&  \frac{1}{8 \pi^2}\,  \bigintssss \left[  \left( r_1 \sin \frac{\theta}{2} - r_2 \cos \frac{\theta}{2} \right)  a^- +  \left( r_1 \cos \frac{\theta}{2} + r_2\sin \frac{\theta}{2} \right)  a^+ \right] \, \Tr( G\wedge G) \nonumber \\
&=& \frac{1}{8 \pi^2} \,  \bigintssss \left[\tilde a^+ + \tilde a^-\right]\,  \Tr( G\wedge G).
\end{eqnarray}
The second equation follows by rescaling the axions such that the anomalous coupling is rewritten in a purely topological form (i.e.~in terms of representation scheme 2 of appendix~\ref{A:ConsAxions}):
\begin{eqnarray}
\tilde{a}^{-}\equiv\left( r_1 \sin \frac{\theta}{2} - r_2 \cos \frac{\theta}{2} \right)  a^-, \qquad \tilde{a}^+ \equiv \left( r_1 \cos \frac{\theta}{2} + r_2\sin \frac{\theta}{2} \right) \, a^+ 
\end{eqnarray}
Through the combination of equations (\ref{Eq:KineticTermsDiagonal}) and (\ref{Eq:PotKahlerMixingPhysBasis}) the axion decay constants for the rescaled version of the physical axions ($\tilde a^-$, $\tilde a^+$) can be read off\footnote{In case $\text{gcd}\,(r_1,\,r_2)\neq 1$, a subtlety arises in defining the axion decay constant. Namely, both axion decay constants in (\ref{Eq:metric-decay-const}) have to be divided by $\text{gcd}\,(r_1,\,r_2)$ to obtain the shortest periodicity. The vacuum configuration resulting from the instantons then consists of $\text{gcd}\,(r_1,\,r_2)$ consistent and independent vacua, separated from each other over a distance $2 \pi f_{\tilde a^\pm}$ respectively by domain walls.\label{Foot:GCDr1r2}}: 
\begin{equation}\label{Eq:metric-decay-const}
f_{\tilde a^-} = \frac{\sqrt{\lambda_-}}{|r_1 \sin \frac{\theta}{2} - r_2 \cos \frac{\theta}{2}|}, \qquad f_{\tilde a^+} = \frac{\sqrt{\lambda_+}}{|r_1 \cos \frac{\theta}{2} + r_2 \sin \frac{\theta}{2}|}.
\end{equation}
At this point, we should also pay attention to the consistency of the change of axion basis with respect to the initial discrete shift symmetry of~(\ref{Eq:shift}). With respect to the physical basis $(\tilde{a}^-,\,\tilde{a}^+)$ this discrete shift symmetry translates into the following shift symmetry:
\begin{eqnarray}
\tilde{a}^- &\rightarrow &\tilde{a}^-  +2\pi\left(r_1\,\text{sin}^2\frac{\theta}{2}-\frac{r_2}{2}\,\text{sin}\,\theta\right)\nu^1-2\pi\left(\frac{r_1}{2}\,\text{sin}\,\theta -r_2\text{cos}^2\frac{\theta}{2}\right)\nu^2,\\
\tilde{a}^+ &\rightarrow &\tilde{a}^+  +2\pi\left(r_1\,\text{cos}^2\frac{\theta}{2}+\frac{r_2}{2}\,\text{sin}\,\theta\right)\nu^1+2\pi\left(\frac{r_1}{2}\,\text{sin}\,\theta +r_2\,\text{sin}^2\frac{\theta}{2}\right)\nu^2.
\end{eqnarray}
Applying this result to the instanton coupling term in equation (\ref{Eq:PotKahlerMixingPhysBasis}), one observes that this topological term, undergoes a shift proportional to the Pontryagin index multiplied by an integer and $2\pi$,  
\begin{eqnarray}
{\cal S}^{\rm anom}_{axion} \rightarrow {\cal S}^{\rm anom}_{axion} + \frac{1}{8 \pi^2} \, 2\pi\left(r_1\,\nu^1+r_2\,\nu^2\right) \int \Tr( \,G\wedge G),
\end{eqnarray}
which leaves the path integral invariant (see also appendix~\ref{A:ConsAxions}). Hence, even expressed in terms of the physical basis $(\tilde{a}^-,\,\tilde{a}^+)$, the full theory remains consistent under the initial shift symmetry (\ref{Eq:shift}).

In order to explore the physical range of the axion decay constants given in (\ref{Eq:metric-decay-const}) we consider a numerical example, satisfying the constraints in (\ref{Eq:constraint0}). Let us consider a configuration where the entries in the metric (\ref{Eq:MetricG}) express a large fraction of metric mixing,
\begin{equation}
{\cal G}_{11} \simeq {\cal G}_{22} \simeq 16 \times 10^{32} \text{ GeV}^2, \qquad {\cal G}_{12} \simeq 9 \times 10^{32} \text{ GeV}^2,
\end{equation}
such that the angle $\theta$ can be approximated by the value $\theta \simeq \frac{\pi}{2} - 10^{-3}$. For this parameter choice and setting $r_1= - r_2 =1$, the respective axion decay constants in the physical basis are given by:
\begin{equation}
f_{\tilde a^-} = 1.87 \times 10^{16} \text{ GeV}, \qquad  f_{\tilde a^+}  = 7.07 \times 10^{19} \text{ GeV } \simeq 30 M_{Pl},
\end{equation}
where $M_{Pl} = (8 \pi G_N)^{-1} \sim 2.4 \times 10^{18}\, \text{GeV}$ corresponds to the reduced Planck mass.
Hence, for a sufficiently large mixing in the moduli space metric, i.e.~${\cal O}({\cal G}_{12}) \simeq {\cal O}({\cal G}_{11}, {\cal G}_{22}) $, and when both axions couple anomalously to the same non-Abelian gauge group with $|r_1| = |r_2|$, one of the physical axions can acquire a super-Planckian decay constant and a hierarchy among the axion decay constants emerges, i.e.~$f_{\tilde a^+} \gg f_{\tilde a^-}$. 

Obviously, one is inclined to contemplate whether this large axion decay constant has any chance to prevail and determine the characteristics  
of the inflationary potential such that trans-Planckian field excursions can take place during inflation. In order to answer this question, we have to expand the action around the instanton background (\ref{Eq:PotKahlerMixingPhysBasis}), by which the axions acquire their mass. The mass generating effects of instanton contributions can be captured by a cosine-type potential for the axions:
\begin{equation}
V_{axion}^{\rm eff} (\hat a^-, \hat a^+) = \Lambda^4 \left[ 1 - \cos \left( \frac{ \hat a^-}{f_{\tilde a^-}} +\frac{ \hat a^+}{f_{\tilde a^+}} \right) \right] ,
\end{equation}
where we have rescaled the axions (to operate in representation scheme 1)
\begin{eqnarray}
\hat a^+ \equiv f_{\tilde a ^+}\,\tilde a^+ ,\qquad \hat a^- \equiv f_{\tilde a ^-}\,\tilde a^+,
\end{eqnarray}
and the full lagrangian is written as:
\begin{equation}
{\cal S}_{axion} = - \bigintssss \left[  \frac{1}{2} d  \hat a^-\wedge \star_4 d  \hat a^-   + \frac{1}{2} d  \hat a^+\wedge \star_4 d  \hat a^+  + V_{axion}^{\rm eff} ( \hat a^-, \hat a^+) \star_4 \1 \right] .
\end{equation} 
One observes that the axion basis for which the kinetic terms are diagonalized does not yet correspond to the proper basis which diagonalizes the mass matrix associated to $V_{axion}^{\rm eff}$:
\begin{equation}\label{Eq:MassMatrixMetricMixing}
M^2_{ij} = \frac{\partial^2 V_{axion}^{\rm eff}  }{\partial \hat a^i \partial \hat a^j} \Bigg|_{\rm min}= \Lambda^4  \left( \begin{array}{cc} f_{\tilde a^+}^{-2} & f_{\tilde a^+}^{-1} f_{\tilde a^-}^{-1} \\ f_{\tilde a^-}^{-1} f_{\tilde a^-}^{-1} & f_{a^-}^{-2}  \end{array} \right).
\end{equation} 
One can diagonalize this mass matrix through an additional $SO(2)$ rotation:
\begin{equation}\label{Eq:TransfoDiagonalInteraction}
\left(\begin{array}{c}\xi \\ \zeta \end{array} \right) = \frac{1}{\sqrt{f_{\tilde a^+}^2 + f_{\tilde a^-}^2}}  \left( \begin{array}{cc} f_{\tilde a^+}  &  - f_{\tilde a^-} \\  f_{\tilde a^-} & f_{\tilde a^+} \end{array}\right)   \left(\begin{array}{c}  \hat a^+ \\ \hat a^- \end{array} \right),
\end{equation}
under which the full lagrangian reduces to the form,
\begin{equation}
{\cal S}_{axion} = - \bigintssss \left[  \frac{1}{2} d \xi \wedge \star_4 d \xi   + \frac{1}{2} d \zeta \wedge \star_4 d \zeta  + V_{axion}^{\rm eff} (\zeta) \star_4 \1 \right],
\end{equation}
and where the effective axion potential only depends on one of the two axions:
\begin{equation}
V_{axion}^{\rm eff} (\zeta) = \Lambda^4 \left[ 1 - \cos\left( \frac{\sqrt{f_{\tilde a^+}^2 + f_{\tilde a^-}^2}}{f_{\tilde a^+} f_{\tilde a^-}} \zeta \right)  \right].
\end{equation}
The absence of the axion $\xi$ in the potential can be traced back to the zero eigenvalue of the mass matrix $M_{ij}^2$ in (\ref{Eq:MassMatrixMetricMixing}), while the effective axion decay constant $f_{\rm eff}$,
\begin{equation}\label{f_zeta}
f_{\rm eff} =  \frac{f_{\tilde a^+} f_{\tilde a^-} }{\sqrt{f_{\tilde a^+}^2 + f_{\tilde a^-}^2}}  ,
\end{equation}
has the correct form to match the other mass eigenvalue. From this expression one can also see that the smallest of the two axion decay constants $(f_{\tilde a^+}, f_{\tilde a^-})$ sets the scale for $f_{\rm eff}$, such that the axion $\zeta$ is not allowed to undertake trans-Planckian excursions.\footnote{Observe that the shift symmetries for the original basis (\ref{Eq:shift}) translate into the desired shift symmetry for $\zeta$ and a more involved one for $\xi$ in the axion basis $(\xi, \zeta)$:
\begin{equation}
\begin{array}{lcl}
\zeta &\rightarrow& \zeta   +  2 \pi \,  f_{\rm eff} (r_1 \nu^1 + r_2 \nu^2),\\
\xi &\rightarrow& \xi + \frac{2 \pi}{\sqrt{f_{\tilde a^+}^2 + f_{\tilde a^-}^2}} \left[ \nu^1 r_1 \left( f_{\tilde a^+}^2 \cos^2 \frac{\theta}{2} - f_{\tilde a^-}^2 \sin^2 \frac{\theta}{2}   \right) + \nu^2 r_2 \left( f_{\tilde a^+}^2 \sin^2 \frac{\theta}{2} - f_{\tilde a^-}^2 \cos^2 \frac{\theta}{2}   \right)   \right] \\
&& \qquad+ \pi \sqrt{f_{\tilde a^+}^2 + f_{\tilde a^-}^2} \sin \theta \left( r_2 \nu^1 + r_1 \nu^2 \right) .
\end{array}
\end{equation}
Note however that the $\xi$-direction does not couple anomously to the non-Abelian gauge group. In this respect the axion $\xi$ corresponds to a flat direction whose shift symmetry is not broken by the envisioned gauge instanton.
}
Considering for instance the case $f_{\tilde a^+} \gg  f_{\tilde a^-}$, for which $f_{\rm eff} \simeq f_{\tilde a^-}$ obviously, one has to conclude that both axion decay constants have to be sufficiently large in order for $f_{\rm eff}$ to be trans-Planckian.
In order to write down the trans-Planckian constraints, we introduce the ratio $\varepsilon$ of the two metric eigenvalues, with
\begin{eqnarray}\label{ratio}
\varepsilon =\sqrt{\frac{\lambda _-}{\lambda _+}}.
\end{eqnarray}
Under the assumption that the largest eigenvalue $\sqrt{\lambda_+}$ lies below the reduced Planck mass, i.e.~$\sqrt{\lambda_+} \ll M_{Pl}$, both dimensionless  pre-factors in the expressions of the decay constants (\ref{Eq:metric-decay-const}) are required to be sufficiently large:
\begin{eqnarray}\label{cond}
\frac{\varepsilon}{|r_1\,\text{sin}\,\frac{\theta}{2}-r_2\,\text{cos}\,\frac{\theta}{2}|}\gg 1,\qquad \frac{1}{|r_1\,\text{cos}\,\frac{\theta}{2}+r_2\,\text{sin}\,\frac{\theta}{2}|}\gg 1,
\end{eqnarray}
in order for the effective decay constant $f_{\rm eff}$ to be super-Planckian.
For a small hierarchy between the eigenvalues of the metric (i.e.~$\varepsilon \simeq 1$), these conditions 
cannot be satisfied simultaneously.\footnote{The argumentation goes as follows: in the limit where $|r_1\,\text{sin}\,\frac{\theta}{2}-r_2\,\text{cos}\,\frac{\theta}{2}|\rightarrow 0$, one has $\text{tan}\,\frac{\theta}{2}\rightarrow \frac{r_2}{r_1}$ given that both axions couple to the instanton contribution in the original basis, i.e.~$r_i\neq 0$. This implies for the other constraint $|r_1\,\text{cos}\,\frac{\theta}{2}+r_2\,\text{sin}\,\frac{\theta}{2}|\rightarrow \left(r_1+\frac{r_2^2}{r_1}\right)\text{cos}\,\frac{\theta}{2}$. Obviously, $\text{cos}\,\frac{\theta}{2}$ cannot be arbitrarly small, otherwise the other constraint $|r_1\,\text{sin}\,\frac{\theta}{2}-r_2\,\text{cos}\,\frac{\theta}{2}|\ll 1$ cannot be satisfied.} 
And also for a large hierarchy between the eigenvalues of the metric (i.e.~$\varepsilon \ll 1$), it is not possible to satisfy both constraints simultaneously, indicating that a super-Planckian axion decay constant $f_{\rm eff}$ is excluded.
Turning the r\^oles of $f_{\tilde a^-}$ and $f_{\tilde a^+}$ around or taking both axion decay constants of the same order $f_{\tilde a^+} \simeq  f_{\tilde a^-}$  does not alter the constraints nor the argumentation. Hence, we can safely conclude $f_{\rm eff} < M_{Pl}$.

This simple two-axion model enables us to draw some interesting conclusions regarding axions and their decay constants. The expressions in equation (\ref{Eq:metric-decay-const}) suggest a splitting between the axion decay constants due to metric kinetic mixing, when the off-diagonal entries in the moduli space metric are of the same order as the diagonal ones. Nonetheless, despite the potential presence of a large axion decay constant, there is only one axionic direction $\zeta$ that couples effectively to the nonperturbative correction and the shape of its potential is set by the smallest axion decay constant eliminating the possibility of trans-Planckian displacements for the axion $\zeta$. This behavior can be awarded to the fact that axionic couplings scale inversely with the axion decay constant. 

Meanwhile, the orthogonal axionic direction $\xi$ corresponds to a flat direction whose shift symmetry remains unbroken. This observation forms the keystone for the remainder of our story. That is to say, if we interpret the axion $\xi$ as the inflaton candidate, we would have to invoke additional physical effects to create a proper inflationary potential for $\xi$. At this point, we envision three plausible and distinguishable  physical effects which could generate a potential for $\xi$ allowing for trans-Planckian displacements:    
\begin{itemize}
\item[(1)] {\it Monodromy effects:} A monomial potential of the form $V(\xi) \sim \xi^p$ can be generated through torsional monodromy effects ($p=2$)~\cite{Marchesano:2014mla} or through fluxed induced monodromies ($p\geq2$)~\cite{Marchesano:2014mla,McAllister:2014mpa}, such that the potential takes the simple chaotic inflation form ($p=2$) or even more generic forms. In order to generate a linear type of potential ($p=1$) one could also resort to D-term monodromies~\cite{Silverstein:2008sg,McAllister:2008hb}. 
\item[(2)] {\it Alignment effects:} Adding a second strongly coupled non-Abelian gauge group to which both axions (in the initial basis) couple anomalously provides for an additional mass contribution to the potential, reminiscent of the Kim-Nilles-Peloso proposal~\cite{Kim:2004rp}. This second instanton contribution is able to generate a potential for $\xi$ provided that the axion decay constants do not perfectly align. We will come back to this case in more detail in section~\ref{Sss:AlignedNaturalInflation}. 
\item[(3)] {\it Abelian $U(1)$ gauge symmetry:} A third alternative consists in adding an Abelian gauge symmetry under which both axions are charged. Due to St\"uckelberg couplings, one of the axions turns into the longitudinal component of the gauge field while the remaining axion will acquire a mass by virtue of the non-perturbative correction. We will study this scenario in detail in sections~\ref{Ss:U1mixing} and \ref{Ss:GenericKineticMixing}.
\end{itemize}

%%%%%%%%%%%%%%%%%%%%%%%%%%%%%%%%%%%%%%%%%%%%%%%%%%%%%%%%%%%%%%%%%%%%%%%%%%%%%%%%%%%%
\subsubsection{Aligned Natural Inflation}\label{Sss:AlignedNaturalInflation}
Two-axion models have already been considered in the past for inflationary purposes, but the minimal set-up given above seems to be rather suitable to realize (and generalize) the alignment mechanism~\cite{Kim:2004rp} explicitly through metric mixing. 
Though aligned natural inflation is not the main point of our paper, we make a digression here to illustrate how kinetic mixing can relax the fine-tuning needed for alignment.
To this end, we consider a two-axion system with non-trivial kinetic terms as in (\ref{Eq:KineticTermsOnlyMetricG}), uncharged under local $U(1)$ symmetries, and coupling anomalously to two distinguishable non-Abelian gauge groups:
\begin{equation}
{\cal S}^{\rm anom}_{axion} =\bigintssss\left[  \frac{1}{8\pi ^2}\left(  r_1 a^1 + r_2 a^2 \right) \text{Tr}( G^{(1)} \wedge G^{(1)}) +  \frac{1}{8\pi ^2}\left(  s_1 a^1 + s_2 a^2 \right) \text{Tr} ( G^{(2)} \wedge G^{(2)})\right] . 
\end{equation} 
Following the same steps as above to diagonalize the metric and integrating out the strongly coupled gauge sector reproduces effectively the Kim-Nilles-Peloso potential~\cite{Kim:2004rp}:
\begin{equation}
V_{axion}^{\rm eff} = \Lambda_1^4 \left[ 1 - \cos \left( \frac{\hat a^-}{f_1} + \frac{\hat a^+}{g_1}  \right) \right] + \Lambda_2^4 \left[ 1 - \cos \left( \frac{\hat a^-}{f_2} + \frac{\hat a^+}{g_2}  \right) \right],
\end{equation}
with the axion decay constants given by,
\begin{equation}
\begin{array}{l@{\hspace{0.2in}}l}\label{f and g}
\vspace{0.1in}f_1 = \frac{\sqrt{\lambda_-}}{|r_1 \sin \frac{\theta}{2} - r_2 \cos \frac{\theta}{2}|}, & g_1 = \frac{\sqrt{\lambda_+}}{|r_1 \cos \frac{\theta}{2} + r_2 \sin \frac{\theta}{2}|},\\
f_2 = \frac{\sqrt{\lambda_-}}{|s_1 \sin \frac{\theta}{2} - s_2 \cos \frac{\theta}{2}|},  & g_2 = \frac{\sqrt{\lambda_+}}{|s_1 \cos \frac{\theta}{2} + s_2 \sin \frac{\theta}{2}|}.
\end{array}
\end{equation}
The hatted axion fields $\hat a^\pm \equiv \sqrt{\lambda_\pm} a^\pm$ are introduced to rewrite the kinetic part (\ref{Eq:KineticTermsDiagonal}) in terms of representation scheme 1, in correspondence with~\cite{Kim:2004rp,Kappl:2014lra}. In the case of perfect alignment we obtain the condition:
\begin{equation}
\frac{f_1}{g_1} = \frac{f_2}{g_2} \qquad \Rightarrow \qquad \left|\frac{r_1 \cos \frac{\theta}{2} + r_2 \sin \frac{\theta}{2}}{r_1 \sin \frac{\theta}{2} - r_2 \cos \frac{\theta}{2}}\right| = \left|\frac{s_1 \cos \frac{\theta}{2} + s_2 \sin \frac{\theta}{2}}{s_1 \sin \frac{\theta}{2} - s_2 \cos \frac{\theta}{2}}\right|, 
\end{equation}   
while deviation from perfect alignment is measured~\cite{Kappl:2014lra} by the parameter $\alpha_{dev}$: 
\begin{equation}\label{dev}
\alpha_{dev} \equiv g_2 - \frac{f_2}{f_1} g_1= \frac{\sqrt{\lambda_+} \left( s_1 r_2 - r_1 s_2 \right)}{ \left( \frac{s_1^2 - s_2^2}{2} \sin \theta - s_1 s_2 \cos \theta \right) \left( r_1 \cos \frac{\theta}{2} + r_2 \sin \frac{\theta}{2} \right)}.
\end{equation}
In order for the alignment of the axionic directions to work, $\alpha_{dev}$ has to be tuned appropriately to small values (in comparison to the magnitude of the individual decay constants). In settings where metric kinetic mixing is not taken into account, one is only able to tune discrete parameters (such as $r_i$ and $s_i$), in order to fix the value of $\alpha_{dev}$, see e.g.~\cite{Ben-Dayan:2014zsa,Long:2014dta,Choi:2014rja}. However, due to kinetic metric mixing in the two-axion model an additional continuous parameter $\theta$, is at our disposal and can be used to alleviate the earlier fine-tuning issue of $\alpha_{dev}$.

 Let us consider a numerical example to clarify the previous statements. For simplicity, we assume that both scales of the nonperturbative effects are of the same order, i.e.~$\Lambda _1=\Lambda _2=\Lambda$, such that the effective axion decay constant for the almost flat direction is given by~\cite{Kappl:2014lra},
\begin{eqnarray}\label{f_align}
f_{\rm eff}=\frac{f_2\,g_1\,\sqrt{(f_{1}^2+f_{2}^2)(f_{1}^2+g_{1}^2)}}{f_{1}^2\,|\alpha _{dev}|},
\end{eqnarray}
up to leading order in $\alpha _{dev}^{-1}$. In order for the parameter $\alpha _{dev}$ in equation (\ref{dev}) to be small, we see that the integers $r_i$ and $s_i$ should make the denominator as large as possible and make the numerator as small as possible (i.e.~$|s_1\,r_2-s_2\,r_1|=1$). Let us for the sake of argument choose values for $r_i$ and $s_i$ of the order ${\cal O}(1-10)$:
\begin{eqnarray}
r_1=9,\qquad r_2=1, \qquad \qquad s_1=10,\qquad s_2=1.
\end{eqnarray}
For this parameter choice the denominator of $\alpha _{dev}$ can be at most of order $\mathcal{O}(10^3)$.  By tuning the continuous parameter $\theta$, say for instance,
\begin{eqnarray}\label{Eq:ParameterTuningTheta}
\text{sin}\,\frac{\theta}{2}\approx 0.2195, \qquad \text{cos}\,\frac{\theta}{2}\approx 0.9756,
\end{eqnarray}
we do find a sufficiently small deviation parameter (with respect to the square root of the metric eigenvalue $\lambda _+$),
\begin{eqnarray}
\alpha _{dev}\approx  0.009 \,\sqrt{\lambda _+}.
\end{eqnarray}
Moreover, the hierarchy between the eigenvalues $\lambda_+$ and $\lambda_-$ of the axion metric can be made small, given the tuned value of the continuous parameter $\theta$ in (\ref{Eq:ParameterTuningTheta}), by ensuring that the diagonal entries of the metric do not differ too much from each other, namely when ${\cal G}_{11}/{\cal G}_{22} \sim {\cal O}(1)$. 
Under these assumptions the individual axion decay constants $f_i$ and $g_i$ take the following expressions,
\begin{equation}
\begin{array}{l@{\hspace{0.4in}}l}
f_1\sim \sqrt{\lambda _+},&
f_2\sim \sqrt{\lambda _+}\times 0.8201,\\
g_1\sim  \sqrt{\lambda _+}\times 0.1111,& 
g_2\sim  \sqrt{\lambda _+}\times 0.1002.
\end{array}
\end{equation}
The eigenvalue $\sqrt{\lambda_+}$ can only take values around mass scales lower than the reduced Planck mass $M_{Pl}$.  If we consider the window $\sqrt{\lambda_+} \sim {\cal O}(10^{16}-10^{17} {\rm GeV})$, the effective axion decay constant in (\ref{f_align}) can become trans-Planckian:
\begin{eqnarray}
f_{\rm eff}&\sim& 132 \sqrt{\lambda_+}\sim 10\,M_P.
\end{eqnarray}
This numerical example shows that aligned natural inflation occurs as a consequence of metric kinetic mixing due to a non-trivial metric on the axion moduli space.   
Furthermore, for reasonable choices of the discrete parameters $(r_i, s_i)_{i=1,2}$ and a mild tuning of the continuous parameter $\theta$ the effective axion decay constant can take on super-Planckian values effortlessly.

%%%%%%%%%%%%%%%%%%%%%%%%%%%%%%%%%%%%%%%%%%%%%%%%%%%%%%%%%%%%%%%%%%%%%%%%%%%%%%%%%%%%%%%%%%%%%%%%%%%%%%%%%%%%%%%%%%%%%%%%%%%%%%%%%%%%%%%%%%%%%%%%%%%%%%%%%%%%%%%%%%%%%%%%
\subsection{$U(1)$ Kinetic Mixing}\label{Ss:U1mixing}
An alternative mechanism inducing kinetic mixing among axions relies on their potentially charged nature with respect to the same $U(1)$ gauge symmetry, as expressed by the St\"uckelberg terms in (\ref{Eq:GeneralLagrangian}). We will see in section~\ref{sec:string axions} that these St\"uckelberg couplings emerge naturally in string compactifications (with D-branes)\footnote{Such St\"uckelberg couplings can provide a portal between the Standard Model and the hidden sector \cite{Feng:2014eja,Feng:2014cla}. They are also part of the $U(1)$ lagrangian for milli-charged dark matter scenarios \cite{Shiu:2013wxa}.}, such that the relevant physical effects of this type of mixing on the axion decay constant deserve their own separate analysis. To this end, we consider a two-axion system with diagonal metric $\mathcal{G}$ (${\cal G}_{12} = 0$) and both axions charged under the same local $U(1)$ symmetry, i.e.~$k^1 \neq 0 \neq k^2$. With these assumptions the kinetic terms for the axions in (\ref{Eq:GeneralLagrangian}) read:
\begin{eqnarray}\label{Eq:kin-U(1)}
{\cal S}^{\rm kin}_{axion} &=& - \bigintssss \left[  \frac{1}{2}\, {\cal G}_{11}\, (\ud a^1 - k^1 A) \wedge \star_4 (\ud a^1 - k^1 A) +  \frac{1}{2}\, {\cal G}_{22}\, (\ud a^2 - k^2 A) \wedge \star_4 (\ud a^2 - k^2 A) \right] \notag\\
\end{eqnarray}
Given that both axions are charged under the same $U(1)$ symmetry, the axion eaten by the gauge field is a linear combination of $a^1$ and $a^2$. We can rewrite the kinetic terms as,
\begin{equation} 
{\cal S} ^{\rm kin}_{axion} = -\bigintssss\left[ \frac{ M_A^2}{2} ( \ud a'^2 - A) \wedge \star_4 ( \ud a'^2 - A) +   \frac{ M_A^2}{2} \ud a'^1 \wedge \star_4 \ud a'^1 \right]  ,
\end{equation}  
by identifying the linear combination of axions eaten by the gauge field $A_{\mu}$, as well as the mass of the gauge boson:
\begin{eqnarray}
a'^2 &=& \frac{{\cal G}_{11}\,k^1\,a^1+{\cal G}_{22}\,k^2\,a^2}{{\cal G}_{11}\,(k^1)^2 +{\cal G}_{22}\,(k^2)^2},\label{a'2}\\
M^{2}_{A}&=& {\cal G}_{11}\,(k^1)^2 + {\cal G}_{22}\,(k^2)^2\label{mass}.
\end{eqnarray}
The linear combination $a'^1$ of the axions $(a^1,a^2)$ orthogonal to the direction $a'^2$, 
\begin{eqnarray}
a'^1 = \frac{\sqrt{{\cal G}_{11} {\cal G}_{22}}(k^2\,a^1 -k^1\,a^2)}{{\cal G}_{11}\,(k^1)^2 +{\cal G}_{22}\,(k^2)^2}\label{a'1},
\end{eqnarray}
corresponds to an axion-like state uncharged under the $U(1)$ gauge symmetry. The original Abelian gauge transformation (\ref{Eq:tran})
can be recast in a gauge transformation for the eaten axion $a'^2$ (with charge $k'^2=1$), while the other $a'^1$-direction remains invariant under the gauge transformation ($k'^1=0$):
\begin{eqnarray}
A\rightarrow A+\ud \eta,\qquad  a'^2\rightarrow a'^2 + \eta, \qquad  a'^1\rightarrow a'^1.
\end{eqnarray}
The new axion basis is related to the old axion basis through a rescaling followed by an $SO(2)$ rotation:
\begin{eqnarray}\label{U(1)_tran}
\left(\begin{array}{c}a'^1 \\
a'^2\end{array}\right)=\frac{1}{M_A}\left(\begin{array}{cc}\text{cos}\, \varphi & -\text{sin}\,\varphi\\
\text{sin}\,\varphi & \text{cos}\,\varphi\end{array}\right)\left(\begin{array}{cc}\sqrt{{\cal G}_{11}} & 0\\
0 &\sqrt{{\cal G}_{22}}\end{array}\right)\left(\begin{array}{c} a^1\\
a^2\end{array}\right),
\end{eqnarray}
where we introduced the parametrization:
\begin{eqnarray}\label{Eq:U1mixingparametrisation}
\text{cos}\,\varphi = \frac{\sqrt{{\cal G}_{22}}\,k^2}{M_A},\qquad \text{sin}\, \varphi = \frac{\sqrt{{\cal G}_{11}}\,k^1}{M_A}.
\end{eqnarray}
Under the assumption that both axions couple anomalously to the same non-Abelian strongly coupled gauge group, as presented in (\ref{Eq:GeneralLagrangian}), one should carefully rewrite this topological coupling in terms of the physical axion basis $(a'^1, a'^2)$ as well: 
\begin{eqnarray}\label{anom}
{\cal S}^{\rm anom}_{axion}  
&=&   \frac{1}{8\pi ^2} \,M_A\, \bigintssss \biggl[\biggl(\frac{r_1}{\sqrt{{\cal G}_{11}}} \,\text{cos}\,\varphi -\frac{r_2}{\sqrt{{\cal G}_{22}}}\,\text{sin}\,\varphi\biggr)\,a'^1  +\biggl(\frac{r_1}{\sqrt{{\cal G}_{11}}} \,\text{sin}\,\varphi +\frac{r_2}{\sqrt{{\cal G}_{22}}}\,\text{cos}\,\varphi\biggr)\,a'^2\biggr] \,\text{Tr} ( \, G \wedge G ) \notag\\
 &=&  \frac{1}{8\pi ^2} \, \bigintssss \left[ \tilde a^1 + \tilde a^2 \right] \, \text{Tr} ( G \wedge G ) \label{Eq:PotentTermU1mixing}
\end{eqnarray}
where we opted once more to rewrite the instantonic coupling in representation scheme~2, with the axion fields $\tilde a^1$ and $\tilde a^2$ given by,
\begin{eqnarray}\label{tilde a}
\tilde{a}^1&=& M_A\biggl(\frac{r_1}{\sqrt{{\cal G}_{11}}} \,\text{cos}\,\varphi -\frac{r_2}{\sqrt{{\cal G}_{22}}}\,\text{sin}\,\varphi\biggr)\,a'^1, \notag \\
\tilde{a}^2&=&M_A\biggl(\frac{r_1}{\sqrt{{\cal G}_{11}}} \,\text{sin}\,\varphi +\frac{r_2}{\sqrt{{\cal G}_{22}}}\,\text{cos}\,\varphi\biggr)\,a'^2.
\end{eqnarray}
This allows us now to read off the effective axion decay constants for the basis ($\tilde a^1$, $\tilde a^2$), purely from the pre-factors in the kinetic terms for ($\tilde a^1$, $\tilde a^2$):\footnote{A similar comment as in footnote~\ref{Foot:GCDr1r2} on page~\pageref{Foot:GCDr1r2} is in order here: in case $\text{gcd}\,(r_1, \,r_2)\neq 1$, the axion decay constants have to be divided by $\text{gcd}\,(r_1,\,r_2)$ to account for the smallest periodicity.}
\begin{equation}\label{Eq:AxionDecayConstantU1}
\begin{array}{rcll}
f_{\tilde{a}^1}&=& \biggl(\frac{r_1}{\sqrt{{\cal G}_{11}}} \,\text{cos}\,\varphi -\frac{r_2}{\sqrt{{\cal G}_{22}}}\,\text{sin}\,\varphi\biggr)^{-1} &= \frac{\sqrt{{\cal G}_{11} {\cal G}_{22}}\,\sqrt{{\cal G}_{11}\,(k^1)^2 +{\cal G}_{22}\,(k^2)^2}}{|r_1\,k^2\,{\cal G}_{22}-r_2\,k^1\,{\cal G}_{11}|},\\ 
f_{\tilde{a}^2}&=& \biggl(\frac{r_1}{\sqrt{{\cal G}_{11}}} \,\text{sin}\,\varphi +\frac{r_2}{\sqrt{{\cal G}_{22}}}\,\text{cos}\,\varphi\biggr)^{-1}&=\frac{\sqrt{{\cal G}_{11}\,(k^1)^2+{\cal G}_{22}\,(k^2)^2}}{|r_1\,k^1+r_2\,k^2|}.
\end{array}
\end{equation}
As a consistency check we now translate the periodicity (\ref{Eq:shift}) of the axions in the original basis into a discrete shift symmetry for the physical axionic states $(\tilde{a}^1,\,\tilde{a}^2)$,
\begin{eqnarray}
\tilde{a}^1&\rightarrow & \tilde{a}^1 + \frac{2\pi\left(r_1\,{\cal G}_{22} \,k^2-r_2\,{\cal G}_{11}\,k^1\right)\left(k^2\,\nu^1-k^1\,\nu^2\right)}{M_A^2},\notag\\
\tilde{a}^2 &\rightarrow & \tilde{a}^2 +\frac{2\pi\left(r_1\,k^1+r_2\,k^2\right)\left({\cal G}_{11} \,k^1\,\nu^1+{\cal G}_{22} \,k^2\,\nu^2\right)}{M_A^2}, \label{Eq:ShiftU1mixing}
\end{eqnarray}
and verify that the anomalous couplings to the non-Abelian gauge group remain invariant under this shift symmetry:
\begin{eqnarray}
{\cal S}^{\rm anom}_{axion} \rightarrow {\cal S}^{\rm anom}_{axion} + \frac{1}{8 \pi^2} \, 2\pi\left(r_1\,\nu^1+r_2\,\nu^2\right) \bigintssss \Tr( \,G\wedge G) .
\end{eqnarray}
A similar argument as the one presented in section~\ref{Ss:MetricMixing}, based on the Pontryagin index in the path integral, is valid here to prove the invariance. Recalling that one of the axions ($\tilde a^2$) is charged under a local $U(1)$ symmetry, we notice that the potential term~(\ref{Eq:PotentTermU1mixing}) might no longer be invariant under local $U(1)$ transformations, raising questions about unitarity properties of this simple two-axion system. In the next section, we will see how this conundrum can be solved by introducing chiral fermions charged under the $U(1)$ gauge group and/or by generalized Chern-Simons terms.

Before doing so, let us first see whether a trans-Planckian axion decay constant can be realised in this setting by appropriate choices of the parameters $({\cal G}_{11}, {\cal G}_{22}, r_i, k^i)$.
Without loss of generality, we assume ${\cal G}_{11} >{\cal G}_{22}$, such that we can relate the two eigenvalues ${\cal G}_{11}$ and ${\cal G}_{22}$ through the parameter $\varepsilon$ introduced in (\ref{ratio}), which reduces in the absence of metric kinetic mixing to: 
\begin{eqnarray}
\varepsilon ^2 = \frac{{\cal G}_{22}}{{\cal G}_{11}},\,\,\,\,\,\text{with }0<\varepsilon<1.
\end{eqnarray}
Inserting this parameter into (\ref{Eq:AxionDecayConstantU1}) allows us to write the axion decay constant $f_{\tilde a^1}$ as,
\begin{eqnarray}
f_{\tilde a^1}=\frac{\varepsilon\sqrt{{\cal G}_{11}}\sqrt{(k^1)^2+\varepsilon ^2\,(k^2)^2}}{|r_1\,\varepsilon ^2\,k^2-r_2\,k^1|}.
\end{eqnarray}
Next, we assume the following relation between the integer parameters $(r_i, k^i)$, 
\begin{eqnarray}\label{condi_riki}
r_1\,k^2=r_2\,k^1,
\end{eqnarray}
such that a high amount of isotropy between the metric entries ${\cal G}_{11}$ and ${\cal G}_{22}$, i.e.~$\varepsilon \rightarrow 1^-$, can enhance the value of the axion decay constant $f_{\tilde a^1}$:
\begin{eqnarray}
f_{\tilde a^1}\sim \sqrt{{\cal G}_{11}}\,\frac{\varepsilon}{1-\varepsilon^2}\sim \frac{\sqrt{{\cal G}_{11}}}{1-\varepsilon^2}.
\end{eqnarray}
Assuming that the square root of the metric eigenvalue ${\cal G}_{11}$ (thus also the St\"uckelberg mass $M_{A}$ in (\ref{mass})) is of the order $10^{17}\,\text{GeV}$, 
and that the eigenvalue ${\cal G}_{22}$ is exactly of the same order as ${\cal G}_{11}$, say for instance $\varepsilon ^2=0.99$, we find as a numerical estimate for the decay constant $f_{\tilde a^1}$:
\begin{equation} 
f_{\tilde a^1}\sim 10^2\,\sqrt{{\cal G}_{11}}\sim 10^{19}\,\text{GeV}=10\,M_{Pl}.
\end{equation} 
Hence, a small deviation from isotropy between the eigenvalues ${\cal G}_{11}$ and ${\cal G}_{22}$ is able to induce a super-Planckian effective axion decay constant for pure $U(1)$ mixing. 

Let us also point out that to generate a super-Planckian axion decay constant the eigenvalues ${\cal G}_{11} $ and ${\cal G}_{22}$ do not need to be almost perfectly isotropic and imposing (\ref{condi_riki}) is not absolutely necessary, provided that $|r_1\,\varepsilon ^2\,k^2-r_2\,k^1|$ is sufficiently small. For example, when $r_1=k^1=r_2=1,\,k^2=2$, the axion decay constant $f_{\tilde a^1}$  is super-Planckian when the dimensionless ratio $\varepsilon ^2$ asymptotes to $\frac{1}{2}$:
\begin{eqnarray}
f_{\tilde a^1}=\frac{\varepsilon\sqrt{{\cal G}_{11}}\sqrt{ 1+4\varepsilon ^2}}{|2\varepsilon ^2-1|}\sim \frac{\sqrt{\frac{1}{2}}\sqrt{1+2} \sqrt{{\cal G}_{11}} }{10^{-2}}\sim 10^2\,\sqrt{{\cal G}_{11}}
\sim 10M_{Pl},
\end{eqnarray}
where $\varepsilon ^2=\frac{1}{2}+10^{-2}$ and $\sqrt{{\cal G}_{11}}$ is assumed to be of the order ${\cal O}(10^{17}\,\text{GeV})$. More generally, we can say that an enhancement of the axion decay constant occurs when the ratio $\varepsilon^2$ asymptotes to $r_2 k^1 / r_1 k^2$.

%%%%%%%%%%%%%%%%%%%%%%%%%%%%%%%%%%%%%%%%%%%%%%%%%%%%%%%%%%%%%%%%%%%%%%%%%%%%%%%%%%%%
\subsubsection{$U(1)$-invariance \& Generalized Chern-Simons Terms}\label{Sss:GCSterms}
Given that a super-Planckian decay constant can be achieved in this set-up, it is definitely worthwhile  to investigate the setting further and ensure that all the gauge symmetries in the system are preserved at all times.
Therefore, let us for the moment consider the subsystem consisting of the charged axion $\tilde a^2$ and the $U(1)$ gauge symmetry with one-form $A$, captured by the action,
\begin{equation}\label{Eq:SubsystempreGCS}
{\cal S}_{sub} =  \bigintssss \left[ - \frac{f_{\tilde a^2}^2}{2} \left( d\tilde a^2 - \tilde k^2 A\right) \wedge \star_4 \left( d\tilde a^2 - \tilde k^2 A\right) -\frac{1}{g_{1}^{2}} F \wedge \star_4 F  + \frac{1}{8\pi ^2} \tilde a^2  \,\text{Tr} ( \, G \wedge G ) \right], 
\end{equation}
which is manifestly invariant under the local $U(1)$ transformation except for the anomalous coupling $\tilde a^2 \, \Tr(G\wedge G)$, 
\begin{equation}\label{Eq:LocalU1shift}
A \rightarrow A + \ud \eta, \qquad  \tilde a^2 \rightarrow \tilde a^2 + \tilde k^2 \eta, 
\end{equation}
In this axion basis the charge $\tilde k^2$ is given by, 
\begin{equation}\label{Eq:tildek2}
\tilde k^2 = M_A\biggl(\frac{r_1}{\sqrt{{\cal G}_{11}}} \,\text{sin}\,\varphi +\frac{r_2}{\sqrt{{\cal G}_{22}}}\,\text{cos}\,\varphi\biggr) = r_1 k^1 + r_2 k^2.
\end{equation} 
The non-invariance of the anomalous coupling indicates the required presence of chiral fermions charged both under the $U(1)$ as well as under the non-Abelian gauge symmetry. More explicitly, if we consider a set of chiral fermions $\psi_L^i$ and $\psi_R^i$ with $i\in\{1,\ldots, n_F\}$, corresponding to the following representations under the respective gauge groups,
\begin{equation}\label{Eq:ChiralSpectrumFermions}
\begin{array}{c|@{\hspace{0.1in}}c@{\hspace{0.2in}}c}
& SU(N) & U(1)\\
\hline
\psi_L^i & R^i_1 & q^i_L\\
\psi_R^i & R^i_2 & q^i_R\\
\end{array}
\end{equation}
the fermions are chirally rotated under the local $U(1)$ gauge transformation in (\ref{Eq:LocalU1shift}). The non-invariance of the fermionic measure in the path integral under this chiral rotation then leads to an anomalous term, see appendix~\ref{A:ChiralRotations} for a brief explanation, 
\begin{equation}
\delta {\cal S}_{\rm mixed}^{anom}  =   \bigintssss  \frac{1}{8\pi ^2}  {\cal A}^{\rm mix}  \eta  \,\text{Tr} ( \, G \wedge G ), 
\end{equation}
where the anomaly coefficient ${\cal A}^{\rm mix}$ is given by,
\begin{equation}
{\cal A}^{\rm mix} =   \sum_{i=1}^{n_F} \left[ \Tr(q_L^i \{T_a^{R^i_1} , T_b^{R^i_1}  \} ) -  \Tr(q_R^i \{T_a^{\ov{R}^i_2} , T_b^{\ov{R}^i_2}  \} ) \right].
\end{equation}
The term $\delta {\cal S}_{\rm mixed}^{anom}$ is able to compensate the transformation of the anomalous coupling $\tilde a^2 \, \Tr(G\wedge G)$ under the $U(1)$ gauge symmetry, provided that the following relation holds: 
\begin{equation}\label{Eq:U(1)GaugeInvariancenoGCS}
\tilde k^2 + {\cal A}^{\rm mix} = 0.
\end{equation}

Nevertheless, in some models the anomaly coefficient $ {\cal A}^{\rm mix}$ might not suffice to compensate for the $U(1)$ gauge transformation of the anomalous coupling $\tilde a^2 \, \Tr(G\wedge G)$, in which case 
 $U(1)$ gauge invariance can be restored~\cite{Aldazabal:2002py,Andrianopoli:2004sv,Anastasopoulos:2006cz,DeRydt:2007vg} by introducing a generalized Chern-Simons term (or GCS-term) of the form:
\begin{equation}\label{Eq:GCStermU1mixing}
{\cal S}_{sub}^{\rm GCS} = - \bigintssss \frac{1}{8 \pi^2}  {\cal A}^{\rm GCS}    A \wedge \Omega,
\end{equation} 
where $\Omega$ corresponds to the Chern-Simons three-form introduced in appendix~\ref{A:Conventions}. Microscopically, such GCS-terms can be linked to the exchange of massive off-shell closed strings in Type II orientifold models with D-branes~\cite{Anastasopoulos:2006cz}, or emerge due to the presence of internal flux along the internal directions of a six-dimensional manifold suited for string theory compactifications~\cite{Aldazabal:2002py,Andrianopoli:2004sv}. In the presence of a GCS-term, $U(1)$ gauge invariance is guaranteed when the following generalization of relation (\ref{Eq:U(1)GaugeInvariancenoGCS}) is satisfied:
\begin{equation} \label{Eq:U(1)GaugeInvarianceII}
 \tilde k^2 +  {\cal A}^{\rm mix} + {\cal A}^{\rm GCS}  = 0.
\end{equation}
This relation represents only one part of the consistency conditions ensuring $U(1)$ gauge invariance, with the second consistency check played by the vanishing of the pure cubic Abelian $U(1)$ gauge anomaly:
\begin{equation}\label{Eq:U(1)GaugeInvarianceI}
{\cal A}^{U(1)^3} = \sum_{i=1}^{n_F} \left[ (q_L^i)^3 -  (q_R^i)^3  \right] = 0.
\end{equation}
Quantum consistency of the non-Abelian gauge symmetry on the other hand implies two additional constraints: the vanishing of the pure cubic non-Abelian anomaly coefficient,
\begin{equation}\label{Eq:PureNonAbelianAnomaly}
{\cal A}^{SU(N)^3} =  \sum_{i=1}^{n_F} \left[ \Tr(T_a^{R^i_1} \{T_b^{R^i_1} , T_c^{R^i_1}  \}) -  \Tr(T_a^{\ov{R}^i_2} \{T_b^{\ov{R}^i_2} , T_c^{\ov{R}^i_2}  \})  \right] = 0,
\end{equation}
and the vanishing of the mixed Abelian non-Abelian gauge anomaly,
\begin{equation}
 {\cal A}^{\rm  GCS}  - {\cal A}^{\rm mix} = 0, \label{Eq:MixedAbelianNonAbelianAnomaly}
\end{equation}
to which the GCS-term contributes as well if present. In case the mixed anomaly ${\cal A}^{mix}$ does not vanish on its own by virtue of the specific representations of the chiral fermions under the gauge groups, a consistent field theory model requires unequivocally the presence of a GCS-term. For string compactifications with D-branes, the mixed anomaly is canceled by virtue of the generalized Green-Schwarz mechanism and the GCS-term is usually not present. Section~\ref{subsec:ex} contains explicit examples in Type IIA superstring theory with intersecting D6-branes and in Type IIB superstring theory with intersecting D7-branes which do not require GCS-terms and where the cancelation of the mixed anomaly and the preservation of the $U(1)$ gauge symmetry correspond to the same constraint (\ref{Eq:U(1)GaugeInvariancenoGCS}).

Hence, by ensuring gauge invariance for our set-up we extend its field content and reconstruct the most generic lagrangian, including the GCS term, in the basis $(\tilde a^1, \tilde a^2)$: 
\begin{eqnarray}
{\cal S}_{axion}^{\rm full} &= & \bigintssss  \left[-\frac{f_{\tilde a^1}^2}{2} d\tilde a^1  \wedge \star_4  d\tilde a^1 -   \frac{f_{\tilde a^2}^2}{2} \left( d\tilde a^2 - \tilde k^2 A\right) \wedge \star_4 \left( d\tilde a^2 - \tilde k^2 A\right) -\frac{1}{g_{1}^{2}} F \wedge \star_4 F \notag \right.  \\
&&\left. \qquad -\frac{1}{g_{2}^{2}} \Tr(G\wedge \star_4 G)  + \frac{1}{8\pi ^2} \left[ \tilde a^1 + \tilde a^2\right]  \,\text{Tr} ( \, G \wedge G ) -  \frac{1}{8 \pi^2}  {\cal A}^{\rm  GCS}   A \wedge \Omega +\ldots \right], \notag \\
\end{eqnarray}
where the $\ldots$ refer to the terms involving the fermions $\psi_L^i$ and $\psi_R^i$, which will be omitted for the remainder of our story.
The remaining question at this stage concerns the shape of the inflationary potential which has to be extracted from the lagrangian ${\cal S}_{axion}^{\rm full}$. In order to answer this question we have to integrate out the massive $U(1)$ gauge field, as well as the chiral fermions charged under the non-Abelian gauge group. First of all, we adopt the unitary gauge for the gauge potential $A$:
\begin{equation}
A \longrightarrow A + \frac{1}{\tilde k^2} d\tilde a^2,
\end{equation}
such that the lagrangian can be written as:
\begin{eqnarray}\label{Eq:LagrFullU1Unitary}
{\cal S}^{\rm full, unitary}_{axion} &=& \bigintssss \left[ -  \frac{f_{\tilde a^1}^2}{2} d\tilde a^1  \wedge \star_4  d\tilde a^1    - \frac{(f_{\tilde a^2} \tilde k^2)^2}{2} A \wedge \star_4 A - \frac{1}{g_1^2} F \wedge \star_4 F -\frac{1}{g_{2}^{2}} \Tr(G\wedge \star_4 G)  \nonumber \right. \\
 &&  \qquad  + \frac{1}{8\pi^2} \tilde a^1 \Tr(G\wedge G) - \frac{1}{8 \pi^2}  {\cal A}^{\rm  GCS} A \wedge \Omega + A \wedge \star _4{\cal J}_\psi
  \notag \\
 && \left. \qquad + \frac{1}{8\pi^2} \frac{\left(\tilde k^2 +  {\cal A}^{\rm  GCS} + {\cal A}^{\rm mix}\right)}{\tilde k^2} \tilde a^2  \Tr(G\wedge G)  + \ldots
 \right].
\end{eqnarray}
The term related to the anomaly ${\cal A}^{\rm mix}$ arises through a chiral rotation of the chiral fermions, as reviewed in more detail in appendix~\ref{A:ChiralRotations}. The current ${\cal J}_\psi$ consists of the vector and axial-vector coupling of the chiral fermions to the $U(1)$ gauge potential $A$, which can be written in local (flat) coordinates as: 
\begin{equation}\label{Eq:CurrentExpression}
{\cal J}_\psi^\mu = \sum_{i} \left[ (q_L^i) \ov \psi_{L}^i \gamma^\mu   \psi_{L}^i  + (q_R^i) \ov \psi_{R}^i  \gamma^\mu  \psi_{R}^i \right].
\end{equation}
In the unitary gauge, the axion $\tilde a^2$ is eaten by the gauge potential and turns into the longitudinal component of the (massive) gauge potential $A$. By virtue of the $U(1)$ gauge invariance (\ref{Eq:U(1)GaugeInvarianceII}) the anomalous coupling of axion $\tilde a^2$ to the non-Abelian gauge group vanishes in the unitary gauge.  
Under the assumption that the energy scale at which the St\"uckelberg mechanism takes place is much higher than the scale $\Lambda$ associated to the instanton contributions of the strongly coupled gauge group, we can integrate out the Abelian gauge field $A$.
To this end, we determine its equations of motion:
\begin{equation}\label{Eq:EOMgaugefield}
-\frac{1}{g_1^2} d (\star_4 d A) - (f_{\tilde a_2} \tilde k^2)^2 \star_4 A  = \frac{ {\cal A}^{\rm GCS}}{8\pi^2} \Omega - \star_4 {\cal J}_\psi.
\end{equation}
The lefthand side corresponds to the usual Proca equation of motion for a massive gauge boson, while the righthand side can be seen as a combination of source terms. Note however that the Chern-Simons three-form and the current are related to each other through the anomalous continuity relation:
\begin{equation}\label{Eq:ABJAnomaly}
d(\star_4 {\cal J}_\psi) =  - \frac{1}{8\pi^2} {\cal A}^{\rm mix} \, d \Omega =  - \frac{1}{8\pi^2} {\cal A}^{\rm mix}\, \Tr(G\wedge G)  .
\end{equation}
The Lorenz gauge condition for $A$ follows by taking the exterior derivative at both sides of the equation of motion (\ref{Eq:EOMgaugefield}):
\begin{equation}\label{Eq:Lorenzgauge}
(f_{\tilde a^2} \tilde k^2)^2 d (\star_4 A) =  \frac{ {\cal A}^{\rm  GCS} + {\cal A}^{\rm mix}}{{\cal A}^{\rm mix}} d( \star_4 {\cal J}_\psi) .
\end{equation}
 From this Lorenz gauge condition we can extract an expression for $A$ in terms of the current ${\cal J}_\psi$ (up to a closed 1-form):
 \begin{equation}
 A =   \frac{ {\cal A}^{\rm GCS} + {\cal A}^{\rm mix}}{{\cal A}^{\rm mix}} \frac{1}{(f_{\tilde a_2} \tilde k^2)^2} {\cal J}_\psi.
 \end{equation}
 Inserting this expression back into the action (\ref{Eq:LagrFullU1Unitary}) allows us to eliminate the gauge potential $A$ in favour of the current ${\cal J}_\psi$:
 \begin{eqnarray}
{\cal S}^{\rm full, unitary}_{axion} &=& \bigintssss \left[ - \frac{f_{\tilde a^1}^2}{2} d\tilde a^1  \wedge \star_4  d\tilde a^1  + \frac{1}{8\pi^2} \tilde a^1 \Tr(G\wedge G) \right. \nonumber \\
&& \qquad \left.  + \frac{({\cal A}^{\rm  GCS} + {\cal A}^{\rm mix})^2}{2({\cal A}^{\rm mix})^2} \frac{1}{(f_{\tilde a^2} \tilde k^2)^2} {\cal J}_{\psi} \wedge \star_4  {\cal J}_{\psi} + \ldots
 \right], \label{Eq:LagrFullU1UnitaryWithoutA}
\end{eqnarray}
and we are left with one axion $\tilde a^1$, one non-Abelian gauge group and a set of chiral fermions charged under the non-Abelian gauge group. By integrating out the massive $U(1)$ gauge boson, four-point couplings among the chiral fermions emerge, suppressed by the squared mass of the gauge boson. Integrating out the chiral fermions and the non-Abelian gauge bosons, for which the procedure is briefly outlined in appendix~\ref{A:ChiralRotations}, yields a cosine-potential for the remaining axion $\tilde a^1$: 
 \begin{equation}
V_{\rm axion}(\tilde a^1) = \Lambda^4 \left[ 1 - \cos \left(\frac{\tilde a^1}{f_{\tilde a^1}}\right) \right].
\end{equation}
 which provides an explicit realisation of natural inflation with a single axion field. In appendix~\ref{A:Fermion} we propose a method to identify a proper spectrum of chiral fermions satisfying the anomaly constraints.

%%%%%%%%%%%%%%%%%%%%%%%%%%%%%%%%%%%%%%%%%%%%%%%%%%%%%%%%%%%%%%%%%%%%%%%%%%%%%%%%%%%%%%%%%%%%%%%%%%%%%%%%%%%%%%%%%%%%%%%%%%%%%%%%%%%%%%%%%%%%%%%%%%%%%%%%%%%%%%%%%%%%%%%%
\subsection{Generic Kinetic mixing}\label{Ss:GenericKineticMixing}
With the insights gathered in sections~\ref{Ss:MetricMixing} and \ref{Ss:U1mixing} we can now tackle the most generic case: a two-axion system for which the metric on the axion moduli space is non-diagonal and where both axions are charged under the same $U(1)$ gauge symmetry through St\"uckelberg-couplings. The action is given by the most general form (\ref{Eq:GeneralLagrangian}) with $N=2$ and supplemented with the generalized Chern-Simons term to ensure $U(1)$ gauge invariance:
\begin{eqnarray}\label{Eq:GeneralLagrangianWithGCS}
{\cal S}_{axion}^{\rm N=2} &= & \bigintssss  \left[ -\frac{1}{2}\, \sum_{i,j=1}^2 \cG _{ij} (\ud a^{i}-k^{i}A) \wedge \star_4 (\ud a^{j}-k^{j}A) -\frac{1}{g_{1}^{2}} F \wedge \star_4 F  -\frac{1}{g_{2}^{2}} \Tr(G\wedge \star_4 G) \notag \right.\\
&& \left. \qquad + \frac{1}{8\pi ^2}\left( r_1 a^1 +  r_2 a^2 \right) \text{Tr} ( G \wedge G ) -  \frac{1}{8 \pi^2}{\cal A}^{\rm  GCS} \,  A \wedge \Omega \right].
\end{eqnarray}
In order to determine the physical axion basis in which the axion decay constants can be read off properly, one has to combine the manipulations of the previous two sections. \\

\vspace{0.1in} 
{\it Step 1: Diagonalizing the metric $ {\cal G}_{ij}$}\\
In the first place the kinetic mixing due to a non-trivial metric on the axion moduli space has to be disengaged. To this end we use the orthogonal transformation introduced in section~\ref{Ss:MetricMixing}, under which also the charges $(k^1, k^2)$ of the axions are now transformed accordingly:
\begin{equation}
\left(\begin{array}{c} a^- \\ a^+ \end{array} \right)\label{pm}
 =  \left(  \begin{array}{cc} \sin \frac{\theta}{2} & - \cos \frac{\theta}{2} \\  \cos \frac{\theta}{2} & \sin \frac{\theta}{2}   \end{array} \right)   \left(\begin{array}{c}  a^1 \\ a^2 \end{array} \right), \qquad 
 \left(\begin{array}{c}  k^- \\ k^+ \end{array} \right) = \left(  \begin{array}{cc} \sin \frac{\theta}{2} & - \cos \frac{\theta}{2} \\  \cos \frac{\theta}{2} & \sin \frac{\theta}{2}   \end{array} \right)   \left(\begin{array}{c}  k^1 \\ k^2 \end{array} \right),
\end{equation}
with the same parameter $\theta$ defined through the parametrization (\ref{Eq:MetricMixingParametrisation}). 
By virtue of this $SO(2)$ rotation the kinetic terms for the two axions can be written in the following form:
\begin{equation}
{\cal S}_{axion}^{\rm N=2, kin} = -\bigintssss  \left[  \frac{1}{2} \lambda_-  \left(\ud a^- - k^- A \right)\wedge \star_4 \left(\ud a^- - k^- A \right)  + \frac{1}{2} \lambda_+  \left( \ud  a^+ - k^+ A \right) \wedge \star_4  \left( \ud  a^+ - k^+ A \right) \right],
\end{equation}
where the eigenvalues $\lambda_\pm$ are given by (\ref{Eq:MetricMixingMetricEigenvalues}). The kinetic terms for the gauge bosons remain unaltered by this $SO(2)$ rotation and the effects on the anomalous couplings and the generalized Chern-Simons term will be discussed at the end, once the physical axion basis has been found.

\vspace{0.1in} 
{\it Step 2: Identifying the eaten axion direction}\\
In the next step we identify the linear combination of axions $(a^-, a^+)$ eaten by the $U(1)$ gauge field $A$. To this end, we rescale the axions and their respective charges:
\begin{equation}
\begin{array}{l@{\hspace{0.2in}}l}\label{tilde}
a^- \rightarrow \tilde a^- \equiv M_{st}^{-1}\,\sqrt{\lambda_-} a^-, & a^+ \rightarrow \tilde a^+ \equiv M_{st}^{-1}\,\sqrt{\lambda_+} a^+,\\
k^- \rightarrow \tilde k^- \equiv M_{st}^{-1}\, \sqrt{\lambda_-} k^-, & k^+ \rightarrow \tilde k^+ \equiv M_{st}^{-1}\,\sqrt{\lambda_+} k^+,
\end{array}
\end{equation}
where $M_{st}$ is the St\"uckelberg mass of the $U(1)$ gauge boson:
\begin{eqnarray}\label{Eq:StuckMassFullMixing}
M^2_{st} = \lambda _- \,(k^-)^2 + \lambda _+\,(k^+)^2.
\end{eqnarray}
Next, we use a similar parametrization as the one introduced in (\ref{Eq:U1mixingparametrisation}) from section~\ref{Ss:U1mixing}:
\begin{equation}\label{varphi}
\tilde k^- =  \cos \varphi, \qquad \tilde k^+ = \sin \varphi. 
\end{equation}
This parametrization allows us to perform the $SO(2)$ transformation on the axion fields in order to extract the physical axion basis:
\begin{equation}
\left(\begin{array}{c} \zeta \\ \xi \end{array}\right) = \left(\begin{array}{cc} \cos \varphi & \sin \varphi \\ - \sin \varphi & \cos \varphi \end{array}\right)  \left(\begin{array}{c}  \tilde a^- \\ \tilde a^+\end{array}\right),
\end{equation}
where $\zeta$ corresponds to the axion eaten by the $U(1)$ gauge field, and $\xi$ to the orthogonal direction. The resulting kinetic terms for the axions read in the basis $(\zeta, \xi)$:
\begin{equation}
{\cal S}_{axion}^{\rm N=2, kin} = - \bigintssss  \left[  \frac{1}{2} \,M_{st}^2\,\left( \ud \zeta -  A \right)\wedge \star_4 \left( \ud \zeta - A \right)  + \frac{1}{2}\,M_{st}^2\, \ud \xi \wedge \star_4 \ud \xi \right].
\end{equation}
From the original gauge symmetry (\ref{Eq:tran}),
we can deduce that the gauge symmetry in the physical axion basis $(\zeta, \xi)$ can be expressed as:
\begin{equation}
A \rightarrow A + \ud \eta, \qquad \zeta \rightarrow \zeta + \eta, \qquad \xi \rightarrow \xi.
\end{equation}
In summary, the physical axion basis $(\zeta, \xi)$, in which the kinetic terms take a diagonal form and the axion eaten in the St\"uckelberg mechanism can be identified unambiguously, relates to the original basis $(a^1,a^2)$ through a combination of $SO(2)$ rotations and a rescaling:
\begin{equation}\label{Eq:PhysOriginalBasis}
\left(\begin{array}{c} \zeta \\ \xi \end{array}\right) = M_{st}^{-1} \left(\begin{array}{cc} \cos \varphi &  \sin \varphi \\ -  \sin \varphi &   \cos \varphi \end{array}\right) \left(  \begin{array}{cc}  \sqrt{\lambda_-} & 0 \\0 & \sqrt{\lambda_+}   \end{array}\right) \left(  \begin{array}{cc} \sin \frac{\theta}{2} & - \cos \frac{\theta}{2} \\  \cos \frac{\theta}{2} & \sin \frac{\theta}{2}   \end{array} \right)   \left(\begin{array}{c}  a^1 \\ a^2 \end{array} \right).
\end{equation}

\vspace{0.1in} 
{\it Step 3: Rewriting the anomalous couplings in the physical basis}\\
Now that we have identified the physical basis, it is time to express the anomalous couplings and generalized Chern-Simons term in term of the basis $(\zeta, \xi)$. To this end, we invert the set of transformations in equation (\ref{Eq:PhysOriginalBasis}):
\begin{equation}
\left(\begin{array}{c}  a^1 \\ a^2 \end{array} \right) =  M_{st}\left(  \begin{array}{cc} \sin \frac{\theta}{2} &  \cos \frac{\theta}{2} \\  -\cos \frac{\theta}{2} & \sin \frac{\theta}{2}   \end{array} \right)  \left(  \begin{array}{cc}  \frac{1}{\sqrt{\lambda_-}} & 0 \\0 &\frac{1}{\sqrt{\lambda_+}}   \end{array}\right)    \left(\begin{array}{cc}  \cos \varphi & - \sin \varphi \\    \sin \varphi &  \cos \varphi \end{array}\right) \left(\begin{array}{c} \zeta \\ \xi \end{array}\right),
\end{equation}
and plug  these expressions back into the anomalous couplings and the generalized Chern-Simons terms:
\begin{eqnarray}
{\cal S}^{\rm anom}_{axion} &=&  \frac{1}{8\pi ^2} \bigintssss \left[ \Tr (G\wedge G) M_{st} \left[ \frac{\zeta}{f_{\tilde \zeta}} + \frac{\xi}{f_{\tilde \xi}}  \right]  -  {\cal A}^{\rm  GCS}  A \wedge \Omega \right]  \notag  \\
&=&   \frac{1}{8\pi ^2} \bigintssss \left[ \left[ \tilde \zeta + \tilde \xi \right]  \,\text{Tr} ( \, G \wedge G ) - \frac{1}{8 \pi^2}  {\cal A}^{\rm  GCS}   A \wedge \Omega \right],
\end{eqnarray} 
where the second equality results from a rescaling of the axions such that we can read off the axion decay constants in representation scheme 2 for the rescaled physical basis~$(\tilde \zeta, \tilde \xi)$:
\begin{equation}
\begin{array}{rcl}
f_{\tilde \zeta} &= &\frac{\sqrt{\lambda_+ \lambda_-}}{\left|\sqrt{\lambda_+}  \cos \varphi \left( \sin \frac{\theta}{2}  \,r_1 - \cos \frac{\theta}{2}\, r_2 \right) + \sqrt{\lambda_-} \sin \varphi \left( \cos \frac{\theta}{2}  \,r_1  + \sin \frac{\theta}{2} \,r_2 \right) \right|},\\ \label{f_xi}
f_{\tilde \xi} &= &\frac{\sqrt{\lambda_+ \lambda_-}}{\left|\sqrt{\lambda_+}  \sin \varphi \left( \cos \frac{\theta}{2}  \,r_2 - \sin \frac{\theta}{2} \,r_1 \right) + \sqrt{\lambda_-} \cos \varphi \left( \cos \frac{\theta}{2}\,  r_1 + \sin \frac{\theta}{2} \,r_2 \right) \right|}.
\end{array}
\end{equation}
With respect to the physical basis $(\tilde \zeta, \tilde \xi)$ the lagrangian for the full two-axion system can now be written as:
\begin{eqnarray}\label{Eq:GeneralLagrangianWithGCSII}
{\cal S}_{axion}^{\rm N=2} &= & \bigintssss \left[ -\frac{f_{\tilde \zeta}^2}{2} \left( \ud \tilde \zeta - k_{\tilde \zeta}  A \right)\wedge \star_4 \left( \ud \tilde \zeta - k_{\tilde \zeta} A \right)  - \frac{f_{\tilde \xi}^2}{2} \ud  \tilde\xi \wedge \star_4 \ud  \tilde\xi -\frac{1}{g_{1}^{2}} F \wedge \star_4 F  \notag \right. \\
&&\left. \qquad  -\frac{1}{g_{2}^{2}} \Tr(G\wedge \star_4 G) + \frac{1}{8\pi ^2}\left( \tilde \zeta + \tilde \xi \right) \text{Tr} ( G \wedge G ) -  \frac{1}{8 \pi^2} {\cal A}^{\rm  GCS}  \,  A \wedge \Omega \right].
\end{eqnarray}
Also here we wonder how the periodicity (\ref{Eq:shift}) in the original axion basis translates into a discrete shift symmetry for the physical axion basis $(\zeta, \xi)$:
\begin{equation}
\begin{array}{lcl}
\zeta &\rightarrow& \zeta +  2\pi\,\frac{\left(\sqrt{\lambda _-}\,\text{cos}\,\varphi\,\text{sin}\,\frac{\theta}{2}+\sqrt{\lambda _+}\,\text{sin}\,\varphi\,\text{cos}\,\frac{\theta}{2}\right)\nu^1+\left(-\sqrt{\lambda _-}\,\text{cos}\,\varphi\,\text{cos}\,\frac{\theta}{2}+\sqrt{\lambda _+}\,\text{sin}\,\varphi\,\text{sin}\,\frac{\theta}{2}\right)\nu^2}{M_{st}},\\
\xi &\rightarrow& \xi + 2\pi\,\frac{\left(-\sqrt{\lambda _-}\,\text{sin}\,\varphi\,\text{sin}\,\frac{\theta}{2}+\sqrt{\lambda _+}\,\text{cos}\,\varphi\,\text{cos}\,\frac{\theta}{2}\right)\nu^1+\left(\sqrt{\lambda _-}\,\text{sin}\,\varphi\,\text{cos}\,\frac{\theta}{2}+\sqrt{\lambda _+}\,\text{cos}\,\varphi\,\text{sin}\,\frac{\theta}{2}\right)\nu^2 }{M_{st}},
\end{array}
\end{equation}
and investigate how the anomalous couplings to the non-Abelian gauge group transforms under such a shift:
\begin{equation}
\Delta {\cal S}_{anom} =  \frac{1}{8 \pi^2} 2\pi\left(r_1\,\nu^1+r_2\,\nu^2\right)
 \bigintssss \, \Tr (G\wedge G),
 \end{equation}
with other terms cancelling each other out. In this computation we explicitly included the anomalous coupling for the eaten axion as well.

The global consistency of this model requires the introduction of chiral fermions charged under the non-Abelian gauge group and the $U(1)$ gauge group, analogous to the discussion in section~\ref{Sss:GCSterms}. Non-Abelian gauge invariance is guaranteed provided that the anomaly conditions (\ref{Eq:PureNonAbelianAnomaly}) and (\ref{Eq:MixedAbelianNonAbelianAnomaly}) are satisfied, while the vanishing of the $U(1)$ anomalies is secured by conditions (\ref{Eq:U(1)GaugeInvarianceI}) and (\ref{Eq:U(1)GaugeInvarianceII}), upon replacing the $U(1)$ charge $\tilde k_2$ with the $U(1)$ charge $k_{\tilde \zeta} = M_{st} f^{-1}_{\tilde \zeta}$. 
Regarding the massive $U(1)$ boson, we can repeat the same reasoning as in section~\ref{Sss:GCSterms} and integrate out the gauge potential $A$ in favour of the current ${\cal J}_\psi$. Upon integrating out the massive $U(1)$ gauge boson, we are left with the axion $\tilde \xi$ coupling anomalously to the non-Abelian gauge theory. By integrating out the heavy fermions and the non-Abelian gauge bosons we are left with a cosine-potential for the remaining axion $\tilde \xi$, which is interpreted at the end of the road as the inflaton.

\begin{figure}[h]
\begin{center}
\includegraphics[width=5cm,height=5cm]{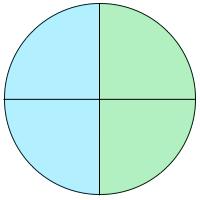} \begin{picture}(0,0) 
\put(0,70){\color{mygr}$0<\Sigma^2 \leq 1$}  
\put(-185,70){\color{myblue}$\Sigma^2 \geq 1$} 
\put(-65,90){${\cal G}_{12} > 0$} \put(-65,40){${\cal G}_{12} < 0$}
\put(-125,90){${\cal G}_{12} > 0$} \put(-125,40){${\cal G}_{12} < 0$}
\put(-85,145){$\pi/2$}\put(-95,-10){$-\pi/2$}
 \end{picture}
\caption{Each quadrant of the unit circle corresponds to a region in the parameter space $(\theta,\Sigma)$, depending on the sign of metric entry ${\cal G}_{12}$ and the relative magnitude between ${\cal G}_{11}$ and ${\cal G}_{22}$, namely $\Sigma^2 \leq 1$ or $\Sigma^2 \geq 1$.\label{Fig:ParamQuadr}}
\end{center}
\end{figure}
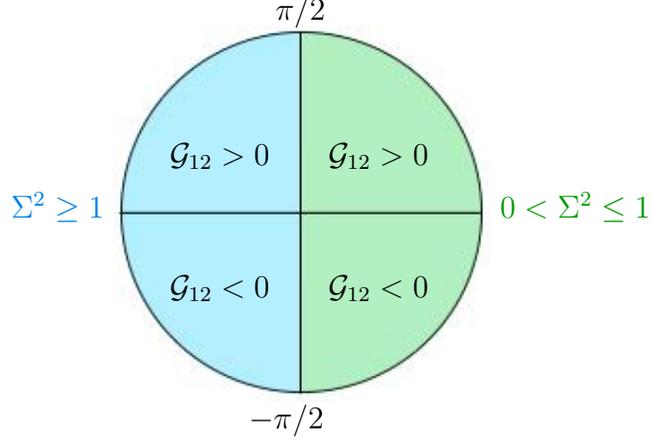

\begin{figure}[h]
\begin{center}
\begin{tabular}{c@{\hspace{0.8in}}c}
\includegraphics[width=5cm, height=5cm]{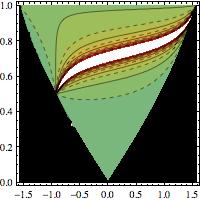} \begin{picture}(0,0) \put(-70,-10){$\theta$}  \put(-165,70){$\Sigma$} \end{picture}  & \includegraphics[width=5cm, height=5cm]{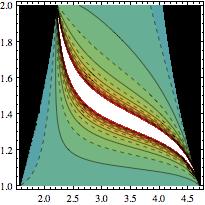} \begin{picture}(0,0) \put(-70,-10){$\theta$}  \put(-165,70){$\Sigma$} \end{picture} 
\end{tabular}
\caption{Contour plot for example 1 with discrete parameters $2 k^1 = k^2 = 2 r_1 = 2 r_2$, and with $0<\Sigma\leq 1$ (left) or $\Sigma\geq1$ (right). Black regions correspond to unphysical values for $f_{\tilde \xi}$, while the physical values follow the color-coding from small (green) to large (red).     \label{Fig:ContEx1}}
\end{center}
\end{figure}

\begin{figure}[h]
\begin{center}
\begin{tabular}{c@{\hspace{0.8in}}c}
\includegraphics[width=5cm, height=5cm]{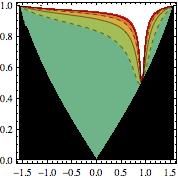} \begin{picture}(0,0) \put(-70,-10){$\theta$}  \put(-165,70){$\Sigma$} \end{picture}  & \includegraphics[width=4.9cm, height=4.9cm]{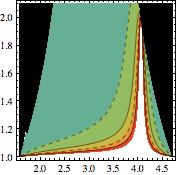} \begin{picture}(0,0) \put(-70,-10){$\theta$}  \put(-165,70){$\Sigma$} \end{picture} 
\end{tabular}
\caption{Contour plot for example 2 with discrete parameters $ k^1 = 2 k^2 =  r_1 = 2 r_2$, and with $0<\Sigma\leq 1$ (left) or $\Sigma\geq1$ (right). Black regions correspond to unphysical values for $f_{\tilde \xi}$, while the physical values follow the color-coding from small (green) to large (red).   \label{Fig:ContEx2}}
\end{center}
\end{figure}

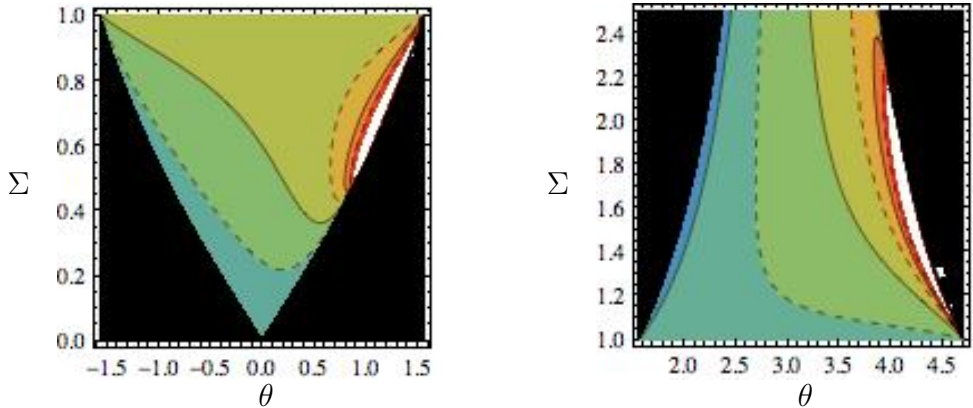
\begin{figure}[h]
\begin{center}
\begin{tabular}{c@{\hspace{0.8in}}c}
\includegraphics[width=5cm, height=5cm]{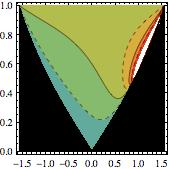} \begin{picture}(0,0) \put(-70,-10){$\theta$}  \put(-165,70){$\Sigma$} \end{picture}  & \includegraphics[width=5cm, height=5cm]{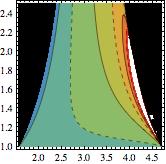} \begin{picture}(0,0) \put(-70,-10){$\theta$}  \put(-165,70){$\Sigma$} \end{picture} 
\end{tabular}
\caption{Contour plot for example 3 with discrete parameters $k^1 = -2 k^2 = r_1 = 2 r_2$, and with $0<\Sigma\leq 1$ (left) or $\Sigma\geq1$ (right). Black regions correspond to unphysical values for $f_{\tilde \xi}$, while the physical values follow the color-coding from small (green) to large (red).   \label{Fig:ContEx3}}
\end{center}
\end{figure}

We end this section by exploring the physical excursion range of this inflaton-axion by virtue of a closer investigation of the axion decay constant $f_{\tilde \xi}$ in equation~(\ref{f_xi}). First intuition regarding the range of this decay constant can be obtained through a numerical examination of the expression in (\ref{f_xi}). To this end, we assume that the larger eigenvalue $\sqrt{\lambda _+}$ of the axion metric (and thus also the U(1) St\"uckelberg mass $M_{st}$) takes values around an energy scale of the order $10^{17}\,\text{GeV}$:
\begin{equation}
\sqrt{\lambda _+} \sim {\cal O}(10^{16}-10^{17})\,\text{GeV}.
\end{equation} 
Then from equations (\ref{pm}), (\ref{tilde}) and (\ref{varphi}) we can deduce the expressions:
\begin{eqnarray}
\text{cos}\,\varphi \sim \varepsilon\left(\text{sin}\,\frac{\theta}{2}\,k^1-\text{cos}\,\frac{\theta}{2}\,k^2\right), \qquad \text{sin}\,\varphi \sim \text{cos}\,\frac{\theta}{2}\,k^1+\text{sin}\,\frac{\theta}{2}\,k^2,
\end{eqnarray}
such that the decay constant $f_{\tilde \xi}$ can be written as,
\begin{equation}\label{f}
f_{\tilde \xi}\sim \frac{\varepsilon \sqrt{\lambda _+}}{\left|\left(\text{cos}\,\frac{\theta}{2}\,k^1+\text{sin}\,\frac{\theta}{2}\,k^2\right)\left(\text{cos}\,\frac{\theta}{2}\,r_2 -\text{sin}\,\frac{\theta}{2}\,r_1\right)+\varepsilon ^2\left(\text{sin}\,\frac{\theta}{2}\,k^1-\text{cos}\,\frac{\theta}{2}\,k^2\right)\left(\text{cos}\,\frac{\theta}{2}\,r_1+\text{sin}\,\frac{\theta}{2}\,r_2\right)\right|},
\end{equation}
where $\varepsilon ^2$ is the ratio between the smaller and the larger eigenvalues of the metric in the axion space as defined in (\ref{ratio}). Based on this expression for the decay constant, we can already discover two regions in the axion moduli where the axion decay constant can enhance to super-Planckian values, namely $\theta = \frac{\pi}{2}$ and $\theta = 0$.
\begin{itemize}
\item[]{\bf Region 1 ($\theta = \frac{\pi}{2}$):}
It is easy to check that for the following choice of discrete parameters:
\begin{eqnarray}\label{wrapping}
r_1=r_2\sim \mathcal{O}(1),\qquad k^1=-k^2\sim \mathcal{O}(1),
\end{eqnarray}
the decay constant reduces to the following simple expression:
\begin{eqnarray}
f_{\tilde \xi}\sim \frac{\varepsilon \,\sqrt{\lambda _+} }{\left(\text{cos}\,\frac{\theta}{2}-\text{sin}\,\frac{\theta}{2}\right)^2+\varepsilon ^2\left(\text{sin}\,\frac{\theta}{2}+\text{cos}\,\frac{\theta}{2}\right)^2}= \frac{\varepsilon \,\sqrt{\lambda _+} }{1+\varepsilon ^2-(1-\varepsilon ^2)\,\text{sin}\,\theta},
\end{eqnarray}
which can grow larger than $\sqrt{\lambda _+} $ when $\varepsilon$ is small enough and $\theta$ asymptotes to $\frac{\pi}{2}$. Indeed when $\theta =\frac{\pi}{2}$, the decay constant scales as,
\begin{eqnarray}
f_{\tilde \xi}\sim \frac{\sqrt{\lambda _+} }{2\varepsilon }.
\end{eqnarray}
If there is a hierarchy between the two eigenvalues $\lambda_+$ and $\lambda_-$, say $\varepsilon \sim {\cal O}(10^{-2})$, then the axion decay constant for $\tilde \xi$ can become super-Planckian, i.e.~$f_{\tilde \xi}\sim 10^2\,\sqrt{\lambda _+}  \sim 10\,M_{Pl}$. Expressed in terms of the entries of the axion moduli space metric, 
\begin{eqnarray}\label{alpha}
{\cal G}_{11} &=&\frac{\lambda _+}{2}\left[1+\frac{\lambda _-}{\lambda _+}+\left(1-\frac{\lambda _-}{\lambda _+}\right)\text{cos}\,\theta\right]\sim \frac{\lambda _+}{2} ,\\ \label{gamma}
{\cal G}_{22} &=&\frac{\lambda _+}{2}\left[1+\frac{\lambda _-}{\lambda _+}-\left(1-\frac{\lambda _-}{\lambda _+}\right)\text{cos}\,\theta\right]\sim\frac{\lambda _+}{2} ,\\ \label{beta}
{\cal G}_{12} &=& \frac{\lambda _+}{2}\left(1-\frac{\lambda _-}{\lambda _+}\right)\text{sin}\,\theta\sim \frac{\lambda _+}{2},
\end{eqnarray}
a hierarchy $\lambda_- \ll \lambda_+$ among the eigenvalues translates into a configuration with large metric mixing. Hence, if the off-diagonal entries are of the same order as the diagonal ones in the metric on the axion moduli space, and the discrete parameters satisfy the relation (\ref{wrapping}), the  decay constant  $f_{\tilde \xi}$ for the axion $\tilde \xi$, orthogonal to the axionic direction devoured by the $U(1)$ gauge boson, becomes trans-Planckian.

\item[]{\bf Region 2 ($\theta=0$):}
Also for configurations where there is a small or no hierarchy between the eigenvalues (i.e.~$\varepsilon \rightarrow 1^-$), one can locate regions of isotropy in the parameter space where the axion decay constant takes super-Planckian values. To see this more explicity, let us rewrite the denominator of (\ref{f}) as follows,
\begin{eqnarray}
\text{Denominator}&=&\left|\text{cos}^2\frac{\theta}{2}\,k^1\,r_2-\frac{\sin \theta}{2}\,\left(k^1\,r_1-k^2\,r_2\right)-\text{sin}^2\frac{\theta}{2}\,k^2\,r_1 \right.\notag\\
&& \qquad +\left. \varepsilon ^2 \left[\frac{\sin \theta}{2}  \left(k^1\,r_1-k^2\,r_2\right)-\text{cos}^2\frac{\theta}{2}\,k^2\,r_1+\text{sin}^2\frac{\theta}{2}\,k^1\,r_2\right]\right|\notag\\
&=&\left| (1-\varepsilon ^2) \left[ -\frac{\text{sin}\,\theta}{2}\left(k^1\,r_1-k^2\,r_2\right)+k^2\,r_1\,\text{cos}^2\frac{\theta}{2} -k^1\,r_2\,\text{sin}^2\frac{\theta}{2} \right] \right. \nonumber\\
 &&  \qquad \qquad \qquad\left.  +  ( k^1 r_2 - k^2 r_1) \phantom{\left[ -\frac{\text{sin}\,\theta}{2}\left(k^1\,r_1-k^2\,r_2\right)+k^2\,r_1\,\text{cos}^2\frac{\theta}{2} -k^1\,r_2\,\text{sin}^2\frac{\theta}{2} \right] } \hspace{-3.5in} \right|. 
\end{eqnarray}
Note that for a small amount of metric kinetic mixing, we are located in a region of the parameter space where  the angle $\theta$ asymptotes to 0, such that in that limit the denominator can be approximated by,
\begin{eqnarray}
&&\left| (1-\varepsilon ^2)  \left[-\frac{\text{sin}\,\theta}{2}\left(k^1\,r_1-k^2\,r_2\right)+k^2\,r_1\,\text{cos}^2\frac{\theta}{2}-k^1\,r_2\,\text{sin}^2\frac{\theta}{2} \right] + k^1 r_2 - k^2 r_1\right| \nonumber \\
&& \qquad \qquad \stackrel{\theta \rightarrow 0}{\longrightarrow} \left| k^1 r_2 - k^2 r_1 \varepsilon^2 \right|.
\end{eqnarray} 
Assuming $k^2 r_1 \sim {\cal O}(k^1 r_2)$ and that the parameter $k^1$ and $r_2$ are not monstrously large, the axion decay constant in~(\ref{f}) thus scales roughly as,
\begin{eqnarray}
f_{\tilde \xi}\sim \frac{\varepsilon \, \sqrt{\lambda _+}}{1-\varepsilon ^2}.
\end{eqnarray} 
From this estimate one notices that the decay constant $f$ reaches large values in the limit $\varepsilon \rightarrow 1^-$. For instance, if the parameter $\varepsilon \approx 0.995$, we find a trans-Planckian decay constant:
$ f_{\tilde \xi}\sim 10^2\,\sqrt{\lambda _+}  \sim 10\,M_{Pl}$.
Naturally, this region of the parameter space resembles the case discussed in section~\ref{Ss:U1mixing}, where no metric mixing occurs at all.
\end{itemize}
To investigate regions of the moduli space where metric mixing occurs and the non-diagonal entries are not of the same order as the diagonal entries (like in region 1), we have to adopt a different strategy. In the first place, we exchange the $\varphi$-parametrization for the charges $(k^-,k^+)$ in the axion decay constant:
\begin{equation}\label{Eq:NewADCFullMixing}
f_{\tilde \xi} = \frac{\sqrt{\lambda_+ \lambda_-} M_{st}}{ \cos \frac{\theta}{2} \left(  \lambda_+  k^+  r_2  + \lambda_- k^- r_1 \right) + \sin \frac{\theta}{2}  \left( \lambda_- k^- r_2 - \lambda_+  k^+  r_1  \right)}.
\end{equation}
Through the expressions~(\ref{pm}) the charges $(k^-,k^+)$ are given in terms of the angle $\theta$ and the original discrete charges $(k^1,k^2)$, implying that the St\"uckelberg mass depends on these parameters as well by virtue of equation~(\ref{Eq:StuckMassFullMixing}). Recall that the continuous parameter $\theta$ measures the amount of metric kinetic mixing through the parametrization:
\begin{equation}
\cos \theta =  {\cal G}_{11} \frac{1 - \Sigma^2 }{\lambda_+ - \lambda_-}, \qquad \sin \theta = \frac{2 {\cal G}_{12}}{\lambda_+ - \lambda_-},
\end{equation}
where we introduced the ratio $\Sigma^2 = {\cal G}_{22}/{\cal G}_{11}$ to measure the relative magnitude between the diagonal entries of the axion metric ${\cal G}_{ij}$. Through this parametrization the eigenvalues $\lambda_+$ and $\lambda_-$ can be written in terms of the continuous parameters $\theta$, $\Sigma^2$ and ${\cal G}_{11}$. Upon fixing the discrete parameters $r_i$ and $k^i$ the axion decay constant~(\ref{Eq:NewADCFullMixing}) in units of~$\sqrt{{\cal G}_{11}}$ can be represented through a two-dimensional contour plot spanned by $\Sigma$ and $\theta$. Based on the sign of the non-diagonal metric entry ${\cal G}_{12}$ and the value of $\Sigma$ we can distinguish four different regions in the parameter space $(\Sigma, \theta)$ and assign to each of them a quadrant of a unit circle as depicted in figure~\ref{Fig:ParamQuadr}. In the two-dimensional plot of the parameter space $(\theta, \Sigma)$ one can depict two quadrants simultaneously, and we have chosen to differentiate the regions in the parameter space based on the values of the ratio $\Sigma^2$: $0<\Sigma^2\leq1$ (green in the unit circle) or $\Sigma^2\geq 1$ (blue in the unit circle). We consider three different examples, distinguishable from each other by the relation among the discrete parameters $(r_i,k^i)$. The contour plots for the three different examples are given in figures~\ref{Fig:ContEx1},~\ref{Fig:ContEx2} and~\ref{Fig:ContEx3} respectively. The black areas in these figures correspond to unphysical regions with a complex decay constant. Physical values for $f_{\tilde \xi}$-magnitude follow the color-coding: small (green) to large (red). The white bands denote the region where the axion decay constant enhances to $f_{\tilde \xi} \geq {\cal O}(20-30) \sqrt{{\cal G}_{11}}$. The shape and position of these white bands in the $(\theta, \Sigma)$-plane clearly depends on the relation among the discrete parameters $(r_i,k^i)$ and the sign of ${\cal G}_{12}$. In their center one can locate regions in the moduli space where trans-Planckian axion decay constants $f_{\tilde \xi}$ are possibly realised, depending on the scale of $\sqrt{{\cal G}_{11}}$.

\clearpage

%%%%%
%%%%%
%%%%%

\section{Implementation in String Theory}\label{sec:string axions}
%%%%%
%%%%%
As already indicated in the introduction, a proper understanding of quantum (gravitational) corrections are crucial to probe the viability and sustainability of any inflationary model. The next chapter in our story of kinetically mixing axions thus consists in embedding the proposed mechanisms of section~\ref{sec:mixing} into string theory, the best framework up-to-date for computing quantum corrections to inflationary models, for recent reviews see e.g.~\cite{McAllister:2007bg,Baumann:2009ni,Cicoli:2011zz,Burgess:2011fa,Baumann:2014nda}. Furthermore, it is also well known~\cite{Witten:1984dg,Barr:1985hk,Choi:1985je,Banks:2003sx,Svrcek:2006yi} by now that axions arise abundantly through compactifications of superstring theories to four spacetime dimensions. In the framework of string inflation,  the axions that are mostly used as candidates for the inflaton field emerge from the dimensional reduction of $p$-forms appearing in the NS-NS and RR-sector, the so-called closed string axions.\footnote{In the case of type II superstring compactifications with D-branes, one can also identify two types of open string axions: the Wilson-line arising from the dimensional reduction of the gauge field living on the D-brane world-volume, and the phase of a complex scalar field within a chiral multiplet located at the intersection of two separate D-branes. We will not discuss open string axions any further, as we will not work with them in this section.} 

In this section we review some well-known facts about Type II compactifications, which will allow us to argue for the string embedding of the ideas presented in section~\ref{sec:mixing}. An important aspect of this review concerns the origin of the closed string axions in Type II compactifications and how their effective action in four dimensions can be spelled out. Observe that our assumptions about the internal manifold for the dimensional reduction are reduced to a minimum to emphasize the generic character of the effective four dimensional action. Last but not least, we end this section by presenting explicit Type II models.

%%%%%%%%%%%%%%%%%%%%%%%%%%%%%%%%%%%%%%%%%%%%%%%%%%%%%%%%%%%%%%%%%%%%%%%%%%%%%%%%%%%%%%%%%%%%%%%%%%%%%%%%%%%%%%%%%%%%%%%%%%%%%%%%%%%%%%%%%%%%%%%%%%%%%%%%%%%%%%%%%%%%%%%%
\subsection{General Observations for Type II Compactifications}
In a first phase we review briefly how closed string axions and the related effective lagrangian in (\ref{Eq:GeneralLagrangian}) arise naturally from the dimensional reduction of type II superstring theory compactified on the product spacetime ${\cal M}_{1,3} \times {\cal X}_6$, where ${\cal M}_{1,3}$ corresponds to a maximally symmetric four dimensional spacetime and ${\cal X}_6$ to a six dimensional internal manifold. Under these assumptions the ten dimensional metric can be factorized as,
\begin{equation}\label{Eq:metricAnsatz}
ds_{10}^2 = \eta_{\mu \nu} (x) dx^\mu dx^\nu + g_{ab} (y) dy^a dy^b,
\end{equation}
where $\{x^\mu\}$ represent the local coordinates of ${\cal M}_{1,3}$ and the coordinates $\{y^a\}$ parametrise the compact manifold ${\cal X}_6$ with metric $g_{ab}$.

\noindent Recall~\cite{Polchinski:1998rr} that the low energy effective action for the ten dimensional massless bosonic string modes of type II superstring theory  is given in the string frame by,
\begin{eqnarray}
&{\cal S}_{\rm bulk} &= {\cal S}_{NS} + {\cal S}_{R}, \notag\\ 
&{\cal S}_{NS}& =  \frac{1}{2 \kappa_{10}^2} \int_{{\cal M}_{1,3}\times {\cal X}_6}  e^{-2 \Phi} \left[ R \star_{10}\1 + 4 d \Phi \wedge \star_{10} d\Phi - \frac{1}{2} H_3 \wedge \star_{10} H_3 \right]  \\
&{\cal S}_{R}& =-  \frac{1}{8 \kappa_{10}^2} \int_{{\cal M}_{1,3}\times {\cal X}_6} \sum_{p } G_{2p} \wedge \star_{10} G_{2p},  \label{Eq:BulkRR}
\qquad \begin{array}{ll}
\text{Type IIA:} & p =1, \ldots, 4\\
\text{Type IIB:} & p =1/2, \ldots, 9/2
\end{array}
\end{eqnarray} 
where we opted for the democratic formulation~\cite{Bergshoeff:2001pv} to express the action for the massless RR-modes. The ten dimensional gravitational coupling $\kappa_{10}$ is related to the string scale $\ell_s$ as expressed in equation (\ref{Eq:10dGravStringLength}). Besides the ten dimensional Einstein-Hilbert term ($R\star_{10}\1$) and the dilaton $\Phi$ kinetic term, the Neveu-Schwarz action ${\cal S}_{NS}$ also contains the kinetic term for NS three-form $H_3$, which derives locally from the NS-NS two-form, i.e.~$H_3 = dB_2$. The Ramond action ${\cal S}_R$ captures the kinetic terms for all differential RR-forms $C_{2p-1}$, namely $(C_1, C_3, C_5, C_7)$ for Type IIA and $(C_0, C_2, C_4, C_6, C_8)$ for Type IIB superstring theories. Their respective gauge-invariant field strengths $G_{2p}$ are defined as:  
\begin{eqnarray}\label{Eq:DefFieldStrength}
G_{1} = dC_{0}, \qquad G_{2} = dC_1, \qquad G_{2p} = d C_{2p - 1} - H_3 \wedge C_{2p-3} \; \; \text{ (otherwise)}.
\end{eqnarray}
The action ${\cal S}_R$ fulfills more the r\^ole of a pseudo-action, as the equations of motion resulting from the action have to be supplemented by the duality constraints:
\begin{equation}\label{Eq:HodgeDualRRforms10D}
G_{m+1} = (-)^{(m+1)/2} \star_{10} G_{9-m} \, \, (\text{IIA}), \qquad G_{m+1} = (-)^{m/2} \star_{10} G_{9-m} \, \, (\text{IIB}),
\end{equation}
effectively reducing the number of physical degrees of freedom. The democratic formulation might seem a bit involved, but it represents the natural formulation to write down the Chern-Simons action for the D-branes including all RR-forms. More explicitly, the (massless) excitations of a (single) D$p$-brane are captured by an effective $p+1$ dimensional action consisting of the Dirac-Born-Infeld (DBI) part and the Chern-Simons action, which read in the string frame:\footnote{For a stack of $N$ coincident D-branes the gauge group on the collective worldvolume enhances to a non-Abelian gauge group, implying that the DBI-action and Chern-Simons action need to be generalized accordingly to capture the non-Abelian features. For our purposes it suffices to replace $e^{2 \pi \alpha' F}$ by $\Tr(e^{2 \pi \alpha' F})$ in the Chern-Simons action~ (\ref{Eq:DbraneCS}) for a non-Abelian gauge group.} 
\begin{eqnarray}
&{\cal S}_{\rm D-brane} &= {\cal S}_{DBI} + {\cal S}_{CS}, \notag \\
&{\cal S}_{DBI} &= - \mu_p \int_{Dp} d^{p+1} \xi \, e^{-\Phi} \sqrt{-\det( \iota^* g +  \iota^* B_{2} - 2 \pi \alpha ' {F}_{MN})},\\
&{\cal S}_{CS} &= \mu_p \int_{Dp} \sum_q \iota^* C_{q} \wedge e^{2 \pi \alpha' F- \iota^* B_2}, \label{Eq:DbraneCS}
\end{eqnarray}
with the parameter $\mu_p$ related to the string lengths $\ell_s$ as given in~(\ref{Eq:DbraneStringLength}). $\iota^*$ represents the pullback of the ten dimensional fields to the D$p$-brane worldvolume parametrized by the local coordinates $\xi$.\footnote{Implicitly, we assume that $p>3$ such that the D$p$-brane wraps $p-3$ cycles along the internal space. And in practice, we have D6-brane configurations for Type IIA  and D7-brane configurations for Type IIB superstring theory in the back of our minds.}  Assuming that the D$p$-brane fills the maximally symmetric spacetime and wraps a $p-3$ dimensional cycle $\Delta_{p-3}$ on the internal space $X_6$ for which $\iota^* B_{2} =0$, we can write the pullback of the ten dimensional metric as follows:
\begin{equation}\label{Eq:Pullbackmetric}
\iota^* g =  \eta_{\mu \nu} dx^\mu dx^\nu + g_{a b} \frac{\partial y^a}{\partial \xi^k}  \frac{\partial \ov y^{\ov\jmath}}{\partial \xi^l} d \xi^k d\xi^l + \text{ D-brane fluctuations },
\end{equation}
and we assume the following decomposition for the gauge field on the D-brane:
\begin{equation}\label{Eq:GaugeFieldDecomposition}
F_{MN} = \left( \begin{array}{cc} F_{\mu \nu} & 0 \\ 0 & {\cal F}_{ab} \end{array}\right).
\end{equation}
Regarding the RR-forms, we assume that the $q$-forms are only turned on along the directions of the D-brane such that the pullback acts trivially, i.e.~$\iota^* C_{q} = C_{q}$.

Dimensional reductions of type II superstring theory with and without D-branes have been investigated in various places, see for instance~\cite{Grimm:2004uq,Grimm:2004ua,Jockers:2004yj,Haack:2006cy,Cicoli:2011yh,Grimm:2011dx,Kerstan:2011dy,Camara:2011jg} for detailed discussions on Calabi-Yau orientifold compactifications. Therefore, it is not our intention to repeat these results in great detail. Instead, we wish to highlight some relations which can be obtained without specifying the geometric properties of the internal space ${\cal X}_6$ too explicitly, similar to the approach considered in~\cite{Svrcek:2006yi}. 

The dimensional reduction of the (bulk) NS-sector ${\cal S}_{NS}$ to four dimensions is completely equivalent for Type IIA and IIB superstring theory such that we do not yet have to differentiate between the two theories at this point. Inserting the metric ansatz~(\ref{Eq:metricAnsatz}) into the kinetic term for the ten-dimensional metric and comparing to the four-dimensional Einstein-Hilbert action with gravitational coupling $\kappa_4^2$ leads to the well-known relation between the reduced Planck mass $M_{Pl}$ and the string mass scale $M_s$:
\begin{equation}\label{Eq:PlanckToString}
\frac{1}{\kappa_4^2} = \frac{1}{\kappa_{10}^2} e^{-2 \langle \Phi \rangle}  \text{Vol}({\cal X}_6) \qquad \leadsto \qquad  \frac{M_{Pl}^2}{M_{s}^2} = \frac{4 \pi}{g_s^2}  \frac{{\cal T}}{6},
\end{equation} 
where the string coupling $g_s$ is set by the {\it vev} of the dilaton, i.e.~$g_s = e^{\langle \Phi \rangle}$. We also introduced the dimensionless volume ${\cal T}$ of the internal space ${\cal X}_6$: ${\cal T} = 6 \ell_s^{-6}  \text{Vol}({\cal X}_6)$.
A controlled compactification is in the first place characterised by a small string coupling $g_s<1$, so let us assume $g_s \sim 10^{-1}$. As a second requirement the characteristic size of the internal space has to be larger than the string scale $\ell_s$ to sustain geometrical control and keep $\alpha'$ corrections small. This means that  the dimensionless volume ${\cal T}$ can lie within the region $10^2 \lesssim {\cal T} \lesssim 10^{30}$, where the (more flexible) upper bound is set by the non-observation of fifth forces assuming an isotropic internal space ${\cal T}$.  
Hence, from equation (\ref{Eq:PlanckToString}) we deduce that the window for the string mass scale in Type II compactifications is roughly given by,
\begin{equation} 
10^{3} \text{ GeV} \lesssim M_s \lesssim 10^{17} \text{ GeV}.
\end{equation} 
A string mass scale larger than the reduced Planck mass would require us to dive into perturbatively uncontrollable regions of the moduli space, with either a large string coupling $g_s>1$ or a small internal volume ${\cal T}<1$.

Also the dimensional reduction of the Dirac-Born-Infeld-action ${\cal S}_{DBI}$ is very analogous for both Type II superstring theories. Only the dimensionality of the cycles wrapped by the D$p$-branes will differ. Under the assumptions of equations (\ref{Eq:Pullbackmetric}) and (\ref{Eq:GaugeFieldDecomposition}), and by ignoring the D-brane fluctuations in the pullback of the metric the DBI-action reduces to a Yang-Mills type action (at leading order in $\alpha'$) with tree-level  gauge coupling given by,
\begin{equation}\label{Eq:GaugeCouplingReduction}
\frac{2\pi}{g^2_{YM}} = \frac{1}{g_s} \frac{1}{\ell_s^{p-3}} \Gamma_\Delta({\cal F}), 
\end{equation}
where we introduce the function $\Gamma_\Delta({\cal F})$ (with a slightly different notation than~\cite{Haack:2006cy}):
\begin{equation}
\Gamma_\Delta({\cal F}) \equiv \int_{\Delta_{p-3}} d^{p-3} \xi \sqrt{\det(\iota^*g_{(6)} + 2 \pi \alpha' { \cal F}_{a b})}.
\end{equation}
In the absence of internal flux ${\cal F}$ the function reduces to 
\begin{equation}
\Gamma_\Delta({\cal F} = 0) = \text{Vol} (\Delta_{p-3}),
\end{equation}
where $\Delta_{p-3}$ represents the $p-3$ dimensional subspace wrapped by the D$p$-brane on the internal manifold ${\cal X}_6$. In supersymmetric compactifications of Type IIB superstring theory with D7-branes, a non-trivial internal flux ${\cal F}$ can give rise~\cite{Jockers:2004yj,Haack:2006cy} to field-dependent D-terms involving the K\"ahler moduli. For Type IIA compactifications with D6-branes, the flux corresponds to a flat connection, such that the function $\Gamma_\Delta({\cal F})$ reduces to the volume of the internal three-cycle wrapped by the D6-brane.

Recalling that the string coupling $g_s$ has to be smaller than one to be in the perturbative regime of Type II string theory, we conclude that the gauge theory on the D-brane worldvolume is weakly (strongly) coupled when the volume of the cycle $\Delta_{p-3}$ wrapped by the D-brane is large (small) in comparison to the string length. Notice, however, that this statement is only true at tree-level.
 Once massive string state contributions are taken into account through gauge threshold corrections at one-loop,  the one-loop gauge kinetic functions can receive positive or negative contributions scaling with other moduli than the volume of the three-cycle $\Delta_{p-3}$. 
In the case of substantial negative contributions, one might even expect the gauge theory to be strongly coupled when a D-brane wraps a (classically) large cycle.\footnote{The observation regarding gauge treshold corrections has been exploited recently to discuss gauge coupling unification~\cite{Gmeiner:2009fb,Honecker:2012qr} and lower bounds on the string mass scale~\cite{Honecker:2013mya} in global intersecting D6-brane models on toroidal orbifolds. In the area of large field inflation, it is the dependence of the gauge threshold corrections on geometric moduli that has prompted the authors of~\cite{Abe:2014pwa,Abe:2014xja} to use them as a building block in the construction of axionic inflation models with a trans-Planckian decay constant.} In this respect clear-cut statements about the coupling strength of the gauge theory on the D-brane worldvolume can only be made for explicit examples of D-brane configurations. 
 If the compactification is not asymmetric, we generically expect 
 the size of the $p-3$ dimensional cycle to be set by the volume of the entire manifold, i.e.~$\text{Vol} (\Delta_{p-3}) \sim \sqrt{{\cal T}}$.
 \footnote{In principle, we should also assume that the internal cycle $\Delta_{p-3}$ wrapped by the D-brane has the smallest volume within its homology class. In mathematical terms, this assumption can be recast in the existence of a calibration form $\phi$ on the internal space such that the volume of the internal cycle $\Delta_{p-3}$ equals the integrated pullback of the calibration form $\phi\big|_{\Delta_{p-3}}$ with respect to $\Delta_{p-3}$. In case the internal space allows for a Calabi-Yau geometry, such calibration forms can be naturally identified by virtue of the K\"ahler two-form or Calabi-Yau three-form and can be used to express the geometric conditions for the D-branes to be supersymmetric, see e.g.~\cite{Koerber:2010bx} for a review.}

%%%%%%%%%%%%%%%%%%%%%%%%%%%%%%%%%%%%%%%%%%%%%%%%%%%%%%%%%%%%%%%%%%%%%%%%%%%%%%%%%%%%%%%%%%%%%%%%%%%%%%%%%%%%%%%%%%%%%%%%%%%%%%%%%
\subsection{The Effective Action for Closed String Axions}\label{Ss:ClosedStringAxions}
Axion-like fields arise abundantly from the various differential $q$-forms in the massless closed string spectrum upon dimensional reduction, which has motivated the extensive use of these states as candidate inflatons in stringy inflationary models. In our discussion we will focus on the closed string axions emerging from the RR-sector through the dimensional reduction of various massless $q$-forms, and 
 for concreteness, we illustrate such reduction with
axions associated to the $C_3$ form in Type IIA and to the $C_4$ form in Type IIB. The (bulk) RR-action ${\cal S}_{RR}$ can be dimensionally reduced for both $q$-forms with field strength defined in equation (\ref{Eq:DefFieldStrength}) in the same manner. Namely, decomposing the $q$-form $C_q$ with respect to a basis of harmonic forms $\alpha_i$ for the cohomology group $H^q({\cal X}_6)$: 
\begin{equation}\label{Eq:DecompPforms}
C_{q} = \frac{1}{2\pi} \sum_{i=1}^{b_q} a^i (x)\, \alpha_i (y) + \ldots ,
\end{equation}
already exposes the coefficients $a^i(x)$ as scalar fields along the four dimensional spacetime ${\cal M}_{1,3}$. In this expression $b_q = b_q ({\cal X}_6) = \text{dim } H^q({\cal X}_6)$ represents the $q^{th}$ Betti number of the internal manifold ${\cal X}_6$, and the factor $2 \pi$ has been introduced to ensure a periodicity of $2\pi$ for the scalar field $a^i$. The $\ldots$ stands for the decomposition with respect to a basis of harmonic forms in $H^k({\cal X}_6)$ with degree $k<q$. Next, we introduce a basis of closed $q$-cycles $\gamma_i$ for the homology group $H_q({\cal X}_6,\Z)$ that is (de Rham) dual to the basis of closed $q$-forms $\alpha_i$:
\begin{equation}\label{Eq:PoincareDual}
\ell_s^{-q} \int_{\gamma_ j} \alpha_i  = \ell_s^{-6} \int_{{\cal X}_6} \alpha_i \wedge \beta^j   = {\delta_{i}}^j.
\end{equation} 
In the second expression we exploit Poincar\'e-duality to introduce a basis $\beta^j$ of $(6-q)$-forms for the cohomology group $H^{6-q}({\cal X}_6)$. Note that the $C_3$ form of Type IIA requires a small adjustment as both the $\alpha_i$ and the $\beta^j$-basis have to fit in the cohomology group $H^{3}({\cal X}_6)$. In that case the indices $i, j$ will run from $1$ to $\frac{1}{2}b_3$ and the basis $(\alpha_i, \beta^j)$ forms a symplectic basis for $H^{3}({\cal X}_6)$. Recalling that the various $q$-forms are related through Hodge duality (\ref{Eq:HodgeDualRRforms10D}) in ten dimensions we can play the same game for the Hodge-dual $C_{8-q}$ form and decompose it with respect to the basis $\beta^i$:
\begin{equation}\label{Eq:DecompPformsII}
C_{8-q} =  \sum_{i=1}^{b_{6-q}} D_{(2)i} \wedge \beta^{i} + \ldots,
\end{equation}
where the $\ldots$ include the decomposition with respect to the bases of other cohomology groups. The two-forms $D_{(2)i}$ can be seen as the four dimensional Hodge duals of the scalar fields $a^i$. Observe that the decomposition of the self-dual $C_4$ form of Type IIB contains both the axions $a^i$ as well  as their Hodge dual 2-forms $D_{(2)i}$.

Let us now focus on the kinetic terms for the $C_q$ form and its dual form in the RR-action (\ref{Eq:BulkRR}), 
\begin{equation}
{\cal S}_{R} = -  \frac{1}{8 \kappa_{10}^2} \int_{{\cal M}_{1,3}\times {\cal X}_6} \left[ G_{q+1} \wedge \star_{10} G_{q+1} + G_{9-q} \wedge \star_{10} G_{9-q} + \ldots \right],
\end{equation}
and perform the dimensional reduction over ${\cal X}_6$ using the decomposition of the forms in (\ref{Eq:DecompPforms}) and (\ref{Eq:DecompPformsII}) respectively:
\begin{equation}\label{Eq:DimRedResultAxions}
{\cal S}_{R} = - \frac{1}{4 \ell_s^2} \int_{{\cal M}_{1,3}}\left[ d a^i \wedge \star_4 d a^j {\cal K}_{ij} + d D_{(2)i} \wedge \star_4 d D_{(2)_j} {\cal K}^{ij}  + \ldots \right],
\end{equation}
where we introduced the moduli-dependent metric ${\cal K}_{ij}$ on the axion moduli space: 
\begin{equation}\label{Eq:MetricAxionModuliSpace}
{\cal K}_{ij} = \frac{1}{2\pi \ell_s^6} \int_{{\cal X}_6} \alpha_i \wedge \star_6 \alpha_j,
\end{equation}
and its inverse ${\cal K}^{ij}$:
\begin{equation}
{\cal K}^{ij} = \frac{2\pi}{\ell_s^6} \int_{{\cal X}_6} \beta^i \wedge \star_6 \beta^j.
\end{equation}
For a generic compact manifold ${\cal X}_6$ with metric $g_{ab}$ it is rather difficult to compute the metric ${\cal K}_{ij}$ on the axion moduli space, as it would require an explicit form for the internal metric $g_{ab}$ as well as knowledge about all possible deformations of the  internal metric. By adding geometric structures to the internal space ${\cal X}_6$, allowing for instance a Calabi-Yau structure, one can provide more details about the metric on the axion moduli space. It is for instance well known that the moduli space of a Calabi-Yau manifold is spanned by two types of deformations: complex structure deformations and K\"ahler deformations, see e.g.~\cite{Candelas:1990pi}. The massless scalars $a^i$ are regrouped with these deformations into complex coordinates which parametrise the moduli space of ${\cal X}_6$.   Locally, the moduli space can be written as the direct product of the two complex submanifolds ${\cal M}_{\text{K\"ahler}} \times {\cal M}_{\rm Complex}$, each with a K\"ahler structure and each parametrized by one type of deformations. An additional orientifold projection along the internal space is required to bring the amount of four dimensional spacetime supersymmetry down to ${\cal N} = 1$ supersymmetry for Type IIA and Type IIB superstring theory. For Type II superstring theory on a Calabi-Yau orientifold, the moduli space can still be written as a direct product $\hat{\cal M}_{\text{K\"ahler}} \times \hat{\cal M}_{\rm Complex}$, with $\hat{\cal M}_{\text{K\"ahler}} \subsetneq {\cal M}_{\text{K\"ahler}}$ and $\hat{\cal M}_{\rm Complex} \subsetneq {\cal M}_{\rm Complex}$. The subspaces $\hat{\cal M}_{\text{K\"ahler}}$ and $\hat{\cal M}_{\rm Complex}$ are not necessarily K\"ahler manifolds, but the metric on these subspaces are inherited from the ${\cal N}=2$ parent spaces upon applying the orientifold projection. Hence, for Type II Calabi-Yau orientifold compactifications the metric ${\cal K}_{ij}$ on the axion moduli space will depend explicitly on the set of deformations tied to the associated axions: complex structure moduli $U^i$ in the case of Type IIA and K\"ahler moduli $T^i$ in the case of Type IIB.
For a consistent embedding of the effective field theory approach in section~\ref{sec:mixing} into superstring theory, one implicitly assumes that the respective moduli have been stabilised at energy scales below the Kaluza Klein-scale and higher than the energy scale at which action~(\ref{Eq:GeneralLagrangian}) is valid.  

The appearance of the field strengths $G_{p}$ is mandated by gauge invariance, such that ten dimensional kinetic terms for the $q$-forms in the RR sector lead upon dimensional reduction to standard kinetic terms for the fields $a^i$ (and their Hodge duals) as presented in equation~(\ref{Eq:DimRedResultAxions}), and thus only yield derivative interactions involving $a^i$ or $D_{(2)i}$. This observation suggests the existence of a shift symmetry for the fields $a^i$ inherited from the remnants of the gauge invariance of the $C_q$ forms and justifies the interpretation of the scalars $a^i$ as axions. The shift symmetry of the axions is, however, broken by nonperturbative effects, such as D-brane instantons and gauge instantons. The strength of a Euclidean D-brane wrapping the $q$-cycle $\gamma_i$ is set by its instanton amplitude~\cite{Blumenhagen:2009qh,Ibanez:2012zz}:
\begin{equation}\label{Eq:StringyInstantonAmplitude}
e^{-S_{E_{q-1}}} = e^{- \frac{2\pi}{\ell_s^{q}} \left(\frac{1}{g_s } {\rm Vol}(\gamma_i) + i\, \bigintssss_{\gamma_i} C_q \right)}  = e^{- \frac{2\pi}{\ell_s^{q}} \frac{1}{g_s } {\rm Vol} (\gamma_i) -i\, a^i },
\end{equation} 
where we inserted the decomposition~(\ref{Eq:DecompPforms}) in the last equality. The amplitude of the D-brane instanton is determined by the volume of the wrapped cycle (measured in units of string length $\ell_s$), while its phase corresponds to the axion $a^i$. The non-perturbative coupling in $g_s$ thus breaks the continuous shift symmetry of the axion $a^i$ to a discrete shift symmetry, which clarifies the assumed periodicity in~(\ref{Eq:shift}). As a direct consequence we can conclude that the moduli space for $b_q$ closed string axions corresponds to a $b_q$~dimensional torus $T^{b_q}$ endowed with metric ${\cal K}_{ij}$ as defined in equation~(\ref{Eq:MetricAxionModuliSpace}). Gauge instantons on the other hand are characterised by an amplitude:
\begin{equation}\label{Eq:GaugeInstantonAmplitude}
e^{-S_{gauge}} = e^{- |I_n| \left( \frac{8 \pi^2}{g_{YM}^2} + i\, \theta   \right)},
\end{equation}
with $I_n$ the topological instanton number as introduced in (\ref{Eq:PontryaginIndex}) and $\theta$ the axionic direction coupling anomalously to the non-Abelian gauge group, thereby breaking the shift symmetry along the $\theta$-direction down to a discrete shift symmetry justifying the assumed periodicity in~(\ref{Eq:CollectiveShiftSymmetry}). 

In string theory, gauge field theory instantons can be interpreted as a particular type of D-brane instantons, namely as Euclidean D$(p-4)$-branes lying on top of the D$p$-branes while wrapping the cycle $\Delta_{p-3}$.
Expression~(\ref{Eq:GaugeCouplingReduction}) allows us to compare the strength between the stringy D-brane instantons in~(\ref{Eq:StringyInstantonAmplitude}) and the gauge instantons in~(\ref{Eq:GaugeInstantonAmplitude}) (if both types of instantons are present), from which we can conclude that the stringy D-brane instanton amplitude on a $(p-3)$-cycle $\gamma_i\neq \Delta_{p-3}$ is subleading with respect to the gauge instanton amplitude provided:
\begin{equation}
\frac{{\rm Vol} (\Delta_{p-3})}{{\rm Vol} (\gamma_i)} < \frac{1}{2} .
\end{equation}
In order to determine the effective contribution of an instantonic effect to an explicit model, one has to integrate over the moduli space of the instanton solution. The integration measure over the instanton moduli space decomposes into bosonic instanton zero-modes (expressing the position, the size and possible deformations of the instanton) and fermionic instanton zero-modes (related to broken supersymmetries, to the superpartners of the deformations and to chiral fermions located at the intersections between instantonic branes and/or D-branes). The instanton corrections will only contribute if all fermionic zero modes can be saturated, which has to be checked explicitly for each instanton in each individual model.  
 
Which linear combinations of closed string axions couple to the gauge instantons can be read off from the dimensional reduction of the D-brane Chern-Simons action (\ref{Eq:DbraneCS}) upon identifying the topological $G\wedge G$ term as introduced in (\ref{Eq:GeneralLagrangian}). For the first time, we will have to distinguish between Type IIA and Type IIB, given that the dimensionality of the D-branes differs for both string theories. On the bright side, the usefulness of the democratic formulation will be truly exposed by the reduction of the D-brane Chern-Simons action to four dimensions. \\

\noindent {\bf D6-branes in Type IIA}\\
For a D6-brane wrapping a three-cycle $\Delta_3$ along ${\cal X}_6$ the relevant terms in the Chern-Simons action are captured by,
\begin{equation}
{\cal S}_{CS}^{D6} = \mu_6 \int_{{\cal M}_{1,3} \times {\Delta}_3}  C_5 \wedge (2 \pi \alpha ') F + \frac{1}{2}C_3 \wedge (2 \pi \alpha ')^2 F \wedge F + \ldots.
\end{equation}
The three-cycle $\Delta_3$ can be decomposed in terms of the closed three-cycles $(\gamma_i, \delta^j)$, serving as the de Rham-duals to the symplectic basis $(\alpha_i, \beta^j)$ respectively:
\begin{equation}
\Delta_{3} = \sum_{i=1}^{b_3/2} \left( r^i \gamma_i + p^i \delta^i \right), \qquad \text{ with } r^i , p^i \in \Z.
\end{equation}
Plugging in both the expansions (\ref{Eq:DecompPforms}) and (\ref{Eq:DecompPformsII}) for the $C_3$ and $C_5$ form respectively, as well as the decomposition of the three-cycle $\Delta_3$, yields the following expression:
\begin{equation}\label{Eq:CSD6reduction}
{\cal S}_{CS} =  \frac{1}{8\pi^2} \sum_{i=1}^{b_3/2} r^i \int_{{\cal M}_{1,3}}  a^i F\wedge F +  \frac{1}{ \ell_s^2} \sum_{j=1}^{b_3/2} p^j \int_{{\cal M}_{1,3}} D_{(2) j} \wedge F + \ldots.
\end{equation}
The first term resembles indeed the non-perturbative coupling of axions to the topological charge density of a gauge group, while the second term corresponds to the dual description of the St\"uckelberg coupling between an Abelian gauge field and CP-odd scalars. Geometrically, a $C_3$-axion $a^i$ couples to the topological term $F\wedge F$ when the three-cycle $\Delta_3$ wraps its associated three-cycle $\gamma_i$ (i.e.~$r^i \neq 0$). And non-vanishing St\"uckelberg charges $p^j\neq0$ under a D6-brane $U(1)$ gauge group arise for those axions $a^i$ whose associated Poincar\'e dual three-cycle $\delta^i$ is wrapped by the D6-brane. 
Note that we have tried to take a minimalistic stance in the dimensional reduction, by assuming as little as possible concerning the geometry of the internal space or the embedding of the D6-brane in ${\cal X}_6$. One can be more explicit by considering type IIA superstring theory on a Calabi-Yau orientifold, for which the axions emerging from $C_3$ form the CP-odd partners of the complex structure moduli. In this Calabi-Yau orientifold setting the dimensional reduction~\cite{Grimm:2011dx,Kerstan:2011dy,Camara:2011jg} is much more involved than presented here, due to the presence of the orientifold projection and of additional moduli describing the position of the D-brane which we ignore here.\\

\noindent {\bf D7-branes in Type IIB}\\
D7-branes are embedded on four-dimensional cycles $\Delta_4$ along ${\cal X}_6$ and can be written in terms of a basis of closed 4-cycles $\gamma_i$, (de Rham) dual to the basis of harmonic 4-forms $\alpha_i$ on ${\cal X}_6$ introduced above:  
\begin{equation}
\Delta_{4} = \sum_{i=1}^{b_4}  r^i \gamma_i , \qquad \text{ with } r^i  \in \Z.
\end{equation}
For axions associated to the RR-form $C_4$ there is only one term in the Chern-Simons part of the D7-brane action of particular interest:
\begin{equation}
{\cal S}^{D7}_{CS} = \frac{\mu_7}{2} \int_{{\cal M}_{1,3} \times {\cal X}_6}  C_4 \wedge (2 \pi \alpha ')^2 F \wedge F, 
\end{equation}
but the term yields both the anomalous coupling and the St\"uckelberg coupling depending on the interpretation of the flux $F$:
\begin{equation}\label{Eq:CSD7reduction}
{\cal S}^{D7}_{CS} =   \frac{1}{8\pi^2} \sum_{i=1}^{b_4} r^i \int_{{\cal M}_{1,3}} a^i F\wedge F +    \frac{1}{\ell_s^2} \sum_{i=1}^{b_2} p^i({\cal F})  \int_{{\cal M}_{1,3}}  D_{(2)i} \wedge  F.
\end{equation}
The first term results from interpreting the flux $F\wedge F$ as the topological charge density along ${\cal M}_{1,3}$, while the second term arises by taking one of the $F$-factors as the flux ${\cal F}$ along the internal direction on $\Delta_4$. This ambiguity is a direct consequence of the self-duality of the four-form $C_4$. 
In analogy with the D6-brane reduction we introduced the symbol $p^i({\cal F})$ which now also depends on the flux ${\cal F}$ apart from the embedding of the 4-cycle $\Delta_4$:
\begin{equation}
p^i({\cal F}) \equiv \frac{1}{2 \pi} \frac{1}{\ell_s^2} \int_{\Delta_4} \beta^i \wedge {\cal F} \in  \Z.
\end{equation}
Hence, when the four-cycle $\Delta_4$ wraps a four-cycle $\gamma_i$ in geometric terms, its associated axion $a^i$ will couple to the topological density $F\wedge F$. For the axion to be charged under the $U(1)$ gauge group supported by the D-brane, the four-cycle $\Delta_4$ has to wrap the four-cycle that is Poincar\'e-dual to the two-cyle supporting the internal flux ${\cal F}$.
Also here we have tried to avoid making particular assumptions about the geometric features of ${\cal X}_6$ or of the four-cycle $\Delta_4$. In case ${\cal X}_6$ is taken to be a Calabi-Yau orientifold various geometric aspects can be expressed in a more explicit way thanks to the virtues of complex geometry~\cite{Jockers:2004yj,Haack:2006cy}. The $C_4$ axions fit within the same ${\cal N}=1$ supermultiplet as the K\"ahler moduli for a compactification set-up where the holomorphic involution maps the Calabi-Yau three-form to minus itself.\footnote{If the orientifold projection leaves the Calabi-Yau three-form invariant, the axions emerging from the reduction of the $C_4$-form recombine with the scalars associated to the reduction of the NS-NS $B_2$ form. Moreover, there are no $O7$-planes whose charges can compensate the D7-brane charges. Hence, such an orientifold projection does not seem to provide a favourable setting for the string embedding of our ideas.}

\noindent Despite the fact that axions emerge from different $q$-forms for type IIA and type IIB superstring respectively, we obtain the same four-dimensional effective field theory for the axions:
\begin{eqnarray}\label{Eq:ReducedEFTAxions}
{\cal S}_{axion} &=&  \frac{1}{2 \ell_s^2} \int_{{\cal M}_{1,3}} \left[ -\frac{1}{2}  d a^i \wedge \star_4 d a^j {\cal K}_{ij}  - \frac{1}{2}  d D_{(2)i} \wedge \star_4 d D_{(2)_j} {\cal K}^{ij} + 2 \sum_{i} p^i  D_{(2) i} \wedge F \right] \nonumber \\
&& + \frac{1}{8\pi^2} \sum_{i} r^i \int_{{\cal M}_{1,3}} a^i F\wedge F. 
\end{eqnarray}
In order to end up with an action written in the form of (\ref{Eq:GeneralLagrangian}), the two-forms $D_{(2)i}$ have to be dualised to their Hodge-dual 0-forms following the procedures outlined in appendix~\ref{A:Dualisation}. By applying these dualization methods on the action in (\ref{Eq:ReducedEFTAxions}), we find the following dual action (with two-forms $D_i$ eliminated):
\begin{equation}\label{Eq:ReducedEFTAxionsFinal}
{\cal S}_{axion} =  -\frac{1}{2 \ell_s^2} \int_{{\cal M}_{1,3}} \left[ \frac{1}{2}  \left(d a^i - 2 p^iA\right)  \wedge \star_4  \left(d a^j - 2 p^j A\right) {\cal K}_{ij} \right]  + \frac{1}{8\pi^2} \sum_{i} r^i \int_{{\cal M}_{1,3}} a^i F\wedge F ,
\end{equation}
which is exactly of the same type as proposed in (\ref{Eq:GeneralLagrangian}). The missing kinetic terms for the gauge fields follow from the dimensional reduction of the DBI-action for the D-brane. For a stack of $N$ coincident D$p$-branes with a non-Abelian gauge group the topological term $F\wedge F$ has to be replaced by $\Tr(G\wedge G)$, with $G$ the field strength of the non-Abelian gauge group. And with this last consideration it is now clear how the effective action in (\ref{Eq:GeneralLagrangian}) emerges from string theory compactifications with moduli space metric ${\cal G}_{ij} = (2 \ell_s^2)^{-1} {\cal K}_{ij}$. Note that the axions $a^i$ are represented as dimensionless fields in (\ref{Eq:ReducedEFTAxionsFinal}) and that the eigenvalues of ${\cal G}_{ij}$ are measured in units of the string mass scale $M_s$. Hence, the numerical examples presented in section~\ref{sec:mixing} should be seen in the light of a high string scale mass $M_s \sim {\cal O} (10^{16}-10^{17} \text{GeV})$.

The couplings in (\ref{Eq:CSD6reduction}) and (\ref{Eq:CSD7reduction}) following from the reduction of the D-bane Chern-Simons action form the building blocks 
for the Green-Schwarz-mechanism in four dimensions by which the mixed Abelian-non-Abelian and cubic Abelian gauge anomalies cancel. The Green-Schwarz terms associated to (\ref{Eq:CSD6reduction}) and (\ref{Eq:CSD7reduction}) usually suffice to cancel these gauge anomalies. Furthermore, the pure non-Abelian gauge anomalies vanish automatically when the RR tadpole cancelation conditions are satisfied. 

In section~\ref{Sss:GCSterms} we indicated that in situations where the anomaly coefficient also contains a non-symmetric part, a generalized Chern-Simons term has to be introduced to ensure $U(1)$ gauge invariance, as discussed in more detail in~\cite{Anastasopoulos:2006cz,DeRydt:2007vg}.
One could wonder whether this generalized Chern-Simon term can be obtained directly from string theory, thereby offering a microscopic explanation for its required presence. To this end, the authors of~\cite{Anastasopoulos:2006cz} derived Chern-Simons terms directly from string theory by computing the appropriate open and closed string amplitudes (for D5-D9 brane modelbuilding scenarios on toroidal orientifolds). Generalized Chern-Simons terms also arise from the D-brane Chern-Simons action in case the internal manifold of a string theory compactification allows for non-vanishing fluxes and the $U(1)$ gauge symmetry descends from the closed string sector, as shown in~\cite{Aldazabal:2002py} by using the descent formalism of Wess and Zumino. Whether or not generalized Chern-Simons terms are required is thus a model-dependent consideration, as is the question how these terms arise microscopically within a string model.

%%%%%%%%%%%%%%%%%%%%%%%%%%%%%%%%%%%%%%%%%%%%%%%%%%%%%%%%%%%%%%%%%%%%%%%%%%%%%%%%%%%%%%%%%%%%%%%%%%%%%%%%%%%%%%%%%%%%%%%%%%%%%%%%%%%%%%%%%%%%%%%%%%%%%%%%%%%%%%%%%%%%%%%%
\subsection{Some Explicit Examples}\label{subsec:ex}

\subsubsection{Factorizable D6-branes in Type IIA on Toroidal Orientifolds}\label{Ss:ExampleFactD6branes}
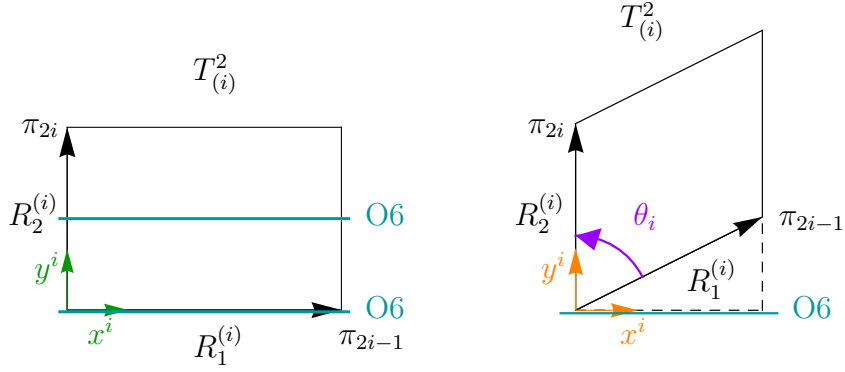
\begin{figure}[ht]
\begin{center}
\begin{tabular}{c@{\hspace{1in}}c}
\includegraphics[width=4cm]{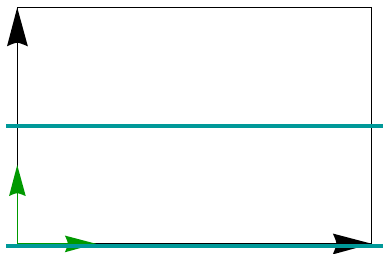} \begin{picture}(0,0) \put(-65,90){$T_{(i)}^2$} \put(-10,-10){$\pi_{2i-1}$} \put(-130,70){$\pi_{2i}$} \put(0,35){{\color{myblue4}O6}} \put(0,0){{\color{myblue4}O6}} \put(-65,-15){$R_1^{(i)}$}   \put(-135,35){$R_2^{(i)}$} \put(-105,-10){{\color{mygr} $x^{i}$}}  \put(-125,15){{\color{mygr} $y^{i}$}}  \end{picture} & \includegraphics[width=3cm]{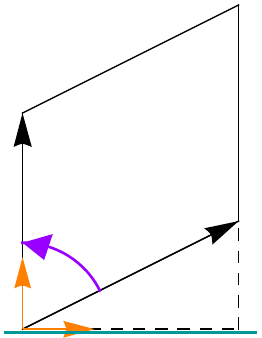}  \begin{picture}(0,0) \put(-65,110){$T_{(i)}^2$} \put(-5,35){$\pi_{2i-1}$} \put(-100,70){$\pi_{2i}$}  \put(0,0){{\color{myblue4}O6}} \put(-40,10){$R_1^{(i)}$}   \put(-105,35){$R_2^{(i)}$}  \put(-65,-10){{\color{myorange} $x^{i}$}}   \put(-95,15){{\color{myorange} $y^{i}$}} \put(-60,35){\color{mypurple} $\theta_i$} \end{picture} \\
& \\
\end{tabular}
\caption{(left) {\bf a}-type lattice for a rectangular two-torus $T_{(i)}^2$ with area $R_1^{(i)} R_2^{(i)}$ and modular parameter $\tau^{(i)} = i\, R_2^{(i)}/ R_1^{(i)}$, and (right) {\bf b}-type lattice for a tilted two-torus $T_{(i)}^2$ with area $R_1^{(i)} R_2^{(i)} \sin \theta_i$ and modular parameter $\tau^{(i)} = R_2^{(i)}/ R_1^{(i)} e^{i\, \theta_i}$. On a rectangular lattice the fixed planes under the $\OR$-projection are located at $\IM(z^i) = 0$ and $\IM(z^i) = 1/2$, while a tilted torus-lattice only has one fixed plane under the $\OR$-projection, namely $\IM(z^i) = 0$. The basic one-cycles $\pi_{2i-1}$ and $\pi_{2i}$ transform as follows under the $\OR$-projection: $\pi_{2i-1} \stackrel{\OR}{\longrightarrow} \pi_{2i-1} - 2b^i \pi_{2i}$ and $\pi_{2i} \stackrel{\OR}{\longrightarrow}  - \pi_{2i}$, where the discrete parameter $b^i$ captures whether the two-torus $T^{2}_{(i)}$ is rectangular $(b^i = 0)$ or tilted $(b^i=1/2)$.  \label{Fig:T2Lattices}}
\end{center}
\end{figure}

The toroidal orientifold $T^6/\OR$ is probably the easiest internal space ${\cal X}_6$ that comes to mind to clarify the set-up in section~\ref{Ss:U1mixing} through explicit examples. To simplify their construction, the six-dimensional torus is taken to be of the factorizable type $T_{(1)}^2 \times T_{(2)}^2 \times T_{(3)}^2$, where each two-torus can be parametrized by a complex coordinate $z^{i = 1,2,3}$ respectively with periodicity relations:
\begin{equation}
z^i \simeq z^i +1, \qquad z^i \simeq z^i + \tau^i,
\end{equation}
and where the parameter $\tau^i$ corresponds to the modular parameter for torus $T_{(i)}^2$.\footnote{To emphasize the structural properties of the background, we simplify the coordinate-dependent expressions by considering dimensionless coordinates $z^i$, i.e.~the dimensionful coordinates have been divided by $\ell_s$, such that also the three-forms $\alpha_i$ and $\beta^i$ are dimensionless.} Considering Type IIA string theory on $T^6$ leads to a four dimensional theory with a maximal amount of supersymmetry, namely ${\cal N}=8$ supersymmetry. To reduce the amount of supersymmetry by a factor $1/2$, one usually introduces an orientifold projection $\OR(-)^{F_L}$, consisting of a worldsheet parity $\Omega$, a projection $(-)^{F_L}$ by the left fermion number and an anti-holomorphic involution ${\cal R}$ acting on the coordinates as:
\begin{equation}
{\cal R} (z^i) = \ov z^i, \qquad \forall\, i = 1,2,3. 
\end{equation}
The orientifold projection has to be a symmetry of the torus lattice, which constrains the torus lattice to be rectangular ({\bf a}-type lattice) or tilted ({\bf b}-type lattice), as depicted in figure~\ref{Fig:T2Lattices}. For a tilted two-torus lattice the angle $\theta_i$ between the two basic one-cycles is set by the ratio $R_2^{(i)}/R_1^{(i)}$, namely $\cos \theta_i = \frac{1}{2} \frac{R_2^{(i)}}{R_1^{(i)}}$.  

When considering D6-branes on type IIA orientifolds it is easier to work with the real coordinates $(x^i, y^i)$ on torus $T_{(i)}^2$, with periodicity conditions:
\begin{equation}
x^i \simeq x^i +n   , \qquad y^i \simeq y^i + b^i n+ m, \qquad n,m \in \Z
\end{equation}
in line with the representation in figure~\ref{Fig:T2Lattices}, and $b^i=0 \left(\frac{1}{2}\right)$ for a rectangular (tilted) lattice. The $\OR$ projection acts as follows on these coordinates:
\begin{equation}
(x^i, y^i) \stackrel{\OR}{\longrightarrow} (x^i, -y^i).
\end{equation} 
In this coordinate system the symplectic basis $(\alpha_i, \beta^i)$ introduced in (\ref{Eq:PoincareDual}) reads:
\begin{equation} \label{Eq:SymplecticBasisT6OR}
\begin{array}{c@{\hspace{0.4in}}c}
\alpha_0 = d x^1 \wedge dx^2 \wedge dx^3, & \beta^0  =  dy^1 \wedge dy^2 \wedge dy^3,\\
\alpha_1 = d x^1 \wedge dy^2 \wedge dy^3, & \beta^1  =   dy^1 \wedge dx^2 \wedge dx^3,\\
\alpha_2 = d y^1 \wedge dx^2 \wedge dy^3, & \beta^2  =   dx^1 \wedge dy^2 \wedge dx^3,\\
\alpha_3 = d y^1 \wedge dy^2 \wedge dx^3, & \beta^3  =   dx^1 \wedge dx^2 \wedge dy^3,
\end{array}
\end{equation}
and the metric on the six-dimensional torus in this coordinate system is given by:
\begin{equation} 
g_{ab} = {\rm diag} \left((R_1^{(1)} \sin \theta_1 )^2,(R_1^{(2)} \sin \theta_2)^2,(R_1^{(3)} \sin \theta_3)^2,(R_2^{(1)})^2,(R_2^{(2)})^2,(R_2^{(3)})^2\right).
\end{equation}
In order to accommodate the D6-branes we have to introduce a proper basis of three-cycles on $T^6/\OR$, which will depend on the shape of the two-torus lattices. For example, for the {\bf aaa} lattice configuration of $T^6/\OR$ the basis of $\OR$-even three-cycles $\gamma_i$ and $\OR$-odd three-cycles $\delta^i$ is given by:
\begin{equation}\label{Eq:SympBasisT6}
\begin{array}{c@{\hspace{0.4in}}c}
\gamma_0 = [\pi_1] [\pi_3] [\pi_5]  ,  & \delta^0 = [\pi_2] [\pi_4] [\pi_6], \\
\gamma_1 = [\pi_1] [\pi_4] [\pi_6]  ,  & \delta^1 = [\pi_2] [\pi_3] [\pi_5], \\
\gamma_2 = [\pi_2] [\pi_3] [\pi_6]  ,  & \delta^2 = [\pi_1] [\pi_4] [\pi_5], \\
\gamma_3 = [\pi_2] [\pi_4] [\pi_5]  ,  & \delta^3 = [\pi_1] [\pi_3] [\pi_6]. 
\end{array}
\end{equation}  
This basis of three-cycles is de Rahm-dual with respect to the symplectic basis of three-forms $(\alpha_i, \beta^j)$:
\begin{equation}
\int_{\gamma_j} \alpha_i = {\delta_{i}}^j, \qquad   \int_{\delta^j} \beta^i = {\delta^{i}}_j.
\end{equation}
The only non-vanishing intersections between the three-cycles are then given by:
\begin{equation}
\gamma_i \cdot \delta^j = - \delta^j \cdot \gamma_i  = {\delta_i}^j.
\end{equation}
As there are four $\OR$-even three-cycles $\gamma^i$ on $T^6/\OR$, the reduction of the $C_3$ form leads to four independent closed string axions $a^i$ with periodicity $2 \pi$, following the discussion in section~\ref{Ss:ClosedStringAxions}. The metric ${\cal K}_{ij}$ of (\ref{Eq:MetricAxionModuliSpace}) on the four-dimensional axion moduli space is diagonal and can be expressed as:
\begin{equation}\label{Eq:MetricExampleT6}
{\cal K}_{ij} = {\rm diag} \left( u_1 u_2 u_3, \frac{u_1}{u_2 u_3}, \frac{u_2}{u_1 u_3}, \frac{u_3}{u_1 u_2} \right),
\end{equation}
by introducing the parameters $u_i$:
\begin{equation}
u_i = \frac{R_2^{(i)}}{R_1^{(i)} \sin \theta_i}.
\end{equation}
Note that $\theta_1 = \theta_2 = \theta_3 = \frac{\pi}{2}$ in case all three two-tori are rectangular.

With all the geometric utensils at hand, we can start considering an explicit model with factorizable D6-branes. The three-cycle $\Pi_x$ wrapped by a D6$_x$-brane along $T^6/\OR$ can be decomposed in terms of the basis $(\gamma_i, \delta^j)$ as:
\begin{equation}\label{Eq:DecompositionBraneORCycles}
\Pi_x = r_x^i \, \gamma_i + s_x^i\, \delta^i.
\end{equation} 
In case of a factorizable three-cycle $\Pi_x$, the integer coefficients $r_x^i$ and $s_{x}^{i}$ can be written in terms of the torus wrapping numbers $(n_x^1,m_x^1;n_x^2, m_x^2; n_x^3, m_x^3)$ describing how the three-cycle wraps each two-torus individually:
\begin{equation}\label{Eq:ExpFact3WrapNum}
\begin{array}{l@{\hspace{0.4in}}l}
r_x^0 = n_x^1 n_x^2 n_x^3, & s_x^0 = m_x^1 m_x^2 m_x^3,\\
r_x^1 = n_x^1 m_x^2 m_x^3, & s_x^1 = m_x^1 n_x^2 n_x^3,\\
r_x^2 = m_x^1 n_x^2 m_x^3, & s_x^2 = n_x^1 m_x^2 n_x^3,\\
r_x^3 = m_x^1 m_x^2 n_x^3, & s_x^3 = n_x^1 n_x^2 m_x^3,
\end{array}
\end{equation}
where $n_x^i, m_x^i \in \Z$ for $i\in \{1,2,3\}$. We consider a D6-brane stack $a$ supporting the gauge group $U(1)_a$ and a D6-brane stack $b$ supporting the gauge group $U(N_b)$ with wrapping numbers presented in table~\ref{tab:TorusWrappingFactModel}.
\mathtab{
\begin{array}{|c||@{\hspace{0.4in}}c@{\hspace{0.4in}}|@{\hspace{0.2in}}c@{\hspace{0.2in}}||c|}
\hline
\multicolumn{4}{|c|}{\text{\bf Factorizable D6-branes on $T^6/\OR$ for rectangular two-tori}} \\
\hline \hline 
\text{stack} & \text{torus wrapping numbers} & r_x^i = 0, \quad s_x^i = 0  &\text{gauge group}\\
\hline \hline
a & (n_a^1, m_a^1; n_a^2, m_a^2; 1,0)&r_a^1 = 0 = r_a^2 , \quad  s_a^0 = 0 = s_a^3  & U(1)_a\\
b& (n_b^1, m_b^1; n_b^2, m_b^2; 0,1)& r_b^0 = 0 = r_b^3, \quad  s_b^1 = 0 = s_b^2   & U(N_b)\\
\hline
\end{array}
}{TorusWrappingFactModel}{Two-stack D6-brane configuration with factorizable three-cycles supporting a gauge factor $U(1)_a\times U(N_b)$ on the toroidal orientifold $T^6/\OR$.}
For the D6-brane configuration given in table~\ref{tab:TorusWrappingFactModel}, the effective action for the four axions can be written as,
\begin{eqnarray}
{\cal S}_{axion} &=& \bigintssss 
\left[ - \frac{1}{2 \ell_s^2} \sum_{i=0,3} {\cal K}_{ii} (d a^i -  N_b s^i_b A_b) \wedge\star_4  (d a^i -  N_b s^i_b A_b) + \frac{1}{8 \pi^2} \left( r_a^0 a^0 + r_a^3 a^3  \right) F_a \wedge F_a  \right. \nonumber \\
&&\qquad -  \frac{1}{2 \ell_s^2} \sum_{l=1,2} {\cal K}_{ll} (d a^l -  s^l_a A_a) \wedge\star_4  (d a^l -  s^l_a A_a) +  \frac{1}{8 \pi^2}  \left( r^1_b a^1 + r^2_b a^2 \right) \Tr(G_b \wedge G_b)   \nonumber\\
&& \left. \qquad+ \frac{1}{8 \pi^2}  \left( r^1_b a^1 + r^2_b a^2 \right) N_b ( F_b \wedge F_b) \right] . \label{Eq:EffAxionModelFactDbranes}
\end{eqnarray} 
With this D6-brane configuration the  axions $(a^0, a^3)$ and $(a^1, a^2)$ form two decoupled systems of axions, such that we can focus only on the second one. A linear combination of the axions $(a^1, a^2)$ couples anomalously to the $U(1)_b$ field strength, but this coupling is ignored as the $U(1)_b$ gauge field acquires a mass due to the St\"uckelberg mechanism involving the other two axions $(a^0, a^3)$ and the Abelian $U(1)_b$ does not give rise to gauge instantons. Hence, only the second line in the action (\ref{Eq:EffAxionModelFactDbranes}) will be considered and matches the set-up discussed in section~\ref{Ss:U1mixing}. Applying the formulae, and in particular equation (\ref{Eq:AxionDecayConstantU1}),  from that section straightforwardly to the current two-axion system yields the following decay constant (in units of the string scale mass $M_s$):
\begin{equation}
f_{\tilde a_1} = \sqrt{\frac{u_1 u_2}{u_3}} \frac{\sqrt{(u_1)^2 (s_a^1)^2 + (u_2)^2 (s_a^2)^2 }}{\left|r_b^1 s_a^2 (u_2)^2 - r_b^2 s_a^1 (u_1)^2 \right|} M_s,
\end{equation}
for the axionic direction $\tilde a^1$ not absorbed by the $U(1)_a$ gauge boson. At the enhancement point, where the denominator of the axion decay constant becomes small, the internal geometry of $T_{(1)}^2\times T_{(2)}^2$ has to be chosen such that the parameters $u_1$ and $u_2$ exhibit a form of isotropy:
\begin{equation}\label{Eq:EnhancementPointT6Rect}
r_b^1 s_a^2 (u_2)^2  \simeq r_b^2 s_a^1 (u_1)^2. 
\end{equation}
In order to appreciate the meaning of this relation, we introduce the axio-dilaton $S$ and the three complex structure moduli $U_i$ defined as:
\begin{equation}
S \equiv e^{-\Phi} \int_{\gamma_0} \Omega_3 + i \, \int_{\gamma_0} C_3, \quad U_i \equiv e^{-\Phi} \int_{\gamma_i} \Omega_3 + i \, \int_{\gamma_i} C_3  \quad  i =1,2,3,
\end{equation}
with $\Phi$ the ten dimensional dilaton, and $\Omega_3$ the Calabi-Yau three-form which reads in terms of the symplectic basis three-cycles on the {\bf aaa} lattice:
\begin{equation} \label{Eq:ComplexStructureModuli}
\Omega_3 =  \prod_{i=1}^3 R^{(i)}_1 \alpha_0 - \sum_{i=1}^3 R^{(i)}_1 R^{(j)}_2  R^{(k)}_2 \alpha_i - i  \prod_{i=1}^3 R^{(i)}_2 \beta^0 + i \sum_{i=1}^3 R^{(i)}_2 R^{(j)}_1  R^{(k)}_1 \beta^i,
\end{equation}
with $(i,j,k)$ an even permutation of $(1,2,3)$. The isotropy condition (\ref{Eq:EnhancementPointT6Rect}) can now be written in terms of the complex structure moduli $U_1$ and $U_2$ as:
\begin{equation}\label{Eq:EnhIsoFactDbranes}
r_b^1 s_a^2 (\RE U_1)^2  \simeq r_b^2 s_a^1 (\RE U_2)^2,
\end{equation}
which should be read as an isotropy relation between $U_1$ and $U_2$ in the complex structure moduli space.

Note however that this point of the moduli space does not correspond to a supersymmetric configuration for the D6-branes wrapping lagrangian three-cycles on the orientifold $T^6/\OR$. It is well-known that factorizable three-cycles wrapped by supersymmetric D6-branes are calibrated with respect to the same Calabi-Yau three-form $\Omega_3$ as the O6-planes, which boils down to the condition:
\begin{equation}\label{Eq:SUSYreqD6branes}
\varphi^1_x + \varphi^2_x + \varphi^3_x = 0 \text{ mod } 2\pi,
\end{equation}
where the angle $\varphi^i_x$ represents the angle between the O6-plane and the D6-brane $\Pi_x$ on two-torus $T^{2}_{(i)}$:
\begin{equation}
\tan \varphi^i_x = \frac{m_x^i + b^i n_x^i}{n_x^i} u_i.
\end{equation}
It is not difficult to show that the enhancement requirement (\ref{Eq:EnhancementPointT6Rect}) is incompatible with the supersymmetry requirement (\ref{Eq:SUSYreqD6branes}) for both D6-brane stacks $a$ and $b$. Let us therefore start from the assumption that the $b$-stack is wrapped along a special lagangian three-cycle and in order to be more explicit we specify the torus wrapping numbers $(n_b^1, m_b^1; n_b^2, m_b^2) = (1,-1;1,-1)$. The angles $\varphi_b^1$ and $\varphi_b^2$ are then chosen such that the $b$-stack corresponds to a supersymmetric three-cycle satisfying (\ref{Eq:SUSYreqD6branes}), which sets the values for  the ratios $u_1$ and $u_2$. As a last step, the torus wrapping numbers of the $a$-stack are chosen in compliance with equation (\ref{Eq:EnhancementPointT6Rect}) in such a way that the respective three-cycle can be seen as a three-cycle slightly deviating from the three-cycle with wrapping numbers $(1,0;1,0;1,0)$ by a rotation over a small angle along $T_{(1)}^2\times T_{(2)}^2$. In table~\ref{tab:ExampleAStackT6Factoris} we list some explicit examples of  $a$-stack configurations obtained through this method.   

\mathtab{
\begin{array}{|@{\hspace{0.2in}}c@{\hspace{0.2in}}|@{\hspace{0.2in}}c@{\hspace{0.2in}}c@{\hspace{0.2in}}|@{\hspace{0.2in}}c@{\hspace{0.2in}}|@{\hspace{0.2in}}c@{\hspace{0.2in}}|}
\hline
\multicolumn{5}{|c|}{\text{\bf $a$-stack configuration for SUSY $b$-stack on $T^6/\OR$}}\\
\hline \hline
(\varphi_b^1,\varphi_b^2,\varphi_b^3) & u_1 & u_2 & (n_a^1, m_a^1; n_a^2, m_a^2)& (\varphi_a^1,\varphi_a^2,\varphi_a^3)  \\
\hline
(-\frac{\pi}{3}, -\frac{\pi}{6}, \frac{\pi}{2}) & \sim\sqrt{3} & \sim 1/\sqrt{3} & (8,1;4,1) & \sim (12^{\circ},8^{\circ},0)\\
(-\frac{\pi}{4}, -\frac{\pi}{4}, \frac{\pi}{2}) & \sim 1 &\sim 1 & (4,1;4,1) & \sim (14^{\circ},14^{\circ},0)\\
(-\frac{\pi}{6}, -\frac{\pi}{3}, \frac{\pi}{2}) & \sim 1/\sqrt{3}  & \sim\sqrt{3} & (4,1;8,1) &  \sim (8^{\circ},12^{\circ},0) \\
\hline
\end{array}
}{ExampleAStackT6Factoris}{Overview of some explicit D-brane configurations for the $a$-stack: the supersymmetric $b$-stack configuration are represented by the angles in the first column, while  the second column provides the parametric values for the ratios $u_1$ and $u_2$. The third column lists the torus wrapping number along $T_{(1)}^2 \times T_{(2)}^2$ for the $a$-stack, corresponding to the non-supersymmetric angles in the fourth column.}

Even though the closed string sector on $T^6/\OR$ preserves ${\cal N} = 4$ supersymmetry, the open string sector associated to the D-brane configurations in tables~\ref{tab:TorusWrappingFactModel} and~\ref{tab:ExampleAStackT6Factoris} do not preserve any supersymmetry, indicating the possible presence of non-vanishing NS-NS tadpoles. The NS-NS tadpoles are an artefact of the wrong vacuum and can be remediated by a shift of the NS-NS background fields. The consistency of the model is rather measured by the vanishing of the RR tadpoles:
\begin{equation} \label{Eq:RRTadpoleCancel}
\sum_x N_x (\Pi_x + \Pi_{x}') = 4 \Pi_{O6}.
\end{equation}
For the D6-brane configurations listed in table~\ref{tab:ExampleAStackT6Factoris} one can easily check that the RR tadpole cancelation conditions are not satisfied. In order for the RR tadpoles to vanish one can introduce additional stacks of D6-branes whose RR charges compensate the RR charges of the $a$-stack, $b$-stack and O6-planes. Given the sum of the RR charges of the latter we expect the additional stacks to be wrapped along non-supersymmetric three-cycles as well. Moreover, there might be additional contributions to the RR tadpoles upon introducing fluxes intended to stabilize the various K\"ahler and complex structure moduli. We postpone the introduction of a consistent moduli stabilization scheme for future research and leave the RR tadpoles uncanceled for now. As an immediate consequence thereof, the cubic $SU(N_b)$ non-Abelian gauge anomalies are not automatically canceled. And unless the spectrum of chiral fermions is particularly constrained, the two-stack set-up is plagued by gauge anomalies.
\begin{table}[h]
\begin{center}
\begin{tabular}{ccc}
\hline \multicolumn{3}{c}{Overview of Chiral Spectrum for Factorizable D6-branes}\\
\hline \hline
sector & $SU(N_b)_{(Q_a,Q_b)}$ & multiplicity \\
\hline
$ab$ & $( {\bf \ov N_b})_{(1,-1)}$  & $\left|(n_a^1 + m_a^1) (n_a^2 + m_a^2)\right|$\\
$ab'$ &  $( {\bf  \ov N_b})_{(-1,-1)}$& $\left|-(n_a^1 - m_a^1) (n_a^2 - m_a^2)\right|$\\
$bb'$&$(\ov \Anti_b)_{(0,-2)}$& $\left|-4 \right|$ \\
$bb'$&$(\Sym_b)_{(0,2)}$ & 4\\
\hline
\end{tabular}
\caption{Chiral Spectrum for the two-stack D6-brane models on $T^6/\OR$ considered in table~\ref{tab:TorusWrappingFactModel} with wrapping numbers $(n_b^1, m_b^1; n_b^2, m_b^2) = (1,-1;1,-1)$ for the $b$-stack. For the explicit examples in table~\ref{tab:ExampleAStackT6Factoris} the relations $n_a^i > m_a^i>0$ with $i=1,2$ are valid.\label{tab:ChiralSpectrum2StackModel}}
\end{center}
\end{table}
Using the generic chiral spectrum listed in table~\ref{tab:ChiralSpectrum2StackModel} the cubic $SU(N_b)$ non-Abelian gauge anomaly coefficient associated to the two-stack models in tables~\ref{tab:TorusWrappingFactModel} and~\ref{tab:ExampleAStackT6Factoris} can be determined:
\begin{equation}
{\cal A}^{SU(N_b)^3} =\Big[ - 2 n_a^1 n_a^2 - 2 m_a^1 m_a^2 \Big] +  \Big[   - (N_b - 4) 4 + (N_b + 4) 4 \Big],
\end{equation}
where the first part on the righthand side comes from the chiral fermions in the anti-fundamental representation and the second part from the chiral fermions in (anti-) symmetric representation of $SU(N_b)$. One can easily check that this anomaly coefficient does not vanish for any of the explicit models considered in table~\ref{tab:ExampleAStackT6Factoris}, implying that supplementary D6-brane stacks intersecting chirally with the $b$-stack are inevitable for the consistency of these models. Next, we can also consider the mixed Abelian non-Abelian anomaly coefficients for the D6-brane set-up in tables~\ref{tab:TorusWrappingFactModel} and~\ref{tab:ExampleAStackT6Factoris} :
\begin{eqnarray}
{\cal A}^{U(1)_a-SU(N_b)^2}  &=&  n_a^1 m_a^2 + n_a^2 m_a^1 
=   - s_a^1 r_b^1 - s_a^2 r_b^2 , \label{Eq:MixedAbeliannonAbelian1}\\
{\cal A}^{U(1)_b-SU(N_b)^2} &=& - n_a^1 n_a^2 - m_a^1 m_a^2 + 16
= - r_a^0 s_b^0 - r_a^3 s_b^3 +16 , \label{Eq:MixedAbeliannonAbelian2}
\end{eqnarray}
and the Abelian anomaly coefficients:
\begin{eqnarray}
{\cal A}^{U(1)_a^3} &=&  N_b \left( 2n_a^1 m_a^2 +2  m_a^1 n_a^2 \right) = - 2 N_b \left( s_a^1 r_b^1 +  s_a^2 r_b^2  \right) =  {\cal A}^{U(1)_a - U(1)_b^2} , \label{Eq:MixedAbelianAbelian1}\\
{\cal A}^{U(1)_b^3} &=& - N_b \left( 2 n_a^1 n_a^2 + 2 m_a^1 m_a^2 -32 \right) = - N_b \left(2 r_a^0 s_b^0 + 2   r_a^3 s_b^3 -32 \right)  , \label{Eq:MixedAbelianAbelian2}\\
{\cal A}^{U(1)_b - U(1)_a^2} &=& - N_b  \left(2 n_a^1 n_a^2 + 2 m_a^1 m_a^2 \right) =  -N_b \left( r_a^0 s_b^0 +   r_a^3 s_b^3 \right)   . \label{Eq:MixedAbelianAbelian3}
\end{eqnarray}
Note that the anomaly coefficient ${\cal A}^{U(1)_a-SU(N_b)^2} $ matches the charge $\tilde k^2$ of the axion $\tilde a^2$ serving as the longitudinal component of the massive $U(1)_a$ gauge boson, namely:
\begin{equation}
{\cal A}^{U(1)_a-SU(N_b)^2} =  - s_a^1 r_b^1 - s_a^2 r_b^2 = - \tilde k^2,
\end{equation}
implying that constraint~(\ref{Eq:U(1)GaugeInvariancenoGCS}) is trivially satisfied in this set-up and that a GCS-term is not required to ensure $U(1)_a$ gauge invariance. This is an immediate consequence of the generalized Green-Schwarz mechanism, by which also the other mixed Abelian non-Abelian and Abelian anomalies vanish. Hence, only the non-vanishing RR tadpoles and related non-Abelian gauge anomalies remain a worrisome element for this set-up.  

One could try to remediate the non-vanishing RR tadpoles by considering the D6-brane setting on the toroidal orbifold $T^6/\Z_2\times \Z_2$ (with discrete torsion $\eta=-1$), for which global intersecting D6-brane models with vanishing RR tadpoles were found~\cite{Blumenhagen:2005tn}. On this background, one can consider fractional three-cycles consisting of a bulk three-cycle (inherited from the ambient space $T^6$) and exceptional three-cycles stuck at the $\Z_2$ fixed loci of the orbifold action. The bulk part of such a fractional three-cycle can easily be played by the D6-brane configuration given in table~\ref{tab:TorusWrappingFactModel}. Given the technicalities of the exceptional three-cycles, we refrain from introducing the required algebraic elements to fully appreciate those fractional three-cycles and postpone the search for global models on $T^6/(\Z_2\times \Z_2\times \OR)$ to future work. Nonetheless, we can already speculate about potential D-brane instanton corrections on  $T^6/(\Z_2\times \Z_2\times \OR)$ (with discrete torsion) coupling to the closed string axions $a^1$ and $a^2$, using reasonings and arguments analogous to~\cite{Cvetic:2007ku}. The axion $a^1$ couples to Euclidean D-branes wrapping the $\OR\Z_2^{(1)}$-plane, while the axion $a^2$ couples to Euclidean D-branes wrapping the $\OR\Z_2^{(2)}$-plane.  
Given that both axions are charged under the $U(1)_a$ symmetry, their respective D-brane instanton amplitude violates the $U(1)_a$ symmetry. The violation of the $U(1)_a$ symmetry can be traced back to the presence of additional charged zero-modes arising at the intersections between the D6$_a$-brane and the Euclidean D2-branes wrapping the $\OR\Z_2^{(1)}$-plane or $\OR\Z_2^{(2)}$-plane respectively. These fermionic zero-modes can be saturated due to interactions with charged matter fields whose collective $U(1)_a$  charge cancels the $U(1)_a$ charge violation by the instanton amplitude~\cite{Blumenhagen:2006xt,Ibanez:2006da,Blumenhagen:2009qh,Ibanez:2012zz}. Moreover, deformation zero modes for the Euclidean D-branes wrapping the $\OR\Z_2^{(1,2)}$-planes are absent due to the rigid nature of the respective three-cycles. Nonetheless, it is the choice of the exotic O6-plane that determines which of the four O6-plane is $\OR$-invariant and supports $O(1)$-instantons, as expressed by the topological condition in table 10 of~\cite{Forste:2010gw} (see also~\cite{Honecker:2011sm}). In case the $\OR$-plane or the $\OR\Z_2^{(3)}$-plane are chosen as the exotic O6-plane, neither the $\OR\Z_2^{(1)}$ nor the $\OR\Z_2^{(2)}$-plane support $O(1)$-instantons, implying that additional effects are needed to lift the universal fermionic zero-modes of the $U(1)$-type D-brane instantons.\footnote{Generically, $U(1)$-type D-brane instanton contributions are also expected from the Euclidean D2-branes wrapping the cycle $\Pi_a$. From the wrapping numbers in table~\ref{tab:TorusWrappingFactModel} one can however deduce that the axions $a^1$ and $a^2$ do not couple to these D-brane instantons.} Based on these considerations, we expect the anomalous coupling of the axions $a^1$ and $a^2$ to the gauge instantons in (\ref{Eq:EffAxionModelFactDbranes}) to be the dominant non-perturbative effect generating the cosine-type potential for $\tilde a_1$.

One might wonder whether the characteristics of the two-stack models in tables~\ref{tab:TorusWrappingFactModel} and~\ref{tab:ExampleAStackT6Factoris} are influenced by the chosen {\bf aaa} lattice configuration of $T^6/\OR$. Let us therefore pick the {\bf aab} lattice configuration, where only the third two-torus is tilted, and investigate whether this lattice configuration offers better perspectives with respect to model building. We still use the orthogonal coordinate system $(x^i, y^i)_{i=1,2,3}$ introduced in the previous section such that the symplectic basis of three-forms $(\alpha_i, \beta^j)$ is still given by~(\ref{Eq:SymplecticBasisT6OR}). The tiltedness of $T_{(3)}^2$ does alter the basis of $\OR$-even and $\OR$-odd three-cycles:
\begin{equation}\label{Eq:BasisT6tiltedT3}
\begin{array}{l@{\hspace{0.4in}}l}
\gamma_0 = 2 [\pi_1] [\pi_3] [\pi_5] - [\pi_1] [\pi_3] [\pi_6]  ,  & \delta^0 = [\pi_2] [\pi_4] [\pi_6], \\
\gamma_1 = [\pi_1] [\pi_4] [\pi_6]  ,  & \delta^1 =  2 [\pi_2] [\pi_3] [\pi_5] - [\pi_2] [\pi_3] [\pi_6], \\
\gamma_2 = [\pi_2] [\pi_3] [\pi_6]  ,  & \delta^2 = 2 [\pi_1] [\pi_4] [\pi_5]- [\pi_1] [\pi_4] [\pi_6], \\
\gamma_3 = 2 [\pi_2] [\pi_4] [\pi_5]  -  [\pi_2] [\pi_4] [\pi_6]  ,  & \delta^3 = [\pi_1] [\pi_3] [\pi_6]. 
\end{array}
\end{equation}  
The basis of three-cycles are still de Rahm-dual to the basis of three-forms:  
\begin{equation}
\int_{\gamma_j} \alpha_i = c_i\, {\delta_{i}}^j, \qquad   \int_{\delta^j} \beta^i = d_i\, {\delta^{i}}_j,
\end{equation}
but an additional constant $c_i$ or $d_i$ slips in: $ c_0 = 2c_1 = 2 c_2 = c_3=2$ and $2 d_0 = d_1 = d_2 = 2 d_3 = 2$. Moreover, the lattice of $\OR$-even and $\OR$-odd three-cycles does no longer form a uni-modular lattice:
\begin{equation}
\gamma_i \cdot \delta^j = - \delta^j \cdot \gamma_i  = 2 {\delta_i}^j.
\end{equation}
The reduction of the $C_3$ form, as reviewed in section~\ref{Ss:ClosedStringAxions}, yields a closed string axion $\xi^i$ for each of the $\OR$-even three-cycles $\gamma^i$ with periodicity $2\pi/c_i$. The decomposition of a factorizable three-cycle $\Pi_x$ according to (\ref{Eq:DecompositionBraneORCycles}) leads to the following coefficients:
\begin{equation}\label{Eq:CoeffDecompT6ORtiltedT3}
\begin{array}{l@{\hspace{0.4in}}l}
r_x^0 = \frac{1}{2} n_x^1 n_x^2 n_x^3, & s_x^0 = m_x^1 m_x^2 (m_x^3 +  \frac{1}{2} n_x^3),\\
r_x^1 = n_x^1 m_x^2 (m_x^3 +  \frac{1}{2} n_x^3), & s_x^1 =  \frac{1}{2} m_x^1 n_x^2 n_x^3,\\
r_x^2 = m_x^1 n_x^2 (m_x^3 +  \frac{1}{2} n_x^3), & s_x^2 = \frac{1}{2} n_x^1 m_x^2 n_x^3,\\
r_x^3 =  \frac{1}{2} m_x^1 m_x^2 n_x^3, & s_x^3 = n_x^1 n_x^2 (m_x^3+  \frac{1}{2} n_x^3),
\end{array}
\end{equation}
with $n_x^i, m_x^i \in \Z$ for $i\in \{1,2,3\}$. One can repeat the steps leading up to action~(\ref{Eq:EffAxionModelFactDbranes}) for the {\bf aab} lattice configuration, yet there are no substantial differences with respect to the {\bf aaa} lattice configuration. More explicitly, the $T_{(3)}^{2}$ tiltedness alters the torus wrapping numbers for the $a$-stack: $a= (n_a^1, m_a^1; n_a^1, m_a^1; 2,-1)$, but leads to the same effective action as in~(\ref{Eq:EffAxionModelFactDbranes}). Furthermore, the discussion below that action remains valid as well, such that physical considerations regarding action~(\ref{Eq:EffAxionModelFactDbranes}) are lattice independent for factorizable D6-brane models on $T^6/\OR$.

%%%%%%%%%%%%%%%%%%%%%%%%%%%%%%%%%%%%%%%%%%%%%%%%%%%%%%%%%%%%%%%%%%%%%%%%%%%%%%%%%%%
\subsubsection{Non-factorizable D6-branes in Type IIA on Toroidal Orientifolds}\label{Ss:ExampleNonFactD6branes}
In the factorizable D6-brane set-up of the previous section it was implicitly assumed that the $U(1)$ gauge group participating in the St\"uckelberg mechanism does not correspond to the center of a $U(N)$ gauge group supported by a stack of $N$ D6-branes. However, if we consider for a moment that the St\"uckelberg $U(1)$ is indeed the center of a non-Abelian gauge group, we might also be able to associate the instanton background responsible for the axion potential to this $U(N)$ gauge group. In this respect, a single stack of $N$ D6-branes would provide a minimal realization of the model discussed in section~\ref{Ss:U1mixing}. Keeping the number of axions charged under this $U(1)$ equal to two, one can deduce from expression~(\ref{Eq:ExpFact3WrapNum}) that the corresponding D6-brane configuration is not realizable using factorizable three-cycles. As an alternative route we investigate whether such a D6-brane configuration can be consistently obtained by using so-called {\it non-factorisable} three-cycles on $T^6/\OR$.

The factorizable three-cycles in~(\ref{Eq:CoeffDecompT6ORtiltedT3}) live in the homology group $[H_1(T^2,\Z)]^3$, which forms an eight-dimensional sublattice $\Lambda_8$ spanned by the basis $(\gamma_i, \delta^i)$ of the homology group $H_3(T^6,\Z)$ of all three-cycles. Note however that the sum $\Pi_c = \Pi_a + \Pi_b$ of two factorizable three-cycles $\Pi_a$ and $\Pi_b$ is not necessarily factorizable, yet the three-cycle $\Pi_c$ is a three-cycle in the sublattice $\Lambda_8$ and can be decomposed in terms of the basis $(\gamma_i, \delta^i)$. This means that for a generic three-cycle $\Pi_x \in \Lambda_8$, its coefficients $r_x^i$ and $s_x^i$ are not necessarily decomposable in terms of one-cycle wrapping numbers $(n_x^i, m_x^i)$ as in expression~(\ref{Eq:CoeffDecompT6ORtiltedT3}) and its coefficients do not necessarily satisfy~\cite{Rabadan:2001mt,Cremades:2002cs} specific relations as is the case for factorizable three-cycles, such as for instance $r_x^0 s_x^3 = r_x^1 s_x^2 = r_x^2 s_x^1 = r_x^3 s_x^0$. Non-factorizable three-cycles on the $\Lambda_8$-lattice can result from a brane recombination process of two factorizable three-cycles, when the volume of the non-factorizable three-cycle is smaller than the volumes of the two factorizable three-cycles (in the same homology class).       

Releasing the geometrically appealing picture of factorizable D6-branes will provide us with some additional freedom, which will allow us to satisfy the constraints from section~\ref{Sss:GCSterms} in an explicit example. Let us thus consider a stack of $N_a$ D6-branes wrapping a non-factorizable three-cycle $\Pi_a$ and a single D6-brane wrapping a non-factorizable three-cycle $\Pi_b$ whose presence is required to ensure vanishing RR tadpoles. In terms of the basis of three-cycles $(\gamma_i, \delta^j)$ from (\ref{Eq:SympBasisT6}) the respective three-cycles can be decomposed as:\footnote{For simplicity, we assumed that none of the axions is charged under the $U(1)$ gauge group supported by the $b$-stack. One could consider a more generic D6-brane configuration where the axions $a^0$ and $a^1$ are charged under $U(1)_b$ through St\"uckelberg charges $s_b^0$ and $s_b^1$ respectively.}
\begin{equation}\label{Eq:NonFact3cyclesExp}
U(N_a): \Pi_a = r_a^2 \gamma_2 + r_a^3 \gamma_3 + s_a^2 \delta^2 + s_a^3 \delta^3,\qquad U(1)_b: 
\Pi_b = r_b^i \gamma_i.
\end{equation}
The coefficients $r_x^i$ associated to the $\OR$-even cycles will be determined later on when discussing the RR tadpole cancelation conditions. The effective action for the four axions is given by:
\begin{eqnarray}
{\cal S}_{axion} &=& \bigintssss 
\left[ - \frac{1}{2 \ell_s^2} \sum_{i=0,1} {\cal K}_{ii} d a^i \wedge\star_4  d a^i  -  \frac{1}{2 \ell_s^2} \sum_{l=2,3} {\cal K}_{ll} (d a^l - N_a s^l_a A_a) \wedge\star_4  (d a^l -  N_a s^l_a A_a)    \right. \nonumber \\
&&\qquad +  \frac{1}{8 \pi^2}  \left( r^2_a a^2 + r^3_a a^3 \right) \Tr(G_a \wedge G_a) +  + \frac{1}{8 \pi^2} \left( r^2_a a^2 + r^3_a a^3  \right) N_a F_a \wedge F_a  \nonumber\\
&& \left. \qquad+  \frac{1}{8 \pi^2} \left( \sum_{i=0}^3 r_b^i a^i \right)  ( F_b \wedge F_b) \right] . \label{Eq:EffAxionModelNonFactDbranes}
\end{eqnarray} 
Due to the last anomalous coupling, the axion system does not perfectly decouple as the previous model, but the absence of $U(1)_b$ gauge instantons allows us to treat the four axions as two decoupled axion systems $(a^0,a^1)$ and $(a^2,a^3)$. Focusing on the second axion system, we observe that a linear combination is absorbed by the $U(1)_a$ gauge boson, by which the latter acquires its mass, while the orthogonal direction remains uncharged under the Abelian gauge symmetry and couples anomalously to the non-Abelian gauge group with axion decay constant:
\begin{equation}\label{Eq:ADCNonFactD6branes}
f_{\tilde a_1} = \sqrt{\frac{u_2 u_3}{u_1}}  \frac{\sqrt{(u_2)^2 (s_a^2)^2 + (u_3)^2 (s_a^3)^2 }}{\left|r_a^2 s_a^3 (u_3)^2 - r_a^3 s_a^2 (u_2)^2\right| } M_s.
\end{equation}
The axion decay constant follows from a straightforward computation by inserting the metric components~(\ref{Eq:MetricExampleT6}) and the $U(1)$ charges into the expression~(\ref{Eq:AxionDecayConstantU1}). The denominator of the axion decay constant acquires a small value in regions of the moduli space where the following relation is valid:
\begin{equation}
r_a^2 s_a^3 (u_3)^2 \simeq r_a^3 s_a^2 (u_2)^2, \quad \text{ or } \quad r_a^2 s_a^3 (\RE U_2)^2 \simeq r_a^3 s_a^2 (\RE U_3)^2,
\end{equation}  
where we used the expressions for the complex structure moduli introduced in (\ref{Eq:ComplexStructureModuli}) to obtain the second relation. Note also the similarities between this isotropy relation and the isotropy relation~(\ref{Eq:EnhIsoFactDbranes}) for the model with factorisable three-cycles.

Next, we focus on RR tadpoles for this two-stack model, which cancel provided the following relations among the coefficients $r_x^i$ are satisfied:
\begin{equation}\label{Eq:RRtadpolesNonFact}
r_b^0 = 16, \quad r_b^1 = 0, \quad N_a r_a^2 = - r_b^2, \quad N_a r_a^3 = - r_b^3. 
\end{equation} 
An immediate consequence of the vanishing RR tadpoles is the cancelation of the non-Abelian anomalies. This can be checked explicitly by determining the chiral spectrum for this two-stack model as in table~\ref{tab:ChiralSpectrum2StackModelNonFact} and by computing the associated cubic anomaly coefficient:
\begin{equation}
{\cal A}^{SU(N_a)^3} =-2 (s_a^2 r_b^2 + s_a^3 r_b^3) - (N_a+4)  (r^2_a s_a^2 + r_a^3 s_a^3) - (N_a-4)  (r^2_a s_a^2 + r_a^3 s_a^3),
\end{equation}
where the first contribution comes from the chiral states in the (anti-)fundamental representation, the second and third term from the chiral states in the symmetric and anti-symmetric representation respectively. The cubic anomaly coefficient vanishes upon imposing the last two relations in~(\ref{Eq:RRtadpolesNonFact}). 
 \begin{table}[h]
\begin{center}
\begin{tabular}{ccc}
\hline \multicolumn{3}{c}{Overview of Chiral Spectrum for Non-factorizable D6-branes}\\
\hline \hline
sector & $SU(N_b)_{(Q_a,Q_b)}$ & multiplicity \\
\hline
$ab$ & $( {\bf N_a})_{(1,-1)}$  & $\left| - s_a^2 r_b^2 - s_a^3 r_b^3\right|$\\
$ab'$ &  $( {\bf  N_a})_{(1,1)}$& $\left| - s_a^2 r_b^2 - s_a^3 r_b^3\right|$\\
$aa'$&$( \Anti_a)_{(2,0)}$& $\left|-r_a^2 s_a^2 - r_a^3 s_a^3 \right|$ \\
$aa'$&$(\Sym_a)_{(2,0)}$ & $\left|-r_a^2 s_a^2 - r_a^3 s_a^3 \right|$ \\
\hline
\end{tabular}
\caption{Chiral Spectrum for the two-stack non-factorizable D6-brane models on $T^6/\OR$ considered in eq.~(\ref{Eq:NonFact3cyclesExp}).\label{tab:ChiralSpectrum2StackModelNonFact}}
\end{center}
\end{table}
To investigate the $U(1)_a$ gauge invariance constraint we have to consider the mixed Abelian non-Abelian anomaly coefficient associated to the $U(1)_a-SU(N_a)^2$ triangle diagram:
\begin{eqnarray}
{\cal A}^{U(1)_a - SU(N_a)^2}
 = - N_a (r_a^2 s_a^2 + r_a^3 s_a^3)  = - \tilde k^2.
\end{eqnarray}
The last equality implies the conservation of $U(1)_a$ gauge invariance as expressed in~(\ref{Eq:U(1)GaugeInvariancenoGCS}), such that a GCS-term is not required for this model. The other anomaly coefficients can be calculated from the chiral spectrum in table~\ref{tab:ChiralSpectrum2StackModelNonFact}, in analogy with the discussion in the previous section. Let us for instance focus on the anomaly coefficient of the mixed $U(1)_a-U(1)_b^2$ triangle diagram:
\begin{eqnarray}
\mathcal A^{U(1)_a-U(1)_b^2}= - 2 N_a(r^2_b s^2_a+r^3_b s^3_a)= 2 N_a^2(r^2_as^2_a+r^3_as^3_a),
\end{eqnarray}
where we have used the tadpole condition (\ref{Eq:RRtadpolesNonFact}) in the second equality. This computation serves as an additional check for the $U(1)_a$ gauge invariance, that is to say, the $U(1)_a$ gauge variation of the $F_b\wedge F_b$ coupling terms in (\ref{Eq:EffAxionModelNonFactDbranes}) is cancelled by the anomalous $U(1)_a-U(1)_b^2$ triangle diagram. Thus, a GCS-term mixing the $U(1)_a$ gauge potential with the $U(1)_b$ field strength is not required for the gauge consistency of this model either. Thus, the generalized Green-Schwarz mechanism acts as the underlying mechanism in the intersecting D6-brane models to ensure the cancelation of Abelian anomalies and mixed Abelian non-Abelian anomalies and thereby also the quantum consistency of the models.

Considerations regarding other instanton contributions apart from the $SU(N_a)$ gauge instantons follow the same line of reasoning as for the factorizable D6-branes in the previous section, upon lifting the D6-brane configuration to the toroidal orbifold $T^6/\Z_2\times \Z_2$ with discrete torsion. The axions $a^2$ and $a^3$ couple to D-brane instantons whose Euclidean worldvolumes wrap the three-cycles parallel to the $\OR\Z_2^{(2)}$-plane and $\OR\Z_2^{(3)}$-plane respectively on this background. If we choose a background configuration where neither the $\OR\Z_2^{(2)}$-plane nor the $\OR\Z_2^{(3)}$-plane are chosen to be the exotic O6-plane, then the Euclidean D-branes do not support $O(1)$-instantons but rather $U(1)$-instantons, which will contribute effectively only if all fermionic zero modes are saturated.   

A last consideration concerns the amount of supersymmetry preserved by the two-stack non-factorizable D6-brane set-up in (\ref{Eq:NonFact3cyclesExp}). In order for the D6-brane to preserve supersymmetry, its corresponding  three-cycle $\Pi_x$ has to wrap a special lagrangian submanifold which can be expressed in geometric terms as, see e.g.~\cite{Forste:2010gw}:
\begin{equation}\label{Eq:SUSYCondGeneric}
\omega_{(1,1)}\big|_{\Pi_x} = 0, \qquad \IM\left(\int_{\Pi_x} \Omega_3 \right) = 0, \qquad \RE\left(\int_{\Pi_x} \Omega_3 \right) > 0,
\end{equation}
where the first relation expresses a condition on the pull-back of the K\"ahler two-form $\omega_{(1,1)}$ to the three-cycle $\Pi_x$, and $\Omega_3$ is the same calibration form as the one used for the O6-planes. It is not difficult to show that these conditions are satisfied for supersymmetric factorizable D-branes, provided relation (\ref{Eq:SUSYreqD6branes}) is satisfied. For non-factorizable D6-branes, it is much harder to compute the pull-back of the K\"ahler two-form to the respective three-cycle due to the non-factorizability. Given that the non-factorizable three-cycles in (\ref{Eq:NonFact3cyclesExp}) correspond to three-cycles within the lattice $\Lambda_8$, we still expect them to wrap lagrangian subspaces. The remaining two calibration conditions on the other hand can be computed straightforwardly for the non-factorizable three-cycles in (\ref{Eq:NonFact3cyclesExp}). Based on the RR tadpole cancelation conditions (\ref{Eq:RRtadpolesNonFact}) one can then deduce that only one of the two stacks preserves the same supersymmetry as the O6-planes, while the other stack violates the third condition in (\ref{Eq:SUSYCondGeneric}).

To end this section, let us have a look at an explicit example with a modest non-Abelian $U(N_a)$ gauge group with gauge factor $N_a=3$. A point in the parameter space for which both the super-Planckian condition~(\ref{Eq:ADCNonFactD6branes}) and the tadpole condition~(\ref{Eq:RRtadpolesNonFact}) are satisfied, is specified in the first place by the wrapping numbers of the three-cycles for both stacks:
\begin{equation}
\begin{array}{l@{\hspace{0.6in}}l}
r^2_a=r^3_a=1, & r^0_b = 16, \, \, r^1_b= 0, \\
s^2_a=2,\, \, s^3_a=3, & r^2_b=r^3_b= - 3.
\end{array}
\end{equation}
For this choice of parameters, the chiral spectrum in line with table~\ref{tab:ChiralSpectrum2StackModelNonFact} contains the following states: $15 \times ({\bf 3})_{(1,-1)} + 15 \times ({\bf 3})_{(1,1)} + 5 \times ({\bf 3}_{A})_{(-2,0)} + 5 \times (\ov{\bf 6}_{S})_{(-2,0)} $ in the respective representations under the gauge factor $SU(3)_{U(1)_a \times U(1)_b}$. In the region of the parameter space where the ratio $u_3/u_2$ asymptotes to the value $\sqrt{2}/\sqrt{3}$, the axion decay constant (\ref{Eq:ADCNonFactD6branes}) reaches trans-Planckian values for a high enough string scale, e.g.~$M_s~\sim~10^{17}~\text{GeV}$:
\begin{equation}
f_{\tilde a^1} \approx \frac{M_{s}}{3}\times 10^3 \sim 10M_{Pl}.
\end{equation}
In conclusion, this two stack set-up with non-factorizable intersecting D6-branes forms an explicit string theory example of the effective field theory model discussed in section~\ref{Ss:U1mixing}.

%%%%%%%%%%%%%%%%%%%%%%%%%%%%%%%%%%%%%%%%%%%%%%%%%%%%%%%%%%%%%%%%%%%%%%%%%%%%%%%%%%%
\subsubsection{D7-branes in Type IIB on Swiss-Cheese Calabi-Yau's}
In order to find explicit stringy examples characterized by metric mixing and $U(1)$ mixing as analyzed in section~\ref{Ss:GenericKineticMixing}, we 
now turn to backgrounds other than toroidal orbifolds. Metric kinetic mixing is expected for Calabi-Yau backgrounds with a Swiss-cheese type structure, where the volume ${\cal T}$ of the internal space is controlled by the volume $\tau_\ell$ of one large four-cycle $D_b$ subtracted by the volumes $\tau_s$ of small four-cycles $D_s$, which arise as blow-up cycles upon resolving the $\Z_n$ singularities in the Calabi-Yau manifold.\footnote{In this set-up the volumes of the four-cycles are measured with respect to the string length $\ell_s$, and we work with conventions for which basis 4-forms $\alpha_i$ and 2-forms $\beta^i$ are dimensionless.} Considering such Swiss-Cheese Calabi-Yau three-folds allows us to kill two birds with one stone by sketching how the set-up from section~\ref{Ss:GenericKineticMixing} can be realized in Type IIB string theory with intersecting D7-branes, as anticipated in section~\ref{Ss:ClosedStringAxions}.

When considering Type IIB string theory on a Calabi-Yau three-fold ${\cal X}_6$, the metric ${\cal K}_{ij}$ on the $C_4$ axion moduli space in (\ref{Eq:MetricAxionModuliSpace}) can be seen~\cite{Candelas:1990pi,Strominger:1985ks,Grimm:2004uq} as the K\"ahler metric resulting from a K\"ahler potential ${\cal K}$ expressed in the volumes $\tau_i$ of the four-cycles with $i\in \{1,\ldots, h^{11}\}$. The volumes of the four-cycles relate to the K\"ahler deformations $t_i$ through the relations:
\begin{equation}\label{Eq:KahlerDefVolume4Cycle}
\tau_i = \frac{1}{2} \int_{\gamma_i} J\wedge J = \frac{1}{2} \kappa_{i j k} t^j t^k, \qquad i,j,k \in \{1, \ldots,h^{1,1}\},
\end{equation} 
where the K\"ahler two-form $J$ is expanded with respect to a basis of harmonic (1,1)-forms $\beta^i$, the Poincar\'e duals to the basis of four-cycles $\gamma_i$: $J=t^\ell \beta^\ell -\sum_{s=1}^{h^{11}-1} t^s  \beta^s$. The coefficients $\kappa_{ijk}$ correspond to the triple intersection numbers for the basis $\beta^i$. The $C_4$ axions are defined as in equation (\ref{Eq:DecompPforms}) with respect to the basis of four-cycles $\gamma_i$. For a Swiss-Cheese type Calabi-Yau the K\"ahler potential then takes the schematic form (in the large volume limit):
\begin{equation}\label{Eq:GenKaehlerSwissCheese}
{\cal K} = -2 \ln {\cal T} =- 2 \ln  \left( \frac{\sqrt{2}}{3} a_\ell {\tau_\ell}^{3/2} - \frac{\sqrt{2}}{3} \sum_{s=1}^{h^{11}-1} b_s {\tau_s}^{3/2} \right), \qquad a_\ell, b_s \in \R.
\end{equation}
This form of the K\"ahler potential results from expressing the K\"ahler deformations in terms of the four-cycle volumes through~(\ref{Eq:KahlerDefVolume4Cycle}) and inserting the inverted expressions into the internal volume ${\cal T} = \frac{1}{6} \kappa_{ijk} t^i t^j t^k$. The effective four dimensional theory upon dimensional reduction preserves ${\cal N}=1$ supersymmetry when considering the orientifold ${\cal X}_6/\OR(-)^{F_L}$ of the Swiss-Cheese Calabi-Yau three-fold. The appropriate orientifold projection $\OR(-)^{F_L}$ consists of a worldsheet parity $\Omega$, a projection $(-)^{F_L}$ involving the left fermion number $F_L$ and an involution ${\cal R}$, which will be chosen here such that $h^{11} = h^{11}_+$ and $h^{11}_- = 0$ for the remainder of our discourse.\footnote{This has as an immediate consequence that the axions associated to the NS-NS 2-form $B_2$ and RR 2-form $C_2$ are projected out from the start. Choosing a different orientifold projection where not all of the $C_2$-axions are projected out, one could consider stringy realizations of the set-up in section~\ref{Ss:GenericKineticMixing} using the $C_2$-axions, as proposed in~\cite{short}.}  

Intuition gathered from sections~\ref{Ss:GenericKineticMixing}, \ref{Ss:ExampleFactD6branes} and~\ref{Ss:ExampleNonFactD6branes} suggests us to consider Swiss-Cheese Calabi-Yau's with Hodge number $h_+^{11}\geq 3$: given that large axion decay constants seem to be connected to isotropy relations among volume moduli and the validity of the large volume limit approach prohibits a vanishing value for the internal volume ${\cal T}$, we are naturally led to consider Swiss-Cheese Calabi-Yau's with 3 or more K\"ahler moduli. An intensive search through databases of constructed Calabi-Yau three-folds reveals that such Swiss-Cheese Calabi-Yau's are abundantly present~\cite{Gray:2012jy,Altman:2014bfa} and some of them were already fruitful in the past to investigate various aspects regarding D7-brane model building, see e.g.~\cite{Blumenhagen:2007sm,Collinucci:2008sq,Cicoli:2011qg}. Let us thus consider a Swiss-Cheese type Calabi-Yau with $h^{11}_+ = 3$ and for simplicity we assume the presence of a certain amount of isotropy among the volumes of the small four-cycles $\tau_1$ and $\tau _2$:
\begin{equation}\label{Eq:IsotropySwissCheeseCY}
 \tau_1  \sim  \tau_2, \qquad b_1 \sim b_2.
\end{equation}   
With these two assumptions we can expand the metric on the axion moduli space in powers of $\varepsilon^2 \equiv \tau_1/\tau_\ell$:
\begin{equation}
{\cal K}_{ij} = \frac{\partial^2  {\cal K} }{\partial \tau_i \partial \tau_j} = \frac{1}{\tau_\ell^2} \left(\begin{array}{ccc}3 & -\frac{9 b_1}{2a_\ell} \varepsilon &  -\frac{9 b_1}{2a_\ell} \varepsilon \\
 -\frac{9 b_1}{2a_\ell} \varepsilon &  \frac{3 b_1}{2a_\ell} \frac{1}{\varepsilon} & {\cal O}(\varepsilon^2) \\
  -\frac{9 b_1}{2a_\ell} \varepsilon & {\cal O}(\varepsilon^2) &  \frac{3 b_1}{2a_\ell} \frac{1}{\varepsilon}    \end{array}\right),
\end{equation}
where we neglect entries of order ${\cal O}(\varepsilon^2)$ and higher in the limit where $\tau_1 \ll \tau_\ell$. With this simple setting a small amount of metric kinetic mixing among axions can be built in from the start.  

Next, we introduce a single D7-brane supporting the $U(1)_a$ gauge group under which the axions $a^1$ and $a^2$, associated to the four-cycles $\gamma_1$ and $\gamma_2$ respectively, are charged. In order for the axions to acquire St\"uckelberg charges, we have to turn on an internal flux ${\cal F}_a$ as reviewed in section~\ref{Ss:ClosedStringAxions}:
\begin{equation}
{\cal F}_a =  f_a^1 \beta^1 + f_a^2 \beta^2, \qquad f_a^i \in \Q_0.
\end{equation}
Note that we turn on the internal flux ${\cal F}_a$ along the two-forms that are Poincar\'e dual to the four-cycles wrapped by the $U(1)_a$ stack: \begin{equation}
\gamma_{a} = n_a^1 \gamma_1 + n_a^2 \gamma_2, \qquad n_a^i \in \Q_0.
\end{equation}
With respect to this D7-brane configuration the charge vector $(p^\ell, p^1, p^2)$ is given by:
\begin{equation}
p^\ell =0, \qquad p^1 = \frac{1}{4 \pi} \frac{f_a^1 n_a^1}{b_1^2} , \qquad p^2 = \frac{1}{4 \pi} \frac{f_a^2 n_a^2}{b_1^2},
\end{equation}
where we used the triple intersection numbers adapted to the basis in which the K\"ahler potential (\ref{Eq:GenKaehlerSwissCheese}) has been expressed:
\begin{equation}\label{Eq:TripleIntersectionNum}
I_3 = \frac{1}{a_\ell^2} \gamma_\ell^3 + \frac{1}{b_1^2} \gamma_1^3  + \frac{1}{b_2^2} \gamma_2^3  \stackrel{(\ref{Eq:IsotropySwissCheeseCY})}{=} \frac{1}{a_\ell^2} \gamma_\ell^3 + \frac{1}{b_1^2} \gamma_1^3  + \frac{1}{b_1^2} \gamma_2^3,
\end{equation}
with $a_\ell^{-2}, b_1^{-2}, b_2^{-2} \in \Z$. Notice that the D7-brane configuration is chosen in such a way that only the two axions associated to the small four-cycles $\gamma_1$ and $\gamma_2$ are charged under the $U(1)_a$ gauge group supported by the $a$-stack.
\footnote{Intuitively one might expect a D7-brane stack to wrap only a single four-cycle. Note that the expression for the internal volume ${\cal T}$ in (\ref{Eq:GenKaehlerSwissCheese}) is closely related to the explicit form of the triple intersection numbers in (\ref{Eq:TripleIntersectionNum}) and is thus only valid in this particular basis of four-cycles. This particular basis of four-cycles consists of linear combinations of $\OR$-even four-cycles used to define the resolved Calabi-Yau manifold in terms of toric geometry. In this respect a D7-brane wrapping a linear combination of four-cycles $\gamma_1$ and $\gamma_2$ results naturally from a configuration where the D7-brane wraps a single $\OR$-even four-cycle in the original toric geometry basis.  
 }

Thirdly, we also introduce a stack of $N_b$ D7-branes wrapping a four-cycle $\gamma_b$ such that both axions $a^1$ and $a^2$ couple anomalously to the $U(N_b)$ gauge group supported by the D7-brane stack:
\begin{equation}
\gamma_b = m_b^1 \gamma_1 + m_b^2 \gamma_2, \qquad m_b^i \in \Q_0.
\end{equation}
In order to prevent that the axions $a^1$ and $a^2$ are charged under the $U(1)_b$ center of the non-Abelian gauge group, we assume that the vector bundle along the internal directions is flat. Taking all these elements into account, we find that the effective action for the three axions $a^l$, $a^1$ and $a^2$ in this set-up is given by:
\begin{eqnarray}
{\cal S}_{axion} &=& \bigintssss 
\left[ - \frac{1}{2 \ell_s^2} \sum_{i,j\in\{\ell,1,2\}} {\cal K}_{ij} (d a^i - p^i A_a) \wedge\star_4  (d a^j - p^j A_a)   +  \frac{1}{8 \pi^2}  \left( n_b^1 a^1 + n_b^2 a^2 \right) (F_a \wedge F_a)  \right. \nonumber\\
&& \left.  \qquad +  \frac{1}{8 \pi^2}  \left( m_b^1 a^1 + m_b^2 a^2 \right) \Tr(G_b \wedge G_b) +  \frac{1}{8 \pi^2}  \left( m_b^1 a^1 + m_b^2 a^2 \right) N_b (F_b \wedge F_b)  \right].
\end{eqnarray}
Determining the decay constants for the axions and the axionic direction eaten by the $U(1)_a$ gauge boson requires us to apply the same steps as presented in section~\ref{Ss:GenericKineticMixing} for the three-axion system. As a first step, we diagonalize the metric ${\cal K}_{ij}$ on the axion moduli space, whose eigenvalues are given by:
\begin{equation}
\lambda_+ \simeq \frac{3}{\tau_\ell^2} + {\cal O}(\varepsilon^3) , \qquad \lambda_- \simeq \frac{3 b_1}{ 2 a_\ell}  \frac{1}{\varepsilon \tau_\ell^2} + {\cal O}(\varepsilon^3), \qquad \lambda_3 = \frac{3 b_1}{ 2 a_\ell} \frac{1}{\varepsilon \tau_\ell^2} ,
\end{equation}
such that the kinetic terms for the axions can be written as,
\begin{eqnarray}
{\cal S}_{axion}^{\rm kin} &\ni& \bigintssss 
\left[ - \frac{1}{2 \ell_s^2} \lambda_+ \left( d a^+ - \frac{p^+}{\pi} A_a \right) \wedge \star_4 \left( d a^+ - \frac{p^+}{\pi} A_a \right) \right. \nonumber\\
&& \qquad \quad  - \frac{1}{2 \ell_s^2} \lambda_- \left( d a^- - \frac{p^-}{\pi} A_a \right) \wedge \star_4 \left( d a^- - \frac{p^-}{\pi} A_a \right)   \nonumber\\
&& \qquad \quad \left. - \frac{1}{2 \ell_s^2} \lambda_3 \left( d a^3 - \frac{p^3}{\pi}  A_a \right) \wedge \star_4 \left( d a^3 - \frac{p^3}{\pi} A_a \right)  \right],
\end{eqnarray}
where also the St\"uckelberg charges have to be expressed in terms of the new axion basis: 
\begin{equation}
\left(
\begin{array}{c} p^+ \\ p^- \\ p^3 \end{array} \right) =   \frac{1}{\sqrt{2}\sqrt{1+ 18 \varepsilon^4}} \left(\begin{array}{ccc}  \sqrt{2} & 3\sqrt{2}\varepsilon^2 & 3\sqrt{2}\varepsilon^2\\
0& - \sqrt{1+18 \varepsilon^4}&\sqrt{1+18 \varepsilon^4}  \\
 -6 \varepsilon^2& 1& 1\\
 \end{array} \right)  \left(\begin{array}{c} p^\ell \\ p^1 \\ p^2 \end{array} \right) = \left( \begin{array}{c} {\cal O}(\varepsilon^2) \\ \frac{p^2-p^1}{\sqrt{2}} \\ \frac{p^2+p^1}{\sqrt{2}} \end{array}\right).
\end{equation}
From the righthand side we deduce that the charge of the axion $a^+$ under the $U(1)_a$ gauge symmetry is negligible, such that only the axions $a^-$ and $a^3$ are characterized by St\"uckelberg charges.   
In order to determine the axion decay constants, we also have to write down the anomalous couplings to the gauge groups in terms of the new axion basis:
\begin{eqnarray}
{\cal S}_{axion}^{anom} &=& \frac{1}{8 \pi^2} \bigintssss \left[  \left( \frac{m_b^2-m_b^1}{\sqrt{2}} a^- + \frac{m_b^2+m_b^1}{\sqrt{2}}  a^3 \right) \Tr(G_b \wedge G_b)       \right. \nonumber \\
 && \qquad \quad + \left( \frac{n_b^2-n_b^1}{\sqrt{2}} a^- + \frac{n_b^2 + n_b^1}{\sqrt{2}}a^3 \right) (F_a \wedge F_a) \nonumber \\
  && \qquad \quad \left.  +\left( \frac{m_b^2-m_b^1}{\sqrt{2}} a^- + \frac{m_b^2+m_b^1}{\sqrt{2}}  a^3 \right) N_b(F_b \wedge F_b)   \right].
\end{eqnarray}
Once the $SU(N_b)$ instanton background is taken into consideration, a cosine-type potential for the axions will be generated, and therefore only the anomalous coupling to the non-Abelian gauge group deserves our attention in the remainder of this discussion. Combining the kinetic part and the potential terms for the axions $(a^-, a^3)$ we notice that the effective action for this two-axion system matches the set-up from section~\ref{Ss:U1mixing}. Applying the analysis from that section to this two-axion system, we can identify a linear combination $\tilde a^2$ of the axions $(a^-, a^3)$ as the axionic direction absorbed by the $U(1)_a$ gauge boson which acquires a St\"uckelberg mass of the order:
\begin{equation}
M_a = \sqrt{ \frac{3 b_1}{ 2 a_\ell} \frac{1}{\varepsilon \tau_\ell^2} } \sqrt{(p^1)^2 + (p^2)^2} M_s .
\end{equation}  
The orthogonal linear combination $\tilde a^1$ of $(a^-, a^3)$ then survives as the inflaton candidate coupling to the non-Abelian gauge group with an axion decay constant (\ref{Eq:AxionDecayConstantU1}) given by:
\begin{equation}
f_{\tilde a^1} = \sqrt{ \frac{3 b_1}{ 2 a_\ell} \frac{1}{\varepsilon \tau_\ell^2} } \frac{\sqrt{(p^1)^2 + (p^2)^2}}{ \left| p^1 m_b^2 - m_b^1 p^2 \right|} M_s.
\end{equation} 
Large axion decay constants ($f_{\tilde a^1} \gg M_s$) can be found in regions of the parameter space where $p^1 m_b^2 - m_b^1 p^2$ asymptotes to zero:
\begin{equation}\label{Eq:SwissCheeseACDIsotropy1}
f_a^1 n_a^1 m_b^2 \simeq m_b^1 f_a^2 n_a^2.
\end{equation}
At first sight this condition seems rather restrictive, but it should actually be combined with the isotropy relations in (\ref{Eq:IsotropySwissCheeseCY}). By relaxing the latter conditions, a trans-Planckian decay constant is realized for a sufficiently high string scale $M_s \sim {\cal O}(10^{16}-10^{17})$ GeV, provided that the following isotropy relation between the volumes $\tau_1$ and $\tau_2$ of the small four-cycles is valid:
\begin{equation}\label{Eq:SwissCheeseACDIsotropy2}
\frac{\tau_1}{\tau_2} \simeq \frac{b_1^2\, (m_b^2 + m_b^1)^2\, (f_a^2 n_a^2 -f_a^1 n_a^1)^2}{b_2^2\, (m_b^2 - m_b^1)^2\, (f_a^2 n_a^2 +f_a^1 n_a^1)^2}.
\end{equation}
In this expression $b_1$ and $b_2$ are constants fixed by the geometry of the internal space, $m_b^i$, $n_a^i$ and $f_a^i$ are integer (or at most rational) parameters which can be freely chosen. This latter isotropy condition seems more flexible and easier to satisfy from a model building perspective than the one in (\ref{Eq:SwissCheeseACDIsotropy1}), but we have to keep in mind that the K\"ahler moduli are constrained to lie within the K\"ahler cone such that the volumes of all curves and four-cycles on ${\cal X}_6$ are positive. It has to be verified for a specific Swiss-Cheese Calabi-Yau whether the isotropy condition in (\ref{Eq:SwissCheeseACDIsotropy2}) is compatible with the K\"ahler cone constraints. 

Similar to the D6-branes models in the previous sections, the quantum consistency of these D7-brane models relies on the vanishing of the RR tadpoles (D3-brane and D7-brane tadpoles) and the cancelation of mixed anomalies by virtue of the generalized Green-Schwarz mechanism~\cite{Plauschinn:2008yd,Blumenhagen:2008zz}. Discussing the quantum consistency is facilitated when considering an explicit Swiss-Cheese Calabi-Yau background with specific orientifold projection, which we will postpone for future work. Nevertheless, we can already speculate that for the considered D7-brane configurations all anomalies involving an Abelian gauge factor can be canceled through the generalized Green-Schwarz mechanism and a GCS-term is not required to restore $U(1)_a$ gauge invariance. More explicitly, at the intersections between the $a$-stack and the $b$-stack we expect the presence of chiral matter in the bifundamental representation under the respective gauge groups with multiplicity:
\begin{equation}
I_{ab} = \bigintssss_{{\cal X}_6} \Big(c_1({\cal F}_a) - c_1({\cal F}_b)\Big) \wedge [\gamma_a] \wedge [\gamma_b] = \frac{f_a^1 n_a^1 m_b^1}{b_1^2} + \frac{f_a^2 n_a^2 m_b^2}{b_2^2} \in \Z  ,
\end{equation}
where $c_1({\cal F}_{a,b})$ corresponds to the first Chern-class of the respective gauge bundles ${\cal F}_{a,b}$ and $[\gamma_{a,b}]$ denote the Poincar\'e dual two-forms to the respective four-cycles $\gamma_{a,b}$. Similar expressions can be written down for the $ab'$, $aa'$ and $bb'$ sectors. The anomaly coefficients for the triangle diagrams associated to the chiral anomaly match the couplings for the Green-Schwarz diagrams, such that the sum of both types of diagram equals zero for the mixed Abelian non-Abelian and the pure Abelian anomalies. The cubic non-Abelian $SU(N_b)$ gauge anomaly on the other hand vanishes provided that the RR tadpoles vanish.   

The presence of a $U(1)$ bundle along $\gamma_a$ also induces a moduli-dependent Fayet-Iliopoulos term $\xi_a$:
\begin{equation}
\xi_a = \frac{1}{{\cal T}} \bigintssss [\gamma_a] \wedge c_1({\cal F}_a) \wedge J =  \frac{1}{{\cal T}} \left( \frac{n_a^1 f_a^1}{b_1^2} t^1 +  \frac{n_a^2 f_a^2}{b_2^2} t^2 \right).
\end{equation}
In combination with the scalar fields $\phi^{(i)}$ from the chiral D7-brane sector charged under $U(1)_a$ gauge group with charge $q^{(i)}_a$, the associated D-term potential scales as,
\begin{equation}
V_D \sim \left( \sum_{i} q^{(i)}_a \left| \phi^{(i)} \right|^2 - \xi_a \right)^2.
\end{equation}
In order for this D-term to vanish, there exist two possible options: either the FI-term $\xi_a$ vanishes, or there is a scalar field (singlet under the $SU(N_b)$ gauge group) whose vacuum expectation value matches the FI-term. In case neither of the two options can be met, the non-vanishing D-term potential might indicate that the considered D7-brane configuration is not supersymmetric. A second issue, absent in the intersecting D6-brane picture but instrumental for the consistency of the D7-brane models, concerns the presence of Freed-Witten anomalies, whenever the D7-branes are wrapped on four-cycles which do not admit a spin structure~\cite{Freed:1999vc}. In order to cancel the Freed-Witten anomalies associated to non-spin four-cycles, the internal flux supported by the D-branes has to contain a contribution that is half-integer quantized. 

Once a specific Swiss-Cheese Calabi-Yau is chosen and a consistent D7-brane model is constructed according to the set-up given above, one has to determine whether there exist (rigid) Euclidean D3-brane instantons  wrapping the four-cycles associated to the axions $a^1$ and $a^2$ and verify that their instanton amplitudes are suppressed with respect to the gauge instanton responsible for the axion potential. Yet, the biggest and most exciting challenge in this framework will consist in tying the aforementioned D7-brane configuration to the mechanisms responsible for stabilizing the volumes $\tau_i$ of the four-cycles, allowing us to find a dynamical explanation for the isotropy relation (\ref{Eq:SwissCheeseACDIsotropy2}).

%%%%%%%%%%%%%%%%%%%%%%%%%%%%%%%%%%%%%%%%%%%%%%%%%%%%%%%%%%%%%%%%%%%%%%%%%%%%%%%%%%%%%%%%%%%%%%%%%%%%%%%%%%%%%%%%%%%%%%%%%%%%%%%%%%%%%%%%%%%%%%%%%%%%%%%%%%%%%%%%%%%%%%%%%%%%%%%%%%%%%%%%%%%%%%%
\section{Conclusion}\label{sec:con}
In this paper, we proposed a new mechanism to obtain an effective
super-Planckian axion decay constant in theories where the axion periodicities are intrinsically sub-Planckian. Our mechanism involves neither monodromy nor 
alignment of the axion decay constants, but kinetic mixing effects among 2 or more axions. The simplicity of our approach brings several virtues.
First of all, the field range enhancement we obtained with kinetic and $U(1)$ mixings is not tied to the number of low energy degrees of freedom (such as the number of axions or the rank of non-Abelian gauge groups).
In fact, the simplest model we presented involves only two axions, a $U(1)$ and a small rank non-Abelian gauge group\footnote{
Some chiral fermions charged under it are also needed but they are there in any case for anomaly cancellation (their presence is implicit in models that invoke non-perturbative instanton potentials).}.
The simplicity of our scenario further enables us to {\it explicitly} integrate out the heavy fields to obtain an effective single axion lagrangian, providing a minimal realization of natural inflation.
Unlike  the alignment mechanism \cite{Kim:2004rp}, 
the effective field range in our scenario is enhanced not by a fine-tuning of discrete parameters, but rather by fine-tuning continuous moduli-dependent quantities.
Thus, the requirement enhancement $f_{eff}/f \gtrsim 100$ can be  satisfied much more readily.
Our mechanism is also different from monodromy inflation in that there is only a single branch of the potential. Hence, there is no additional requirement on model building for the tunneling between different branches to be suppressed. While our scenario applies generally to field theories with multiple axions, it is most naturally realized in string theory, as exposed by the explicit examples consisting of intersecting D6-branes in Type IIA and intersecting D7-branes in Type IIB string theory.
As is inevitable in string inflation models, moduli stabilization is a major challenge. While our scenario may seem to impose additional requirements on moduli stabilization, it is interesting to note that the effective axion decay constant is enhanced (or reduced) at symmetric points in the moduli space and thus the tuning needed in fact may be natural from a moduli stabilization standpoint.

The kinetic and $U(1)$ mixings invoked in this work appear rather generically in string compactifications. Axions in string theory are typically mixed kinetically (at tree level) and St\"uckelberg couplings are in fact a necessity for anomaly cancellation in string theory.
The lagrangian for the multi-axion system considered here is thus more general, and subsumes that of previous proposals. The general lagrangian presented in appendix \ref{A:Generalisation}  thus provides a well-motivated starting point for further  studies of multi-axion inflation, both in terms of model building and statistical studies. It would be interesting to carry out a 
random matrix analysis of an ensemble of lagrangians of the form of eq.~(\ref{Eq:CompleteGeneralAxionGauge}).
Other than inflation, a broader range of axion decay constants made possible by axion mixings may find applications in other contexts~\cite{short}. It would also be interesting to find explicit string compactifications realizing the inflationary conditions outlined in this work, perhaps in conjunction with realistic particle physics features.
We hope to return to these issues in future work.

%%%%%
%%%%%
\acknowledgments
%%%%%
%%%%%
We would like to thank Kiwoon Choi, 
Michele Cicoli, Daniel Junghans,
Luis Ib\'a\~nez,  Fernando Marchesano, Francisco Pedro, Jan Rosseel, Pablo Soler and Angel Uranga for useful discussions and suggestions.  G.S. and F.Y. are supported in part by the DOE grant DE-FG-02-95ER40896 and the HKRGC grants 
HUKST4/CRF/13G, 604231 and 16304414.
W.S. was initially supported by the {\it Cluster of Excellence `Precision Physics, Fundamental Interactions and Structure of Matter' (PRISMA)} DGF no. EXC 1098 and
the DFG research grant HO 4166/2-1, but is now supported by the ERC Advanced Grant SPLE under contract ERC-2012-ADG-20120216-320421, by the grant FPA2012-32828 from the MINECO, and the grant SEV-2012-0249 of the ``Centro de Excelencia Severo Ochoa" Programme. W.S.~would also like to thank the European COST action MP1210 ``The String Theory Universe" for a Short Term Scientific Mission Grant.

%%%%%%%%%%%%%%%%%%%%%%%%%%%%%%%%%%%%%%%%%%%%%%%%%%%%%%%%%%%%%%%%%%%%%%%%%%%%%%%%%%%%%%%%%%%%%%%%%%%%%%%%%%%%%%%%%%%%%%%%%%%%%%%%%%%%%%%%%%%%%%%%%%%%%%%%%%%%%%%%%%%%%%%%%%%%%%%%%%%%%%%%%%%%%%%%%%%%%%%%%%%%%%%%%%%%%%%%%%%%%%%%%%%%%%%%%%%%%%%%%%%%%%%%%%%%%%%%%%%%%%%%%%%%%%%%%%%%%%%%%%%%%%%%%%%%%%%%%%%%%%%%%%%%%%%%%%%%%%%%%%%%%%%%%%%%%%%%

\appendix

%%%%%%%%%%%%%%%%%%%%%%%%%%%%%%%%%%%%%%%%%%%%%%%%%%%%%%%%%%%%%%%%%%%%%%%%%%%%%%%%%%%%%%%%%%%%%%%%%%%%%%%%%%%%%%%%%%%%%%%%%%%%%%%%%%%%%%%%%%%%%%%%%%%%%%%%%%%%%%%%%%%%%%%%
\section{Scales, Conventions and Notations}\label{A:Conventions}
We provide here an overview of various scales appearing throughout the paper. In the first place,  there is the reduced Planck mass $M_{Pl}$, i.e. $M_{Pl}  = (8 \pi G_N)^{-1} \sim 2.4 \times 10^{18}\,\text{GeV} $ in natural units. Secondly, there are the string mass scale $M_s$ and string length $\ell_s$, which are both related to the $\alpha'$ parameter: $M_s^{-1}= \ell_s = 2 \pi \sqrt{\alpha'}$. The ten-dimensional gravity coupling $\kappa_{10}^2$, expansion parameter for the bulk NS-NS action, is in its turn set by the string length $\ell_s$:
\begin{eqnarray}\label{Eq:10dGravStringLength}
\kappa_{10}^2 = \frac{1}{4 \pi} (4 \pi^2 \alpha')^4 = \frac{\ell_s^8}{4 \pi}.
\end{eqnarray}
The parameter appearing in the Dirac-Born-Infeld action for a D$p$-brane, related to the D-brane tension and charge, is also set by the string length $\ell_s$: 
\begin{equation}\label{Eq:DbraneStringLength}
\mu_p = \frac{1}{(2 \pi)^p (\alpha')^{(p+1)/2}} = \frac{2 \pi}{\ell_s^{p+1}},
\end{equation}
where we used the conventions of~\cite{Ibanez:2012zz,Blumenhagen:2006ci}.

Let us also present our conventions regarding differential $p$-forms defined on an $n$-dimensional differentiable manifold ${\cal M}$.
A differential $p$-form (or simply $p$-form) $C_{(p)} \in \Omega^p({\cal M})$ is a totally antisymmetric tensor of type $(0,p)$:
\begin{equation}
C_{(p)} = C_{\mu_1 \ldots \mu_p} \ud x^{\mu_1} \ldots \ud x^{\mu_p} = \frac{1}{p!} C_{\mu_1 \ldots \mu_p} \ud x^{\mu_1} \wedge \ldots \wedge \ud x^{\mu_p},
\end{equation}
where we introduced the local coordinates $\left(x^{\mu= 1, \ldots, n}\right)$ on ${\cal M}$. The differential operator acting on $p$-forms is offered by the exterior derivative $\ud: \Omega^{p} ({\cal M}) \rightarrow \Omega^{p+1} ({\cal M})$, acting as follows in local coordinates:
\begin{equation}
\ud C_{(p)}  = \frac{1}{p!} \partial_\mu C_{\nu_1 \ldots \nu_p} \ud x^\mu \wedge \ud x^{\nu_1} \wedge \ldots \wedge \ud x^{\nu_p}.
\end{equation}
Some useful properties regarding differential form calculus are:
\begin{equation}
\begin{aligned}
&C_{(p)} \wedge D_{(r)} = (-)^{pr} D_{(r)} \wedge C_{(p)},\\
&\ud (C_{(p)} \wedge D_{(r)}) = \ud C_{(p)} \wedge D_{(r)} + (-)^p C_{(p)} \wedge \ud D_{(r)}.
\end{aligned}
\end{equation}
In case the differentiable manifold is equipped with a metric $ds^2 = g_{\mu \nu} \ud x^\mu\otimes \ud x^\nu$ and $g = \det(g_{\mu \nu})$, we can introduce the Hodge star $\star_{(n)}$ as the linear map $\Omega^p ({\cal M}) \rightarrow \Omega^{n-p} ({\cal M})$ between the space of $p$-form and space of $(n-p)$-form on ${\cal M}$,
\begin{equation}
\star_{(n)} ( \ud x^{\mu_1} \wedge \ldots \wedge \ud x^{\mu_p}) = \frac{\sqrt{|g|}}{(n-p)!} g^{\mu_1 \rho_1} \cdots g^{\mu_{p} \rho_{p}} \varepsilon^{(n)}_{\rho_1 \ldots \rho_p \nu_{p+1} \ldots \nu_n} \ud x^{\nu_{p+1}} \wedge \ldots \wedge \ud x^{\nu_{n}},
\end{equation}
such that the action of the Hodge star on a $p$-form leads to the following expression for the $(n-p)$-form in local coordinates:
\begin{equation}
\star_{(n)} C_{(p)} = \frac{\sqrt{|g|}}{p! (n-p)!} C_{\mu_1 \ldots \mu_p} g^{\mu_1 \rho_1} \cdots g^{\mu_{p} \rho_{p}} \varepsilon^{(n)}_{\rho_1 \ldots \rho_p \nu_{p+1} \ldots \nu_n}  \ud x^{\nu_{p+1}} \wedge \ldots \wedge \ud x^{\nu_{n}}.
\end{equation}
The invariant volume element can also be written using the Hodge star:
\begin{equation}
\star_{(n)} \1 = \frac{\sqrt{|g|}}{n!} \varepsilon_{\mu_1 \ldots \mu_n} \ud x^{\mu_1} \wedge \ldots \wedge \ud x^{\mu_n} = \sqrt{|g|} \ud x^1 \wedge \ldots \wedge \ud x^n .
\end{equation}
The Hodge star and the differential form language is most convenient to write down the kinetic parts for $p$-forms in a compact way:
\begin{equation}
\int_{{\cal M}} C_{(p)} \wedge \star_n C_{(p)} = \int_{{\cal M}} \frac{1}{p! (n-p)!} C_{\mu_1 \ldots \mu_p} C^{\mu_1 \ldots \mu_p} \sqrt{|g|} \ud^n x.
\end{equation}

The language of differential forms is extremely suited for gauge theories, both Abelian as well as non-Abelian. We will limit our expressions to four dimensions, but they can be generalized without  any problem to other dimensions. For the Abelian gauge symmetry in this paper we denote the field strength by $F$ and the gauge potential by $A$, such that $F=\ud A$. The kinetic terms for the Abelian gauge fields read in differential form language:
\begin{equation}
\int_{{\cal M}} F \wedge \star_4 F = \int_{{\cal M}} \ud^4 x \sqrt{|g|}  \,  \frac{1}{4} F_{\mu \nu} F^{\mu \nu}. 
\end{equation} 
For a non-Abelian gauge group we first introduce a set of generators $T_a$ spanning a Lie algebra $[T_a, T_b] = i {f_{ab}}^c \, T_c$ and satisfying $\Tr(T_a T_b) = \frac{1}{2} \delta_{ab}$. The gauge potential $B = B^a T_a$ corresponds to a set of one-forms transforming in the adjoint representation of the Lie algebra. The field strength $G$ for a non-Abelian gauge theory can also be defined in terms of the gauge potential $B$ as:  
\begin{equation}
G = \ud B + B \wedge B, 
\end{equation}
or equivalently in local coordinates $G_{\mu \nu} = \partial_\mu B_\nu - \partial_\nu B_\mu + [B_\mu, B_\nu]$. The kinetic terms for the non-Abelian gauge fields can also be expressed elegantly by using differential forms:
 \begin{equation}
\int_{{\cal M}} \Tr(G\wedge \star_4 G) = \int_{{\cal M}} \ud^4 x \sqrt{|g|}  \, \frac{1}{4} \Tr (G_{\mu \nu} G^{\mu \nu})
 \end{equation}
For a non-Abelian gauge theory in four dimensions we can introduce the Chern-Simons three-form $\Omega$, defined as
\begin{equation}\label{Omega}
\Omega \equiv  \Tr\left( G\wedge B - \frac{1}{3} B \wedge B \wedge B \right),
\end{equation}
such that the closure of the three-form corresponds to the second Chern character associated to the gauge potential $B$: 
\begin{equation}\label{Eq:ClosureCS3form}
\ud \Omega = \Tr\left( G \wedge G  \right).
\end{equation} 
We can also treat these expressions in local coordinates, for which the Chern-Simons 3-form is given by, 
\begin{equation}
\Omega_{\mu \nu \rho} = \frac{1}{3} \Tr \left( \left\{G_{\mu \nu} B_\rho  - \frac{1}{6} [B_\mu, B_\nu] B_\rho \right\} + \text{cyclic permutations in } (\mu, \rho, \nu)\right).
\end{equation}

We finish this section with a couple of relations which will allow us to expose the relation between the topological charge density $\Tr(G\wedge G)$ and the generalized Chern-Simons term introduced in section~\ref{Sss:GCSterms}. The topological charge density reads in local coordinates:  
\begin{eqnarray}
\int_{{\cal M}}  \Tr (G\wedge G) &=&\int_{{\cal M}}  \ud^4 x \, \frac{1}{4}  \varepsilon^{\mu \nu \rho \sigma} \Tr(G_{\mu \nu} G_{\rho \sigma})\\
&=& \int_{{\cal M}}  \ud^4 x \,  \varepsilon^{\mu \nu \rho \sigma} \partial_\mu \Tr\left( B_\nu \partial_\rho B_\sigma + \frac{2}{3} B_\nu B_\rho B_\sigma \right).
\end{eqnarray}
The generalized Chern-Simons term (\ref{Eq:GCStermU1mixing}) on the other hand reads in local coordinates:
\begin{eqnarray}
\int_{{\cal M}}   A \wedge \Omega &=&\int_{{\cal M}}  d^4 x\, \varepsilon^{\mu \nu \rho \sigma} A_\mu \Omega_{\nu \rho \sigma}\\
&=& \int_{{\cal M}}  \ud^4 x\, \varepsilon^{\mu \nu \rho \sigma} A_\mu  \frac{1}{3} \times 3 \times  \Tr \left( G_{\nu \rho} B_{\sigma} - \frac{1}{3} [B_\nu, B_\rho]  B_\sigma \right)\\
&=& 2 \int_{{\cal M}}  \ud^4 x\, \varepsilon^{\mu \nu \rho \sigma} A_\mu \Tr\left( B_\nu \partial_\rho B_\sigma + \frac{2}{3} B_\nu B_\rho B_\sigma \right),
\end{eqnarray}
from which one can see the resemblance with the expression for the topological charge density in local coordinates. 

For the reader not accustomed to the differential form language, we spell out the action in (\ref{Eq:GeneralLagrangian}) in local coordinate form:
\begin{eqnarray}
{\cal S}^{\rm eff}_{axion} &=& \bigintssss d^4 x \sqrt{|g|} \left[ - \frac{1}{2} \sum_{i,j=1}^N {\cal G}_{ij} g^{\mu \nu} \left(\partial_\mu a^ i - k^ i A_\mu  \right) \left( \partial_\nu a^ j - k^ j A_\nu  \right) - \frac{1}{4g_1^2} F_{\mu \nu} F^{\mu \nu} \right. \nonumber \\
&& \qquad \qquad \left. -  \frac{1}{4g_2^2} \Tr (G_{\mu \nu} G^{\mu \nu})   + \frac{1}{32 \pi^2} \left(\sum_{i=1}^N r_i a^i \right) \varepsilon^{\mu\nu\rho\sigma} \Tr(G_{\mu \nu} G_{\rho \sigma}) \right],
\end{eqnarray}
where $g_{\mu \nu}$ now represents the metric on the four-dimensional spacetime with metric signature $(-+++)$.

%%%%%%%%%%%%%%%%%%%%%%%%%%%%%%%%%%%%%%%%%%%%%%%%%%%%%%%%%%%%%%%%%%%%%%%%%%%%%%%%%%%%%%%%%%%%%%%%%%%%%%%%%%%%%%%%%%%%%%%%%%%%%%%%%%%%%%%%%%%%%%%%%%%%%%%%%%%%%%%%%%%%%%%%
\section{Some Considerations about Axions}\label{A:ConsAxions}
An axion is a CP-odd scalar degree of freedom with a classical continuous shift symmetry $a\rightarrow a +\epsilon$, with $\epsilon \in \R$. Non-perturbative effects are expected to break the shift symmetry, in which case the continuous symmetry reduces to a discrete shift symmetry. In the vast literature on axions one will find that there exist two different ways of representing the lagrangian for an axion:
\begin{equation}\label{Eq:RepSchemesAxions}
\begin{array}{lll}
\text{rep (1):} & {\cal S}^{(1)} \supset \bigintsss \left[ -\frac{1}{2} \ud a \wedge \star_4 \ud a + \frac{a}{8 \pi^2 f_a} \Tr(G\wedge G) \right] & \text{with: } a \rightarrow a + 2 \pi f_a,  \\
\text{rep (2):} & {\cal S}^{(2)} \supset \bigintsss \left[ - \frac{1}{2}  f_{a}^2\, \ud \alpha \wedge \star_4 \ud \alpha  + \frac{\alpha}{8\pi^2} \Tr(G\wedge G) \right] & \text{with: } \alpha \rightarrow \alpha + 2 \pi.   \\
\end{array}
\end{equation} 
Representation scheme (1) is very characteristic for field theory discussions, while representation scheme (2) is inherent to four dimensional reductions of string theories. Nevertheless, the specific form of the action is always determined by the shift symmetry. Classically an axion can couple to matter only through derivative terms of the form $J^\mu (X)\partial_\mu a$, where $J^\mu(X)$ corresponds to a pseudo-vector depending on other matter fields $X$, as imposed by the shift symmetry. The topological term $\Tr(G\wedge G)$, responsible for the breaking of the shift symmetry, is characteristic for non-perturbative effects in non-Abelian gauge theories and can be related to the Pontryagin index (in case of strong interactions it corresponds to the QCD instanton number):\footnote{The conventions are chosen in correspondence with appendix~\ref{A:Conventions} such that:
\begin{equation}
G = \frac{1}{2} G_{\mu \nu} dx^\mu \wedge dx^\nu,  \qquad \tilde G_{\mu \nu} =  \frac{1}{2} \varepsilon_{\mu \nu \alpha \beta} G^{\alpha \beta} \qquad \Tr_G(T_a T_b) = \frac{1}{2} \delta_{ab}.
\end{equation}}
\begin{equation}\label{Eq:PontryaginIndex}
I_n  = \frac{1}{8 \pi^2} \int \ud^4 x \Tr(G\wedge G) = \frac{1}{16 \pi^2} \int \ud^4 x\, \varepsilon^{\mu \nu \alpha \beta} \Tr(G_{\mu \nu} G_{\alpha \beta}) \in \Z.
\end{equation}
The generating functional (or partition function) for the gauge theory coupled to the axion is given by (with external sources set to zero):
\begin{equation}
\begin{array}{ll}
\text{rep (1):} & \bigintsss {\cal D}a {\cal D} A_\mu\, e^{ i\, {\cal S}^{(1)}_{kin} +i \bigintssss \, \frac{a}{8 \pi^2 f_a} \Tr(G\wedge G)}, \\
\text{rep (2):} &\bigintsss {\cal D}\alpha {\cal D} A_\mu \, e^{ i\, {\cal S}^{(2)}_{kin} +i \bigintssss \, \frac{\alpha}{8 \pi^2} \Tr(G\wedge G)}, 
\end{array}
\end{equation}
where ${\cal S}_{kin}$ denotes the kinetic part of the action for the gluon field as well as the axion:\
\begin{equation}
\begin{array}{ll}
\text{rep (1):} & {\cal S}_{kin}^{(1)} = \bigintsss \left[ -\frac{1}{g^2} \Tr(G\wedge \star_4 G) - \frac{1}{2}\ud a \wedge \star_4 \ud a \right],  \\
\text{rep (2):} &  {\cal S}_{kin}^{(2)} = \bigintsss \left[ -\frac{1}{g^2} \Tr(G\wedge \star_4 G) - \frac{1}{2} f_a^2 \ud \alpha \wedge \star_4 \ud \alpha \right].
\end{array}
\end{equation}
The form of the shift symmetry in both representation schemes (\ref{Eq:RepSchemesAxions}) now follows from considerations regarding the required invariance of the path integral: the discrete shift symmetry has to be defined in such a way that the exponent in the path integral transforms as $e^{i \ldots} \rightarrow e^{i \ldots + i\, 2 \pi I_n}$, implying the invariance of the partition function.
One can easily switch between the two representations through the rescaling:
\begin{equation}
a =  f_a\,  \alpha,
\end{equation}
which has obviously consequences for the mass dimensions of the fields:
\begin{equation}
[a] = M, \qquad [f_a] = M, \qquad  [\alpha] = M^0.
\end{equation}
Both representations schemes yield the same axion decay constant and are fully equivalent to each other.
%%%%%%%%%%%%%%%%%%%%%%%%%%%

Reading off the correct axion decay constant can become rather tricky in the presence of kinetic mixing among axions, as discussed in section~\ref{sec:mixing} where we are required to perform a set of $SO(2)$ transformations to obtain a diagonalized form for the kinetic terms of the axions.  It is therefore of utmost importance to keep the representation schemes in mind and to ensure that expressions are written in the same representation scheme to read off the axion decay constant correctly.  One can easily argue that the axion measure remains invariant under an $SO(2)$ rotation, such that the path integral does not alter when changing the axion basis through an $SO(2)$ rotation. When determining the axion decay constants in case of kinetic mixing one should however assume one of the representation schemes above and stick with it all the way to the end. 

In summary, the (effective) axion decay constant can only be consistently determined in an axion basis where both kinetic terms and mass terms are diagonalized, see for instance~\ref{Ss:MetricMixing}. And depending on which representation scheme assumed, the axion decay constant can be read off as the eigenvalue of the diagonalized metric in the kinetic term (representation scheme 2), or as the dimensionful coupling suppressing the anomalous coupling of an axion to a nonperturbative instanton correction (representation scheme 1). In the presence of multiple instanton corrections, as in section~\ref{Sss:AlignedNaturalInflation}, one can distinguish an axion decay constant for each nonperturbative contribution. In this situation representation scheme 1 is the preferred representation scheme to read off the axion decay constants per instanton seperately. Of course, in order to determine the effective axion decay constants, one has to find an axion basis in which the mass matrix arising from the multiple instanton corrections is diagonalized.

%%%%%%%%%%%%%%%%%%%%%%%%%%%%%%%%%%%%%%%%%%%%%%%%%%%%%%%%%%%%%%%%%%%%%%%%%%%%%%%%%%%%%%%%%%%%%%%%%%%%%%%%%%%%%%%%%%%%%%%%%%%%%%%%%%%%%%%%%%%%%%%%%%%%%%%%%%%%%%%%%%%%%%%%
\section{Chiral Rotations and Axion Potentials}\label{A:ChiralRotations}
In this appendix we gather some useful properties involving chiral rotations in the path integral and additional information about the emergence of the cosine-potential of the axion. These useful background results have been used in the main text, but were omitted in order to keep up with the main storyline of the paper. More details can be found in textbooks~\cite{Peskin:1995ev} and review literature on anomalies~\cite{Harvey:2005it}.   

Let us first have a look at the effects of chiral rotations in the path integral.
Considering the chiral spectrum in equation (\ref{Eq:ChiralSpectrumFermions}) we can write down the gauge-invariant lagrangian for the chiral fermions (in Minkowski spacetime) as follows,
\begin{equation}
{\cal L}_{\rm fermion} = \ov \psi_L^i\, i \left( \slashed{\partial}  -i q_L^i \slashed{A}  -i \slashed{B}^a T_a^{R^i_1} \right) \psi_L^i + \ov \psi_R^i \, i \left( \slashed{\partial} -i q_R^i \slashed{A} -i\slashed{B}^a T_a^{R^i_2}  \right) \psi_R^i,
\end{equation}
where we used the common notation $\slashed{C} = \gamma^\mu C_\mu$ to denote the contraction with the Dirac $\gamma$-matrices.
Under a chiral transformation of the type,
\begin{equation}
\psi_L^i \rightarrow e^{i\, q^i_L \tilde a^2} \psi_L^i, \qquad \psi_R^i \rightarrow e^{i\,  q^i_R  \tilde a^2} \psi_R^i,
\end{equation}
the fermionic path integral measure is not invariant and by using Fujikawa's method~\cite{Fujikawa:1979ay} one can show that the non-invariance corresponds to the inclusion of an anomaly term:
\begin{equation}
{\cal L}_{\rm fermion}^{\rm eff}  = {\cal L}_{\rm fermion} + \tilde a^2 \left( \partial_\mu {\cal J}^\mu_\psi + \frac{1}{32 \pi^2} {\cal A}^{\rm mix} \varepsilon^{\mu\nu\rho\sigma}\Tr(G_{\mu\nu} G_{\rho \sigma})  \right)
\end{equation}
besides the ${\cal J}_\psi$ current term for which the current is given in equation~(\ref{Eq:CurrentExpression}). These considerations have been used to derive the action~(\ref{Eq:LagrFullU1Unitary}) in the unitary gauge. Notice that the second part corresponds to the Adler-Bell-Jackiw anomaly equation given in equation~(\ref{Eq:ABJAnomaly}), which expresses the violation of a chiral $U(1)$ symmetry at the quantum level due to instanton contributions.

In the configurations of sections~\ref{Ss:U1mixing} and~\ref{Ss:GenericKineticMixing} the r\^ole of the chiral $U(1)$ symmetry is played by the Abelian symmetry under which the axions and the chiral fermions are charged. This set-up is very reminiscent of QCD, where the axial $U(1)_A$ symmetry is spontaneously broken due to the presence of the chiral anomaly. In this respect we can exploit many of the well known results  and translate them to our set-up. Assuming that the non-Abelian gauge theory develops a strong gauge coupling, the chiral fermions with spectrum given in (\ref{Eq:ChiralSpectrumFermions}) condense into mesonic-like states. The condensate is also responsible for the generation of effective masses for the mesons. Indeed the four-point couplings in (\ref{Eq:LagrFullU1UnitaryWithoutA}) among the fermions, following from integrating out the massive $U(1)$ gauge field, yield effective masses of the order $\Xi^3/f_{\tilde a^2}^2$, where $\Xi$ is the characteristic scale of the condensate.  
The calculation of the axion potential for the axion $\tilde a^1$ not absorbed by the gauge boson can now be done in analogy with the computation of the QCD axion potential. Namely, by integrating out the heavy mesons using non-linear sigma-models techniques, one finds a cosine-potential of the form:
\begin{equation}
V_{\rm axion}(\tilde a^1) = \Lambda^4 \left[ 1 - \cos \left(\frac{\tilde a^1}{f_{\tilde a^1}}\right) \right],
\end{equation}
with the axion decay constant $f_{\tilde a^1}$ given in (\ref{Eq:AxionDecayConstantU1}) and the scale $\Lambda$ depending on the dimensionful ratio $\Xi^3/f_{\tilde a^2}^2$. A more detailed analysis of an explicit model is however required to determine the functional dependence of the scale $\Lambda$  on $\Xi^3/f_{\tilde a^2}^2$.

Nonperturbative corrections associated to gaugino condensation share various physical properties with the previous setting. In case a supersymmetric gauge theory runs to strong coupling, the vacuum also consists of a condensate but now formed by bilinears of the gaugini. This gaugino condensate breaks the $U(1)_R$ symmetry and the Adler-Bell-Jackiw anomaly equation is now valid for the R-current with the anomaly coefficient ${\cal A}$ proportional to the rank $N$ of the $SU(N)$ gauge group. By integrating out the non-Abelian $SU(N)$ gauge bosons and the gaugini one obtains a nonperturbative correction to the superpotential:
\begin{equation}
{\cal W} = {\cal W}_{per} + A\, e^{- \frac{2\pi}{N} T},
\end{equation}
where $T$ is a chiral superfield appearing in the gauge kinetic function. In the case of rigid supersymmetry this field is considered as a background superfield, whereas in string theory the field $T$ is promoted to a chiral superfield in which one of the moduli (K\"ahler or complex structure moduli fields) resides. The coefficient $A$ is then a function of other moduli that are already assumed to be stabilised. More explicitly, we write the scalar components of the superfield $T$ as $t+i\,  a$, where $t$ is the CP-even real modulus field and $a$ is its CP-odd scalar partner, and assume for simplicity that the modulus field has a no-scale K\"ahler potential:
\begin{equation}
K(T, \ov T) = -3 \ln (T + \ov T).
\end{equation}
Inserting the superpotential and no-scale K\"ahler potential in the F-term scalar potential of ${\cal N} = 1$ supergravity one obtains a cosine-type potential:
\begin{equation}
V_{\rm axion}(a) = \frac{8\pi}{ N} \frac{\langle t\rangle}{{\cal T}^2} \left|A\right| \left|{\cal W}_{per}\right| e^{-2 \frac{\pi}{N} t} \cos \left( \frac{2\pi}{N} a + i \gamma\right), 
\end{equation}  
with $\gamma = {\rm Arg} \left( {\cal W}_{per} A^* \right)$, ${\cal T}$ the dimensionless volume of the internal space as introduced in equation (\ref{Eq:PlanckToString}) and $\langle t \rangle$  the stabilised vev of the modulus $t$.

%%%%%%%%%%%%%%%%%%%%%%%%%%%%%%%%%%%%%%%%%%%%%%%%%%%%%%%%%%%%%%%%%%%%%%%%%%%%%%%%%%%%%%%%%%%%%%%%%%%%%%%%%%%%%%%%%%%%%%%%%%%%%%%%%%%%%%%%%%%%%%%%%%%%%%%%%%%%%%%%%%%%%%%%
\section{Dualization Procedure for Two-forms in Four Dimensions}\label{A:Dualisation}
In this appendix, we review the dualization procedure for the St\"uckelberg mechanism expressed in terms of the two-forms $D_{(2)i}$ as they arise from dimensional reduction in section~\ref{Ss:ClosedStringAxions}. Let us for purposes of clarity consider such systems in a somewhat simplified set-up, with one axion $a$ and one dual 2-form $D$ and in the absence of a gauge field. For such a system the original action can be written as:
\begin{equation}
{\cal S}_{orig} =-  \bigintssss  \frac{1}{2c} \ud D \wedge \star_4 \ud D  + \frac{c}{2} \ud a \wedge \star_4 \ud a.
\end{equation} 
The two-form and axion are known to be related by the Hodge-duality relation:
\begin{equation}
\star_4 \ud D = c\, \ud a, \qquad \text{ or equivalently,}\quad \ud D = c \star_4 \ud a.
\end{equation}
Turning to a first order formalism, we have to write the action in terms of a three-form~$H$ and we do it in such a way that the field $a$ serves as a Lagrange multiplier:
\begin{equation}\label{Eq:ParentActionNoGaugeField}
{\cal S}_{(1)} = -  \bigintssss \frac{1}{2c} H \wedge \star_4 H  -\frac{1}{2} a\,  \ud H - \frac{1}{2} H \wedge \ud a,
\end{equation}
Varying the first order action yields the following equations:
\begin{equation}
\begin{array}{ll}
\delta a:& \ud H = 0  \leadsto H = \ud D\quad (\text{locally}), \\
\delta H:& \star_4 H = c\, \ud a  \leadsto \ud(\star_4 \ud a) = 0\quad (\text{e.o.m.~for $a$}).
\end{array}
\end{equation}
The first equation of motion corresponds to the Bianchi identity for $H$, which can be solved locally in terms of a two-form $D$. The second equation of motion expresses the Hodge-duality relation.
Imposing first the Bianchi identity for $H$ allows us to reproduce the original action:
\begin{equation}
{\cal S}_{orig} = - \int  \frac{1}{2c} \ud D \wedge \star_4 \ud D - \frac{1}{2} \ud D \wedge \ud a,
\end{equation}
provided we impose the Hodge-duality relation $\ud D = c \star_4 \ud a$ afterwards as well.
Imposing on the other hand first the Hodge-duality relation eliminates the three-form from the action in favor of the axion $a$:
\begin{equation}
{\cal S}_{dual} = -  \bigintssss  \frac{c}{2} \ud a \wedge \star_4 \ud  a.
\end{equation}
The Bianchi identity for $H$ reduces to the equations of motion for $a$ upon imposing the duality relation.

Next, we add a source term to the original action involving a $U(1)$ gauge symmetry with gauge potential $A$:
\begin{equation}
{\cal S}_{orig} = -  \bigintssss  \frac{1}{2c} \ud D \wedge \star_4 \ud D  + \frac{c}{2} \ud a \wedge \star_4 \ud a - m H\wedge  A.
\end{equation}
A straightforward generalization of (\ref{Eq:ParentActionNoGaugeField}) yields the following parent action in the first order formalism upon inclusion of the gauge potential:
\begin{equation}
{\cal S}_{(1)} = -  \bigintssss \frac{1}{2c} H \wedge \star_4 H  -\frac{1}{2} a\,  \ud H - \frac{1}{2} H \wedge \ud a  - m H\wedge  A
\end{equation}
for which the equations of motion now read:
\begin{equation}
\begin{array}{ll}
\delta a:& \ud H = 0  \leadsto H = \ud D\quad (\text{locally}), \\
\delta H:& \star_4 H = c ( \ud a + m A)  \leadsto \ud (\star_4 ( \ud a + m A)) = 0 \quad (\text{e.o.m.~for $a$}).
\end{array}
\end{equation}
The first equation of motion corresponds again to the Bianchi identity for $H$, while the second one expresses a more involved Hodge-duality relation.
Imposing the Bianchi identity first brings us back to the original action:
\begin{equation}
{\cal S}_{orig} = - \bigintssss  \frac{1}{2c} \ud D \wedge \star_4 \ud D  + \frac{c}{2} \ud a \wedge \star_4 \ud a - m H\wedge  A - \frac{c}{2} m\, \ud(\star_4 A) a,
\end{equation}
with an additional term which vanishes upon imposing (by hand) the Lorenz gauge condition for the gauge field: $\ud(\star_4 A) = 0$. On the other hand, imposing first the Hodge-duality relation brings us to the dual action, where the two-form $B$ is eliminated:
 \begin{equation}
{\cal S}_{dual} = -  \bigintssss  \frac{c}{2} \left(\ud a + m A\right)  \wedge \star_4 \left(\ud a + m A \right),
\end{equation}
and the axion is now charged under the $U(1)$ gauge symmetry in the form of a St\"uckelberg mass term.

%%%%%%%%%%%%%%%%%%%%%%%%%%%%%%%%%%%%%%%%%%%%%%%%%%%%%%%%%%%%%%%%%%%%%%%%%%%%%%%%%%%%%%%%%%%%%%%%%%%%%%%%%%%%%%%%%%%%%%%%%%%%%%%%%%%%%%%%%%%%%%%%%%%%%%%%%%%%%%%%%%%%%%%%

\section{Generalization to Multiple $U(1)$'s and Multiple Gauge Instantons}\label{A:Generalisation}
In this appendix, we generalize the $N$-axion system (\ref{Eq:GeneralLagrangian}) by including $M$ $U(1)$ gauge fields $A^a$ under which the axions are charged through St\"uckelberg terms and adding anomalous coupling terms associated to  $P$ non-Abelian gauge groups with gauge potentials $B^{A}$. The full action then reads,
\begin{eqnarray}
{\cal S}^{\rm eff}&=&\bigintssss \left[ -\frac{1}{2}\, \sum_{i,j =1}^{N}\cG _{ij}\left(\ud a^i- \sum_{\alpha=1}^M k^i_\alpha\,A^\alpha\right)\wedge \star _{4}\left(\ud a^j- \sum_{\beta=1}^Mk^j_\beta\,A^\beta\right) - \sum_{\alpha, \beta =1}^M f_{\alpha \beta}\,F^\alpha \wedge \star _4 F^\beta \right. \notag\\
&&\qquad - \sum_{A=1}^P \frac{1}{g_{A}^2}\,\text{Tr}\,G^{A}\wedge \star_4 G^{A} + \frac{1}{8\pi ^2}\, \sum_{A=1}^P \text{Tr}\,G^{A}\wedge G^{A}\, \left( \sum_{i=1}^{N} s^i_{A}\,a^i \right) \notag \\
&& \qquad \left. +\frac{1}{8\pi ^2}\, \sum_{\alpha, \beta =1}^M F^{\alpha}\wedge F^{\beta}\, \left( \sum_{i=1}^{N} r^i_{\alpha \beta}\,a^i \right) \right] +{\cal S}_{fermion}+ {\cal S}_{GCS}.\label{Eq:CompleteGeneralAxionGauge}
\end{eqnarray}
The Abelian and non-Abelian field strengths are respectively given by:
\begin{eqnarray}
F^\alpha=\ud A^\alpha,\qquad G^A=\ud B^A+B^{A}\wedge B^A,
\end{eqnarray}
and the Abelian gauge kinetic function $f_{ab}$ is generically non-diagonal. We start in a basis such that the ``axion charges" $k^i_\alpha$, matter charges and the model-dependent parameters $r^{i}_{\alpha \beta}$ and $s^{i}_{A}$ are all integers. The metric $\cG$ on the axion space is of mass dimension 2. Supposing that the $U(1)$ mass (squared) matrix,
\begin{eqnarray}\label{Eq:MassMatrixU(1)Gauge}
(M^2)_{\alpha \beta}=\cG _{ij}\,k^i_\alpha\,k^{j}_\beta=(k^T\cdot \cG \cdot k)_{\alpha \beta},
\end{eqnarray}
has rank $R\leq M $, we identify $R$ massive $U(1)$ gauge fields and $(M-R)$ massless $U(1)$ gauge fields in the mass eigenbasis. Furthermore, we have to assume $N > R$, so that $R$ axions turn into the longitudinal components of the massive $U(1)$ fields and $(N-R)$ axions will remain uncharged under those massive $U(1)$ gauge fields. Note that the anomalous coupling terms are not $U(1)$ gauge-invariant. Thus  generalized Chern-Simons terms (GCS) and chiral fermions are introduced (accompanied by anomalous triangle diagrams) as well to ensure the invariance under $U(1)$ transformations and the non-Abelian gauge invariance as discussed in section~\ref{Sss:GCSterms}. In order to make the St\"uckelberg mechanisms appearing in the action (\ref{Eq:CompleteGeneralAxionGauge}) more explicit and before we can read off the axion field ranges, we need to transform the basis for the axions and $U(1)$ gauge bosons in such a way that:
\begin{itemize}
\item[(a)] the $U(1)$ gauge bosons are expressed in a basis reflecting their mass eigenstates;
\item[(b)] $R$ axions are identified as the axionic directions eaten by the $R$ massive $U(1)$ bosons, while the $(N-R)$ orthogonal axionic directions remain uncharged under the massive $U(1)$ gauge symmetries;
\item[(c)] the kinetic terms for the axions and the $U(1)$ gauge bosons are expressed in a basis for which they take the canonical form.
\end{itemize}
In order to find such a basis, we have to perform four different orthogonal transformations:
\begin{enumerate}
\item Perform an orthogonal transformation on the space of $U(1)$ gauge bosons to bring the abelian gauge kinetic matrix $[f_{\alpha \beta}]$ to a diagonal form:
\begin{eqnarray}
O_1^T\cdot [f_{\alpha \beta}]\cdot O_1 =\text{diag}\left(g_1^{-2},\,...,\,g_{M}^{-2}\right)\equiv D_1^2,
\end{eqnarray}
expressed in the new basis $\vec{A}'$ which is related to the original basis $\vec{A}$ through the $SO(M)$ transformation:
\begin{eqnarray}
\vec{A}' = O_1^T\cdot \vec{A}.
\end{eqnarray}
In order to bring the diagonalized gauge kinetic matrix to the canonical form, we perform a rescaling transformation on the $U(1)$ space:
\begin{eqnarray}
D_1^{-1}\cdot D_1^2\cdot D_1^{-1}=\mathbb{I},
\end{eqnarray}
with
\begin{eqnarray}
\vec{A}''=D_1\cdot \vec{A}'.
\end{eqnarray}
\item Perform an additional orthogonal transformation on the space of $U(1)$ gauge bosons to diagonalize the mass matrix (\ref{Eq:MassMatrixU(1)Gauge}) in terms of the mass eigenstates:
\begin{eqnarray}\label{Eq:DiagonalU(1)MassMatrix}
O_2^T\cdot D_1^{-1}\cdot O_1^{T}\cdot k^T\cdot \cG \cdot k\cdot O_1\cdot D_1^{-1}\cdot O_2=\text{diag}\left(M_1^2,\,...,\,M_M^2\right),
\end{eqnarray}
now expressed in the basis $\vec{A}'''$ related to the previous basis $\vec{A}''$ via the orthogonal $SO(M)$ transformation: 
\begin{eqnarray}
\vec{A}''' = O_2^T\cdot \vec{A}''.
\end{eqnarray}
The diagonal matrix in (\ref{Eq:DiagonalU(1)MassMatrix}) allows to identify the $M-R$ massless $U(1)$ gauge bosons characterised by vanishing mass eigenstates: 
\begin{eqnarray}
M_{R+1}^2=...=M_{M}^2=0,
\end{eqnarray}
and distinguish them from the $R$ massive $U(1)$ gauge bosons with non-vanishing mass eigenstates: 
\begin{eqnarray}
M_i^2 \neq 0, \qquad \forall\,  i  \in \{1, \ldots, R \} .
\end{eqnarray}
\item Perform an orthogonal transformation on the axion space to diagonalize the axion moduli space metric $\cG_{ij}$ with eigenvalues $f_i^2$ along the diagonal entries:
\begin{eqnarray}
O_3^T\cdot \cG \cdot O_3=\text{diag}\left(f_1^2,\,...,\,f_N^2\right)\equiv D_2^2,
\end{eqnarray}
where the axion basis $\vec{a}'$ relates to the original axion basis $\vec{a}$ through the $SO(N)$ transformation:
\begin{eqnarray}
\vec{a}' = O_3^T\cdot \vec{a}.
\end{eqnarray}
The eigenvalues $f_i$ with mass dimension $M^1$ can be interpreted as axion decay constants in the fundamental domain. Notice that at this stage, the axions $\vec{a}'$ are scalar fields with mass dimension $M^0$ expressed in representation scheme 2. A transition to representation scheme 1, where the axions have mass dimension $M^1$, can be made through a rescaling of the basis $\vec{a}'$: 
\begin{eqnarray}
\vec{a}'' = D_2\cdot \vec{a}'.
\end{eqnarray}
\item Perform an additional $SO(N)$ transformation on the axion space such that $R$ axions become the longitudinal components of the $R$ massive $U(1)$ gauge bosons in this new basis  (denoted as St\"uckelberg axions in the following) while the orthogonal $(N-R)$ axionic directions remain as uncharged axions:
\begin{eqnarray}
\vec{a}''' = O_4^T\cdot \vec{a}''.
\end{eqnarray}  
\end{enumerate}
In the bases $\vec{a}'''$ and $\vec{A}'''$ satisfying the conditions (a)-(c) stated above, the kinetic terms for the axions can be written as follows,
\begin{eqnarray}\label{S_kin}
{\cal S}_{axion}^{\rm eff, kin}&\ni& -\bigintssss \frac{1}{2}\,\sum _{i=1}^{N-R}\,\ud a'''^i\wedge \star _4\ud a'''^i \notag\\
&&+\frac{1}{2}\,\sum _{i=N-R+1}^N \left[\ud a'''^i-(O_4^T\cdot D_2\cdot O_3^T\cdot k\cdot O_1\cdot D_1^{-1}\cdot O_2)^i_\alpha \cdot A'''^{\alpha}\right]\notag \\
&& \qquad \qquad \wedge  \star _4 \left[\ud a'''^i-(O_4^T\cdot D_2\cdot O_3^T\cdot k\cdot O_1\cdot D_1^{-1}\cdot O_2)^i_\beta \cdot A'''^{\beta}\right].
\end{eqnarray}
The orthogonal matrix $O_4$ has to satisfy a ``chargeless" condition for each of the $(N-R)$ uncharged axions, i.e.
\begin{eqnarray}\label{chargeless}
(O_4^T\cdot D_2\cdot O_3^T\cdot k\cdot O_1\cdot D_1^{-1}\cdot O_2)^i_\alpha =0, \qquad  \begin{array}{l}\forall\, i \in \{ 1,\,...,\,N-R\}, \\ \forall\, \alpha \in \{1,\,...,\,R\} .\end{array}
\end{eqnarray}
We immediately point out that the orthogonal matrix $O_4$ does not always exist to satisfy the chargeless condition (\ref{chargeless}). More explicitly, in case the number of axions $N$ is smaller than or equal to the number of massive $U(1)$ gauge bosons, it is possible that only the zero matrix solves the chargeless condition (\ref{chargeless}).\footnote{Focusing on the argument in more details, we define an $N\times R$ matrix
\begin{eqnarray}
X^i\,_\alpha\equiv \left(D_2\cdot O_3^T\cdot k\cdot O_1\cdot D_1^{-1}\cdot O_2\right)^i\,_\alpha,\qquad i=1,\,...,\,N,\,\,\,\, \alpha =1,\,...,\,R.
\end{eqnarray}
For a given $i\in \{1,\,2,\,...,\,N-R\}$, the chargeless condition
is now a set of $R$ linear equations with $N$ unknowns,
\begin{eqnarray}\label{chargeless_i}
\left(\begin{array}{cccc} X^1\,_1 &X^2\,_1 & ... & X^N\,_1\\
X^1\,_2&X^2\,_2&...&X^N\,_2\\
& ...& ...&\\
X^1\,_R &X^2\,_R &...&X^N\,_R
\end{array}\right)\left(\begin{array}{c}(O_4)^{1i}\\
(O_4)^{2i}\\
...\\
(O_4)^{Ni}
\end{array}\right)=0.
\end{eqnarray}
The condition to have non-zero solutions to these linear equations is
\begin{eqnarray}
\text{rank}\,(X)<N.
\end{eqnarray}
Since $\text{rank}\,(X_{N\times R})\leq\text{min}\,(R,\,N)$, when $R<N$, (\ref{chargeless_i}) always has nontrivial solutions, but when $R\geq N$ (\ref{chargeless_i}) may only be solved by a trivial solution, namely the zero-matrix for $O_4$.}

If we choose an appropriate gauge for each massive $U(1)$ gauge field, such as the unitary gauge in section~\ref{Sss:GCSterms}, the St\"uckelberg axions will disappear from the spectrum. The remaining uncharged axions are expected to couple to the (non-perturbative) effects, such as the gauge instantonic effects induced by the non-Abelian gauge bosons. In the eigenbases $\vec{a}'''$ and $\vec{A}'''$, the anomalous couplings of the uncharged axions to the  topological terms are given by, 
\begin{eqnarray}\label{S_inst}
{\cal S}_{axion}^{\rm eff, anom}=\bigintssss \frac{1}{8\pi ^2}\,\sum _{A=1}^P\,\text{Tr}\,G^A\wedge G^A\, \left(\sum _{k=1}^{N-R}\sum _{i,j=1}^N  s^i_A\,O_3^{ij}\,\frac{1}{f_j}\,O_4^{jk}\,a'''^{k} \right).
\end{eqnarray}
From this expression, one can now deduce that the $k^{\rm th}$ axionic direction $a'''^k$ couples anomalously to non-Abelian gauge group associated to gauge potential $B^A$ with a decay constant $f'''_{Ak}$ given by:
\begin{eqnarray}\label{f'''}
f'''_{Ak}=\left(\sum _{i,j=1}^N  s^i_A\,O_3^{ij}\,\frac{1}{f_j}\,O_4^{jk}\right)^{-1}.
\end{eqnarray}
By integrating out the non-Abelian degrees of freedom as discussed in appendix~\ref{A:ChiralRotations},  the effective axion potential generated by the non-perturbative effects takes the cosine-form for each separate non-Abelian gauge group: 
\begin{eqnarray}\label{pot}
V=\sum _{A=1}^P \,\Lambda _A^4\left(1-\text{cos}\,\sum _{i=1}^{N-R}\frac{a'''^i}{f'''_{Ai}}\right).
\end{eqnarray} 
On top of this linear combination, higher-order harmonics and cross-terms might arise, which we neglect for the moment using similar arguments as the ones put forward in~\cite{Dimopoulos:2005ac}. If each uncharged axion couples to a single gauge instanton, i.e.~for a diagonal matrix $f'''_{Ak}$, a trans-Planckian axion decay constant might arise in regions of the axion moduli space with a high level of isotropy, analogous to the examples in sections~\ref{Ss:U1mixing},~\ref{Ss:GenericKineticMixing} and~\ref{Ss:ExampleFactD6branes}. 

For configurations where the uncharged axions couple anomalously to the gauge instantons through linear combinations, i.e.~for a non-diagonal matrix $f'''_{Ak}$ with no hierarchy among the sub-Planckian decay constants, we can apply a similar analysis as the one presented in~\cite{Choi:2014rja}. Under the assumptions in that paper,  one linear combination of axions can be formed corresponding to a nearly flat direction with an effective axion decay constant scaling as $f_{\rm eff}\propto \sqrt{(N-R)!} \, n^{N-R-1}$, where the parameter $n$ now depends not only on anomaly coefficients, but also on the discrete $U(1)$ charges inherent to $U(1)$ kinetic mixing and the continuous parameters resulting from kinetic metric mixing. 

In this regard, we should emphasize that the axion decay constants (\ref{f'''}) might not be indicative of the effective axion field ranges, because the axions $a'''^i$ do not correspond to the mass eigenstates for a non-diagonal matrix $f'''_{Ak}$. By expanding the potential (\ref{pot}) around the minimum to the second order, we obtain the mass (squared) matrix for $(N-R)$ uncharged axions,
\begin{eqnarray}\label{mass_NP}
m^2_{NP}=\left(\begin{array}{ccc} \sum _{A}\frac{\Lambda _A^4 }{(f'''_{A1})^2}& \sum _{A}\,\frac{\Lambda _A^4 }{f'''_{A1}\,f'''_{A2}} & ...\\
\sum _{A}\,\frac{\Lambda _A^4 }{f'''_{A2}\,f'''_{A1}} & \sum _{A}\,\frac{\Lambda _A^4 }{(f'''_{A2})^2} & ...\\
 & . ..& \end{array}\right),
\end{eqnarray}
where we assumed for convenience that $P = N-R$.
Hence, a further orthogonal $SO(N-R)$ transformation needs to be performed to transform those uncharged axions into the mass eigenbasis,
\begin{eqnarray}
\vec{a}'''' = O_5^T\cdot \vec{a}''',
\end{eqnarray}
such that mass matrix diagonalizes to:
\begin{eqnarray}
O_5^T\cdot m^2_{NP}\cdot O_5=\text{diag} \left(m_1^2,\,...,\,m_{N-R}^2\right),
\end{eqnarray}
where $m_{i}^2$ are eigenvalues of (\ref{mass_NP}). The smallest eigenvalue $m_{i}^2$ then corresponds to the nearly flat direction with axion decay constant $f_{\rm eff}$, which is supposed to play the r\^ole of the inflaton. Furthermore, the mass eigenvalue is expected to scale inversely proportional to $f_{\rm eff}$, or more explicitly $m^2_{\rm flat} \sim f^{-2}_{\rm eff}$, as discussed in~\cite{Choi:2014rja}.\footnote{We also point out that the potential (\ref{pot}) was recognised in~\cite{Bachlechner:2014gfa} in a different axion basis, where the mass matrix generated by the non-perturbative effects is diagonalized but where the axion kinetic terms are non-canonical. Nevertheless, the authors of~\cite{Bachlechner:2014gfa} did not take kinetic $U(1)$ mixing into consideration. Hence, action (\ref{Eq:CompleteGeneralAxionGauge}) can be seen as the most generic effective action for axions expected to result from string theory compactifications.}

%%%%%%%%%%%%%%%%%%%%%%%%%%%%%%%%%%%%%%%%%%%%%%%%%%%%%%%%%%%%%%%%%%%%%%%%%%%%%%%%%%%%%%%%%%%%%%%%%%%%%%%%%%%%%%%%%%%%%%%%%%%%%%%%%%%%%%%%%%%%%%%%%%%%%%%%%%%%%%%%%%%%%%%%

\section{Anomaly-free Chiral Spectrum}\label{A:Fermion}
In this appendix, we return to the two-axion system in section~\ref{Ss:U1mixing} and present a method to find a consistent field theory model satisfying the constraints~(\ref{Eq:U(1)GaugeInvarianceII}), (\ref{Eq:U(1)GaugeInvarianceI}), (\ref{Eq:PureNonAbelianAnomaly}) and (\ref{Eq:MixedAbelianNonAbelianAnomaly}). To this end, we assume that the chiral left-handed and right-handed fermions in (\ref{Eq:ChiralSpectrumFermions}) correspond to fundamental representations under the non-Abelian $SU(N)$ gauge group, such that the mixed anomaly coefficient reduces to:
\begin{eqnarray}
{\cal A}^{\rm mix} = \frac{1}{2} \sum _{i=1}^{n_F}\left(q^i_L-q^i_R\right)  .
\end{eqnarray}
Provided that the number of left-handed chiral fermions is equal to the number of right-handed chiral fermions, the pure non-Abelian anomaly (\ref{Eq:PureNonAbelianAnomaly}) vanishes trivially. This leaves us with three conditions to be solved explicitly. Our method to find solutions now distinguishes between two cases depending on whether the axion $\tilde a^2$ -- the longitudinal component of the massive $U(1)$ gauge boson -- couples anomalously to the $U(1)$ field strength or not:
\begin{itemize}
\item[\it (1)] {\it In the presence of an anomalous $\tilde a^2 F\wedge F$ term:}\\
Combining the mixed gauge anomaly (\ref{Eq:MixedAbelianNonAbelianAnomaly}) with the $U(1)$ gauge invariance constraint (\ref{Eq:U(1)GaugeInvarianceII}) allows us to reduce the three remaining constraints to two:
\begin{eqnarray}
\label{Eq:NewMixedAnomaly}{\cal A}^{\rm GCS} =\mathcal{A}^{\rm mix}&\stackrel{!}{=}&-\frac{\tilde k^2}{2}=-\frac{r_1\,k^1+r_2\,k^2}{2},\\
\label{pureAbelianAnomaly}\mathcal A^{U(1)^3}&\stackrel{!}{=}&-\tilde k ^2\label{U(1)2'},
\end{eqnarray}
where the cubic $U(1)$ anomaly cancelation condition slightly differs from equation~(\ref{Eq:U(1)GaugeInvarianceI}) due to the presence of the anomalous coupling of $\tilde a^2$ to the Abelian gauge group. Next, we want to determine the $U(1)$ charges $q^i_{L} (q^i_{R})$ of the left(right)-handed chiral fermions for which the remaining two conditions are satisfied. A possible solution consists in choosing equally distributed charges such that:
\begin{eqnarray}
q^i_L-q^i_R=-\frac{r_1\,k^1+r_2\,k^2}{n_F}, \qquad i=1,\,2,\,...,\,n_F,
\end{eqnarray}
for which the mixed anomaly condition (\ref{Eq:NewMixedAnomaly}) is trivially satisfied. In order for the charges to be rational, we assume that $n_F$ is a divisor of $r_1 k^1 + r_2 k^2$, such that $r_1 k^1 + r_2 k^2 = \nu \, n_F $ with $\nu \in \Z_0$. These considerations allow us to write the cubic Abelian anomaly cancelation condition as:
\begin{equation}
\sum_{i=1}^{n_F} \left( 3(q_L^i)^2 + 3 \nu\, q_L^i +  \nu^2   \right) = n_F.    
\end{equation}
If we further also assume that all $U(1)$ charges of the left-handed fermions are equal to each other, i.e.~$q_L^i = q_L$ $\forall \, i \in \{1, \ldots, n_F \}$, we can solve the anomaly constraint for $q_L$ as a function of $\nu$:
\begin{equation}\label{Eq:QLSolutionWithFF}
q_L (\nu) = \frac{-3 \nu \pm \sqrt{12 - 3 \nu^2}}{6}.  
\end{equation}
An overview of integer charges for $q_L$ as a function of $\nu$ is given in table~\ref{tab:OverviewSolWithFF}, including the corresponding charges for the right-handed fermions.

\begin{table}[h]
\begin{center}
\begin{tabular}{|c||c|cc|cc|c|}
\hline
\multicolumn{7}{|c|}{\bf Overview of Integer Charges}\\
\multicolumn{7}{|c|}{\bf in the Presence of $\tilde a^2 F\wedge F$}\\
\hline \hline
$\nu$ & $-2$ & \multicolumn{2}{|c|}{$-1$}&\multicolumn{2}{|c|}{$1$} &  $2$\\
\hline\hline
$q_L^i$ & 1 & 0 & 1 & $-1$ & 0 & $-1$\\
$q_R^i$ & $-1$   & $-1$  & 0 &  0 & 1 & 1 \\
\hline
\end{tabular}
\caption{Summary of all solutions $q_L \in \Z$ for various integer values of $\nu$, based on eq.~(\ref{Eq:QLSolutionWithFF}).\label{tab:OverviewSolWithFF}}
\end{center}
\end{table}
For other integer values of $\nu$ the charges $q_L$ turn out to be complex. Hence, table~\ref{tab:OverviewSolWithFF} gives the full set of integer solutions satisfying our aforementioned assumptions and each choice of charges from table~\ref{tab:OverviewSolWithFF} yields a chiral fermionic spectrum for which the cubic $U(1)$ anomaly vanishes.

\item[\it (2)] {\it In the absence of an anomalous $\tilde a^2 F\wedge F$ term:}\\
Also for this case we can reduce the three remaining constraints to only two conditions:
 \begin{eqnarray}
{\cal A}^{GCS} =\mathcal{A}^{mix}&\stackrel{!}{=}&-\frac{\tilde k^2}{2}=-\frac{r_1\,k^1+r_2\,k^2}{2}, \label{MixedU1NoFF}\\
\mathcal A^{U(1)^3}&\stackrel{!}{=}& 0\label{U(1)4'},
\end{eqnarray}
where the cubic $U(1)$ anomaly condition is given by equation (\ref{Eq:U(1)GaugeInvarianceI}). If we try to apply the same reasoning as the one used above, the charges $q_L^i$ as a function of $\nu$ are all complex numbers. Hence, the assumptions that all charge differences $q_L^i-q_R^i$ are equally distributed and that all charges $q_L^i$ are equal to each other no longer work in the search for a solution without GCS-term.  Instead, we find that $n_F\geq 2$ and assume that the charge differences satisfy the relation:
\begin{equation}
q^i_L-q^i_R= \xi^i,  \qquad \text{with } \sum_{i=1}^{n_F} \xi^i = -\left( r_1\,k^1+r_2\,k^2 \right), \text{ and } \forall\,  i: \, \xi^i \in \Z_0 .
\end{equation}
This ansatz ensures that the mixed anomaly condition~(\ref{MixedU1NoFF}) is satisfied. 
Let us now discuss a method to determine the $U(1)$ charges by looking at the simplest case, namely $n_F=2$. Our reasoning will be based on some basic number theory applied to the charges. To this end, we write the charge differences as,
\begin{eqnarray}\label{charge difference}
q^1_L-q^1_R\equiv n,\qquad q^2_L-q^2_R\equiv m,
\end{eqnarray}
with $n+m=-\left( r_1\,k^1+r_2\,k^2 \right) \neq 0$, and $n,m \in\Z$.
Inserting (\ref{charge difference}) into the cubic anomaly constraint~(\ref{U(1)4'}) leads to: 
\begin{eqnarray}\label{cubicanomaly}
n\left[3\left(q^1_R+\frac{n}{2}\right)^2+\frac{n^2}{4}\right]+m\left[3\left(q^2_R+\frac{m}{2}\right)^2+\frac{m^2}{4}\right]=0.
\end{eqnarray}
In order for this constraint to be satisfied, $n$ and $m$ must have opposite signs, so that both contributions can cancel each other out. Without losing generality, we choose:
\begin{eqnarray}
n>0,\qquad m<0. \label{signs of n and m}
\end{eqnarray}
Note that with some minor algebra equation (\ref{cubicanomaly}) can be re-written as,
\begin{eqnarray}\label{Eq:CubicAnomWriting1}
3n\left(2q^1_R+n\right)^2+3m\left(2q^2_R+m\right)^2=-(m+n)(m^2-mn+n^2),
\end{eqnarray}
from which we can deduce that one of the two factors on the righthand side has to be divisible by three: 
\begin{eqnarray}
3\big|(m+n)\,\,\text{or}\,\,3\big|(m^2-mn+n^2).
\end{eqnarray}
Let us seek a solution for $3|(m+n)$ and write:~\footnote{Of course, one may use similar reasonings to obtain a solution in case $3\big|(m^2-mn+n^2)$.}
\begin{eqnarray}
m+n=3k\neq 0,\qquad k\in \Z.
\end{eqnarray}
Substituting $m$ for $k$ in the equation (\ref{Eq:CubicAnomWriting1}) allows us to write the cubic anomaly constraint in terms of $n$ and $k$ as,
\begin{eqnarray}\label{cubic anomaly_n_k}
q^1_R n\left(q^1_R+n\right)+q^2_R\left(3k-n\right)\left(q^2_R+3k-n\right)=-3k(n^2-3kn+3kn).
\end{eqnarray}
We notice that the left hand side of the equation (\ref{cubic anomaly_n_k}) is always even.\footnote{For arbitrary $a,\,b\in\Z$, it is easy to see that $ab(a+b)$ is always even.} Subsequently, we can check that the right hand side of (\ref{cubic anomaly_n_k}) can only be even if both $n$ and $k$ are even. Now let us summarize the constraints for integers $n$ and $k$:
\begin{eqnarray}
n>0, \qquad k\neq 0,\qquad n>3k, \qquad2\big|n,\qquad2\big|k.
\end{eqnarray}
The minimal $n$ and the maximal $k$ that satisfy the above constrains are:
\begin{eqnarray}\label{n k}
n=2,\qquad k=-2.
\end{eqnarray}
Substituting (\ref{n k}) into the cubic anomaly equation (\ref{cubic anomaly_n_k}) yields:
\begin{eqnarray}
(q^1_R+1)^2-4(q^2_R-4)^2=21,
\end{eqnarray}
which is solved by $4$ integer solutions:
\begin{eqnarray}
\begin{cases}\label{qiR}
q^1_R+1=\pm 11\\
q^2_R-4=\pm 5.
\end{cases}
\end{eqnarray}
Having determined the charges $q_R^i$, the integers $n$ and $k$, we consider the charge difference (\ref{charge difference}) again and find the complete spectra for $n_F=2$ as listed in table~\ref{tab:OverviewSolWOFF}. In all cases, the St\"uckelberg $U(1)$ charge of the eaten axion is given by: 
\begin{eqnarray}
\tilde k^2= - 3k=6\neq 0,
\end{eqnarray}
for the solution considered in equation (\ref{n k}).

\begin{table}[h]
\begin{center}
\begin{tabular}{|@{\hspace{0.17in}}c@{\hspace{0.17in}}|@{\hspace{0.17in}}c@{\hspace{0.17in}}||@{\hspace{0.2in}}c@{\hspace{0.2in}}|@{\hspace{0.2in}}c@{\hspace{0.2in}}|}
\hline \multicolumn{4}{|c|}{\bf Overview of Integer Charges}\\
\multicolumn{4}{|c|}{\bf in the Absence of $\tilde a^2 F\wedge F$}\\
\hline \hline
$q^1_R$ & $q^1_L$ & $q^2_R$&$q^2_L$ \\
\hline
$10$ & $12$&$9$&$1$\\
$-12$ &  $-10$& $9$&$1$\\
$10$&$12$& $-1$& $-9$ \\
$-12$&$-10$ & $-1$&$-9$\\
\hline
\end{tabular}
\caption{Summary of all solutions for $n_F=2$ considering the solution (\ref{n k}) and based upon (\ref{qiR}).\label{tab:OverviewSolWOFF}}
\end{center}
\end{table}

\end{itemize}

%%%%%%%%%%%%%%%%%%%%%%%%%%%%%%%%%%%%%%%%%%%%%%%%%%%%%%%%%%%%%%%%%%%%%%%%%%%%%%%%%%%%%%%%%%%%%%%%%%%%%%%%%%%%%%%%%%%%%%%%%%%%%%%%%%%%%%%%%%%%%%%%%%%%%%%%%%%%%%%%%%%%%%%%

\addcontentsline{toc}{section}{References}
\bibliographystyle{ieeetr}
\bibliography{refs_AxionInflation}

\providecommand{\href}[2]{#2}\begingroup\raggedright\begin{thebibliography}{100}

\bibitem{Lyth:1996im}
D.~H. Lyth, {\it {What would we learn by detecting a gravitational wave signal
  in the cosmic microwave background anisotropy?}},  {\em Phys.Rev.Lett.}, {\bf
  78} (1997)\href{http://dx.doi.org/10.1103/PhysRevLett.78.1861 }{,
  1861--1863}, [\href{http://www.arxiv.org/abs/hep-ph/9606387}{{\tt
  hep-ph/9606387}}].

\bibitem{Senatore:2011sp}
L.~Senatore, E.~Silverstein, and M.~Zaldarriaga, {\it {New Sources of
  Gravitational Waves during Inflation}},  {\em JCAP}, {\bf 1408}
  (2014)\href{http://dx.doi.org/10.1088/1475-7516/2014/08/016 }{, 016},
  [\href{http://www.arxiv.org/abs/1109.0542}{{\tt arXiv:1109.0542}}].

\bibitem{Ozsoy:2014sba}
O.~Ozsoy, K.~Sinha, and S.~Watson, {\it {How Well Can We Really Determine the
  Scale of Inflation?}},  \href{http://www.arxiv.org/abs/1410.0016}{{\tt
  arXiv:1410.0016}}.

\bibitem{Mirbabayi:2014jqa}
M.~Mirbabayi, L.~Senatore, E.~Silverstein, and M.~Zaldarriaga, {\it
  {Gravitational Waves and the Scale of Inflation}},
  \href{http://www.arxiv.org/abs/1412.0665}{{\tt arXiv:1412.0665}}.

\bibitem{Sorbo:2011rz}
L.~Sorbo, {\it {Parity violation in the Cosmic Microwave Background from a
  pseudoscalar inflaton}},  {\em JCAP}, {\bf 1106}
  (2011)\href{http://dx.doi.org/10.1088/1475-7516/2011/06/003 }{, 003},
  [\href{http://www.arxiv.org/abs/1101.1525}{{\tt arXiv:1101.1525}}].

\bibitem{Cook:2011hg}
J.~L. Cook and L.~Sorbo, {\it {Particle production during inflation and
  gravitational waves detectable by ground-based interferometers}},  {\em
  Phys.Rev.}, {\bf D85}
  (2012)\href{http://dx.doi.org/10.1103/PhysRevD.86.069901,
  10.1103/PhysRevD.85.023534 }{, 023534},
  [\href{http://www.arxiv.org/abs/1109.0022}{{\tt arXiv:1109.0022}}].

\bibitem{Barnaby:2012xt}
N.~Barnaby, J.~Moxon, R.~Namba, M.~Peloso, G.~Shiu, {\em et.~al.}, {\it
  {Gravity waves and non-Gaussian features from particle production in a sector
  gravitationally coupled to the inflaton}},  {\em Phys.Rev.}, {\bf D86}
  (2012)\href{http://dx.doi.org/10.1103/PhysRevD.86.103508 }{, 103508},
  [\href{http://www.arxiv.org/abs/1206.6117}{{\tt arXiv:1206.6117}}].

\bibitem{Mukohyama:2014gba}
S.~Mukohyama, R.~Namba, M.~Peloso, and G.~Shiu, {\it {Blue Tensor Spectrum from
  Particle Production during Inflation}},  {\em JCAP}, {\bf 1408}
  (2014)\href{http://dx.doi.org/10.1088/1475-7516/2014/08/036 }{, 036},
  [\href{http://www.arxiv.org/abs/1405.0346}{{\tt arXiv:1405.0346}}].

\bibitem{Ferreira:2014zia}
R.~Z. Ferreira and M.~S. Sloth, {\it {Universal Constraints on Axions from
  Inflation}},  {\em JHEP}, {\bf 1412}
  (2014)\href{http://dx.doi.org/10.1007/JHEP12(2014)139 }{, 139},
  [\href{http://www.arxiv.org/abs/1409.5799}{{\tt arXiv:1409.5799}}].

\bibitem{Freese:1990rb}
K.~Freese, J.~A. Frieman, and A.~V. Olinto, {\it {Natural inflation with pseudo
  - Nambu-Goldstone bosons}},  {\em Phys.Rev.Lett.}, {\bf 65}
  (1990)\href{http://dx.doi.org/10.1103/PhysRevLett.65.3233 }{, 3233--3236}.

\bibitem{Banks:2003sx}
T.~Banks, M.~Dine, P.~J. Fox, and E.~Gorbatov, {\it {On the possibility of
  large axion decay constants}},  {\em JCAP}, {\bf 0306}
  (2003)\href{http://dx.doi.org/10.1088/1475-7516/2003/06/001 }{, 001},
  [\href{http://www.arxiv.org/abs/hep-th/0303252}{{\tt hep-th/0303252}}].

\bibitem{Svrcek:2006yi}
P.~Svrcek and E.~Witten, {\it {Axions In String Theory}},  {\em JHEP}, {\bf
  0606} (2006)\href{http://dx.doi.org/10.1088/1126-6708/2006/06/051 }{, 051},
  [\href{http://www.arxiv.org/abs/hep-th/0605206}{{\tt hep-th/0605206}}].

\bibitem{Silverstein:2008sg}
E.~Silverstein and A.~Westphal, {\it {Monodromy in the CMB: Gravity Waves and
  String Inflation}},  {\em Phys.Rev.}, {\bf D78}
  (2008)\href{http://dx.doi.org/10.1103/PhysRevD.78.106003 }{, 106003},
  [\href{http://www.arxiv.org/abs/0803.3085}{{\tt arXiv:0803.3085}}].

\bibitem{McAllister:2008hb}
L.~McAllister, E.~Silverstein, and A.~Westphal, {\it {Gravity Waves and Linear
  Inflation from Axion Monodromy}},  {\em Phys.Rev.}, {\bf D82}
  (2010)\href{http://dx.doi.org/10.1103/PhysRevD.82.046003 }{, 046003},
  [\href{http://www.arxiv.org/abs/0808.0706}{{\tt arXiv:0808.0706}}].

\bibitem{Palti:2014kza}
E.~Palti and T.~Weigand, {\it {Towards large r from [p, q]-inflation}},  {\em
  JHEP}, {\bf 1404} (2014)\href{http://dx.doi.org/10.1007/JHEP04(2014)155 }{,
  155}, [\href{http://www.arxiv.org/abs/1403.7507}{{\tt arXiv:1403.7507}}].

\bibitem{Marchesano:2014mla}
F.~Marchesano, G.~Shiu, and A.~M. Uranga, {\it {F-term Axion Monodromy
  Inflation}},  {\em JHEP}, {\bf 1409}
  (2014)\href{http://dx.doi.org/10.1007/JHEP09(2014)184 }{, 184},
  [\href{http://www.arxiv.org/abs/1404.3040}{{\tt arXiv:1404.3040}}].

\bibitem{Blumenhagen:2014gta}
R.~Blumenhagen and E.~Plauschinn, {\it {Towards Universal Axion Inflation and
  Reheating in String Theory}},  {\em Phys.Lett.}, {\bf B736}
  (2014)\href{http://dx.doi.org/10.1016/j.physletb.2014.08.007 }{, 482--487},
  [\href{http://www.arxiv.org/abs/1404.3542}{{\tt arXiv:1404.3542}}].

\bibitem{Hebecker:2014eua}
A.~Hebecker, S.~C. Kraus, and L.~T. Witkowski, {\it {D7-Brane Chaotic
  Inflation}},  {\em Phys.Lett.}, {\bf B737}
  (2014)\href{http://dx.doi.org/10.1016/j.physletb.2014.08.028 }{, 16--22},
  [\href{http://www.arxiv.org/abs/1404.3711}{{\tt arXiv:1404.3711}}].

\bibitem{Ibanez:2014kia}
L.~E. Ib‡–\'a\~nez and I.~Valenzuela, {\it {The inflaton as an MSSM Higgs and
  open string modulus monodromy inflation}},  {\em Phys.Lett.}, {\bf B736}
  (2014)\href{http://dx.doi.org/10.1016/j.physletb.2014.07.020 }{, 226--230},
  [\href{http://www.arxiv.org/abs/1404.5235}{{\tt arXiv:1404.5235}}].

\bibitem{Arends:2014qca}
M.~Arends, A.~Hebecker, K.~Heimpel, S.~C. Kraus, D.~Lust, {\em et.~al.}, {\it
  {D7-Brane Moduli Space in Axion Monodromy and Fluxbrane Inflation}},  {\em
  Fortsch.Phys.}, {\bf 62} (2014)\href{http://dx.doi.org/10.1002/prop.201400045
  }{, 647--702}, [\href{http://www.arxiv.org/abs/1405.0283}{{\tt
  arXiv:1405.0283}}].

\bibitem{McAllister:2014mpa}
L.~McAllister, E.~Silverstein, A.~Westphal, and T.~Wrase, {\it {The Powers of
  Monodromy}},  {\em JHEP}, {\bf 1409}
  (2014)\href{http://dx.doi.org/10.1007/JHEP09(2014)123 }{, 123},
  [\href{http://www.arxiv.org/abs/1405.3652}{{\tt arXiv:1405.3652}}].

\bibitem{Franco:2014hsa}
S.~Franco, D.~Galloni, A.~Retolaza, and A.~Uranga, {\it {Axion Monodromy
  Inflation on Warped Throats}},
  \href{http://www.arxiv.org/abs/1405.7044}{{\tt arXiv:1405.7044}}.

\bibitem{Blumenhagen:2014nba}
R.~Blumenhagen, D.~Herschmann, and E.~Plauschinn, {\it {The Challenge of
  Realizing F-term Axion Monodromy Inflation in String Theory}},
  \href{http://www.arxiv.org/abs/1409.7075}{{\tt arXiv:1409.7075}}.

\bibitem{Hebecker:2014kva}
A.~Hebecker, P.~Mangat, F.~Rompineve, and L.~T. Witkowski, {\it {Tuning and
  Backreaction in F-term Axion Monodromy Inflation}},
  \href{http://www.arxiv.org/abs/1411.2032}{{\tt arXiv:1411.2032}}.

\bibitem{Ibanez:2014swa}
L.~E. Ibanez, F.~Marchesano, and I.~Valenzuela, {\it {Higgs-otic Inflation and
  String Theory}},  \href{http://www.arxiv.org/abs/1411.5380}{{\tt
  arXiv:1411.5380}}.

\bibitem{Kaloper:2008fb}
N.~Kaloper and L.~Sorbo, {\it {A Natural Framework for Chaotic Inflation}},
  {\em Phys.Rev.Lett.}, {\bf 102}
  (2009)\href{http://dx.doi.org/10.1103/PhysRevLett.102.121301 }{, 121301},
  [\href{http://www.arxiv.org/abs/0811.1989}{{\tt arXiv:0811.1989}}].

\bibitem{Kaloper:2011jz}
N.~Kaloper, A.~Lawrence, and L.~Sorbo, {\it {An Ignoble Approach to Large Field
  Inflation}},  {\em JCAP}, {\bf 1103}
  (2011)\href{http://dx.doi.org/10.1088/1475-7516/2011/03/023 }{, 023},
  [\href{http://www.arxiv.org/abs/1101.0026}{{\tt arXiv:1101.0026}}].

\bibitem{Kaloper:2014zba}
N.~Kaloper and A.~Lawrence, {\it {Natural chaotic inflation and ultraviolet
  sensitivity}},  {\em Phys.Rev.}, {\bf D90} (2014),
  no.~2\href{http://dx.doi.org/10.1103/PhysRevD.90.023506 }{, 023506},
  [\href{http://www.arxiv.org/abs/1404.2912}{{\tt arXiv:1404.2912}}].

\bibitem{Kim:2004rp}
J.~E. Kim, H.~P. Nilles, and M.~Peloso, {\it {Completing natural inflation}},
  {\em JCAP}, {\bf 0501}
  (2005)\href{http://dx.doi.org/10.1088/1475-7516/2005/01/005 }{, 005},
  [\href{http://www.arxiv.org/abs/hep-ph/0409138}{{\tt hep-ph/0409138}}].

\bibitem{Dimopoulos:2005ac}
S.~Dimopoulos, S.~Kachru, J.~McGreevy, and J.~G. Wacker, {\it {N-flation}},
  {\em JCAP}, {\bf 0808}
  (2008)\href{http://dx.doi.org/10.1088/1475-7516/2008/08/003 }{, 003},
  [\href{http://www.arxiv.org/abs/hep-th/0507205}{{\tt hep-th/0507205}}].

\bibitem{Berg:2009tg}
M.~Berg, E.~Pajer, and S.~Sjors, {\it {Dante's Inferno}},  {\em Phys.Rev.},
  {\bf D81} (2010)\href{http://dx.doi.org/10.1103/PhysRevD.81.103535 }{,
  103535}, [\href{http://www.arxiv.org/abs/0912.1341}{{\tt arXiv:0912.1341}}].

\bibitem{Choi:2014rja}
K.~Choi, H.~Kim, and S.~Yun, {\it {Natural inflation with multiple
  sub-Planckian axions}},  {\em Phys.Rev.}, {\bf D90}
  (2014)\href{http://dx.doi.org/10.1103/PhysRevD.90.023545 }{, 023545},
  [\href{http://www.arxiv.org/abs/1404.6209}{{\tt arXiv:1404.6209}}].

\bibitem{Higaki:2014pja}
T.~Higaki and F.~Takahashi, {\it {Natural and Multi-Natural Inflation in Axion
  Landscape}},  {\em JHEP}, {\bf 1407}
  (2014)\href{http://dx.doi.org/10.1007/JHEP07(2014)074 }{, 074},
  [\href{http://www.arxiv.org/abs/1404.6923}{{\tt arXiv:1404.6923}}].

\bibitem{Tye:2014tja}
S.~H.~H. Tye and S.~S.~C. Wong, {\it {Helical Inflation and Cosmic Strings}},
  \href{http://www.arxiv.org/abs/1404.6988}{{\tt arXiv:1404.6988}}.

\bibitem{Kappl:2014lra}
R.~Kappl, S.~Krippendorf, and H.~P. Nilles, {\it {Aligned Natural Inflation:
  Monodromies of two Axions}},  {\em Phys.Lett.}, {\bf B737}
  (2014)\href{http://dx.doi.org/10.1016/j.physletb.2014.08.045 }{, 124--128},
  [\href{http://www.arxiv.org/abs/1404.7127}{{\tt arXiv:1404.7127}}].

\bibitem{Bachlechner:2014hsa}
T.~C. Bachlechner, M.~Dias, J.~Frazer, and L.~McAllister, {\it {A New Angle on
  Chaotic Inflation}},  \href{http://www.arxiv.org/abs/1404.7496}{{\tt
  arXiv:1404.7496}}.

\bibitem{Ben-Dayan:2014zsa}
I.~Ben-Dayan, F.~G. Pedro, and A.~Westphal, {\it {Hierarchical Axion
  Inflation}},  \href{http://www.arxiv.org/abs/1404.7773}{{\tt
  arXiv:1404.7773}}.

\bibitem{Long:2014dta}
C.~Long, L.~McAllister, and P.~McGuirk, {\it {Aligned Natural Inflation in
  String Theory}},  {\em Phys.Rev.}, {\bf D90}
  (2014)\href{http://dx.doi.org/10.1103/PhysRevD.90.023501 }{, 023501},
  [\href{http://www.arxiv.org/abs/1404.7852}{{\tt arXiv:1404.7852}}].

\bibitem{Higaki:2014mwa}
T.~Higaki and F.~Takahashi, {\it {Axion Landscape and Natural Inflation}},
  \href{http://www.arxiv.org/abs/1409.8409}{{\tt arXiv:1409.8409}}.

\bibitem{Bachlechner:2014gfa}
T.~C. Bachlechner, C.~Long, and L.~McAllister, {\it {Planckian Axions in String
  Theory}},  \href{http://www.arxiv.org/abs/1412.1093}{{\tt arXiv:1412.1093}}.

\bibitem{Burgess:2014oma}
C.~Burgess and D.~Roest, {\it {Inflation by Alignment}},
  \href{http://www.arxiv.org/abs/1412.1614}{{\tt arXiv:1412.1614}}.

\bibitem{Gao:2014uha}
X.~Gao, T.~Li, and P.~Shukla, {\it {Combining Universal and Odd RR Axions for
  Aligned Natural Inflation}},  {\em JCAP}, {\bf 1410} (2014),
  no.~10\href{http://dx.doi.org/10.1088/1475-7516/2014/10/048 }{, 048},
  [\href{http://www.arxiv.org/abs/1406.0341}{{\tt arXiv:1406.0341}}].

\bibitem{Kenton:2014gma}
Z.~Kenton and S.~Thomas, {\it {D-brane Potentials in the Warped Resolved
  Conifold and Natural Inflation}},  {\em JHEP}, {\bf 1502}
  (2015)\href{http://dx.doi.org/10.1007/JHEP02(2015)127 }{, 127},
  [\href{http://www.arxiv.org/abs/1409.1221}{{\tt arXiv:1409.1221}}].

\bibitem{Shiu:2013wxa}
G.~Shiu, P.~Soler, and F.~Ye, {\it {Millicharged Dark Matter in Quantum Gravity
  and String Theory}},  {\em Phys.Rev.Lett.}, {\bf 110} (2013),
  no.~24\href{http://dx.doi.org/10.1103/PhysRevLett.110.241304 }{, 241304},
  [\href{http://www.arxiv.org/abs/1302.5471}{{\tt arXiv:1302.5471}}].

\bibitem{Feng:2014eja}
W.-Z. Feng, G.~Shiu, P.~Soler, and F.~Ye, {\it {Probing Hidden Sectors with
  St\"uckelberg U(1) Gauge Fields}},  {\em Phys.Rev.Lett.}, {\bf 113}
  (2014)\href{http://dx.doi.org/10.1103/PhysRevLett.113.061802 }{, 061802},
  [\href{http://www.arxiv.org/abs/1401.5880}{{\tt arXiv:1401.5880}}].

\bibitem{Feng:2014cla}
W.-Z. Feng, G.~Shiu, P.~Soler, and F.~Ye, {\it {Building a Stückelberg
  portal}},  {\em JHEP}, {\bf 1405}
  (2014)\href{http://dx.doi.org/10.1007/JHEP05(2014)065 }{, 065},
  [\href{http://www.arxiv.org/abs/1401.5890}{{\tt arXiv:1401.5890}}].

\bibitem{BerasaluceGonzalez:2012vb}
M.~Berasaluce-Gonzalez, P.~Camara, F.~Marchesano, D.~Regalado, and A.~Uranga,
  {\it {Non-Abelian discrete gauge symmetries in 4d string models}},  {\em
  JHEP}, {\bf 1209} (2012)\href{http://dx.doi.org/10.1007/JHEP09(2012)059 }{,
  059}, [\href{http://www.arxiv.org/abs/1206.2383}{{\tt arXiv:1206.2383}}].

\bibitem{Chen:2012jg}
M.-C. Chen, M.~Ratz, C.~Staudt, and P.~K. Vaudrevange, {\it {The mu Term and
  Neutrino Masses}},  {\em Nucl.Phys.}, {\bf B866}
  (2013)\href{http://dx.doi.org/10.1016/j.nuclphysb.2012.08.018 }{, 157--176},
  [\href{http://www.arxiv.org/abs/1206.5375}{{\tt arXiv:1206.5375}}].

\bibitem{BerasaluceGonzalez:2011wy}
M.~Berasaluce-Gonzalez, L.~E. Ib\'{a}\~{n}ez, P.~Soler, and A.~M. Uranga, {\it
  {Discrete gauge symmetries in D-brane models}},  {\em JHEP}, {\bf 1112}
  (2011)\href{http://dx.doi.org/10.1007/JHEP12(2011)113 }{, 113},
  [\href{http://www.arxiv.org/abs/1106.4169}{{\tt arXiv:1106.4169}}].

\bibitem{Anastasopoulos:2012zu}
P.~Anastasopoulos, M.~Cveti\v{c}, R.~Richter, and P.~K. Vaudrevange, {\it
  {String Constraints on Discrete Symmetries in MSSM Type II Quivers}},  {\em
  JHEP}, {\bf 1303} (2013)\href{http://dx.doi.org/10.1007/JHEP03(2013)011 }{,
  011}, [\href{http://www.arxiv.org/abs/1211.1017}{{\tt arXiv:1211.1017}}].

\bibitem{Honecker:2013hda}
G.~Honecker and W.~Staessens, {\it {To Tilt or Not To Tilt: Discrete Gauge
  Symmetries in Global Intersecting D-Brane Models}},  {\em JHEP}, {\bf 1310}
  (2013)\href{http://dx.doi.org/10.1007/JHEP10(2013)146 }{, 146},
  [\href{http://www.arxiv.org/abs/1303.4415}{{\tt arXiv:1303.4415}}].

\bibitem{Aldazabal:2002py}
G.~Aldazabal, L.~Ibanez, and A.~Uranga, {\it {Gauging away the strong CP
  problem}},  {\em JHEP}, {\bf 0403}
  (2004)\href{http://dx.doi.org/10.1088/1126-6708/2004/03/065 }{, 065},
  [\href{http://www.arxiv.org/abs/hep-ph/0205250}{{\tt hep-ph/0205250}}].

\bibitem{Andrianopoli:2004sv}
L.~Andrianopoli, S.~Ferrara, and M.~Lledo, {\it {Axion gauge symmetries and
  generalized Chern-Simons terms in N = 1 supersymmetric theories}},  {\em
  JHEP}, {\bf 0404} (2004)\href{http://dx.doi.org/10.1088/1126-6708/2004/04/005
  }{, 005}, [\href{http://www.arxiv.org/abs/hep-th/0402142}{{\tt
  hep-th/0402142}}].

\bibitem{Anastasopoulos:2006cz}
P.~Anastasopoulos, M.~Bianchi, E.~Dudas, and E.~Kiritsis, {\it {Anomalies,
  anomalous U(1)'s and generalized Chern-Simons terms}},  {\em JHEP}, {\bf
  0611} (2006)\href{http://dx.doi.org/10.1088/1126-6708/2006/11/057 }{, 057},
  [\href{http://www.arxiv.org/abs/hep-th/0605225}{{\tt hep-th/0605225}}].

\bibitem{DeRydt:2007vg}
J.~De~Rydt, J.~Rosseel, T.~T. Schmidt, A.~Van~Proeyen, and M.~Zagermann, {\it
  {Symplectic structure of N=1 supergravity with anomalies and Chern-Simons
  terms}},  {\em Class.Quant.Grav.}, {\bf 24}
  (2007)\href{http://dx.doi.org/10.1088/0264-9381/24/20/017 }{, 5201--5220},
  [\href{http://www.arxiv.org/abs/0705.4216}{{\tt arXiv:0705.4216}}].

\bibitem{McAllister:2007bg}
L.~McAllister and E.~Silverstein, {\it {String Cosmology: A Review}},  {\em
  Gen.Rel.Grav.}, {\bf 40}
  (2008)\href{http://dx.doi.org/10.1007/s10714-007-0556-6 }{, 565--605},
  [\href{http://www.arxiv.org/abs/0710.2951}{{\tt arXiv:0710.2951}}].

\bibitem{Baumann:2009ni}
D.~Baumann and L.~McAllister, {\it {Advances in Inflation in String Theory}},
  {\em Ann.Rev.Nucl.Part.Sci.}, {\bf 59}
  (2009)\href{http://dx.doi.org/10.1146/annurev.nucl.010909.083524 }{, 67--94},
  [\href{http://www.arxiv.org/abs/0901.0265}{{\tt arXiv:0901.0265}}].

\bibitem{Cicoli:2011zz}
M.~Cicoli and F.~Quevedo, {\it {String moduli inflation: An overview}},  {\em
  Class.Quant.Grav.}, {\bf 28}
  (2011)\href{http://dx.doi.org/10.1088/0264-9381/28/20/204001 }{, 204001},
  [\href{http://www.arxiv.org/abs/1108.2659}{{\tt arXiv:1108.2659}}].

\bibitem{Burgess:2011fa}
C.~Burgess and L.~McAllister, {\it {Challenges for String Cosmology}},  {\em
  Class.Quant.Grav.}, {\bf 28}
  (2011)\href{http://dx.doi.org/10.1088/0264-9381/28/20/204002 }{, 204002},
  [\href{http://www.arxiv.org/abs/1108.2660}{{\tt arXiv:1108.2660}}].

\bibitem{Baumann:2014nda}
D.~Baumann and L.~McAllister, {\it {Inflation and String Theory}},
  \href{http://www.arxiv.org/abs/1404.2601}{{\tt arXiv:1404.2601}}.

\bibitem{Witten:1984dg}
E.~Witten, {\it {Some Properties of O(32) Superstrings}},  {\em Phys.Lett.},
  {\bf B149} (1984)\href{http://dx.doi.org/10.1016/0370-2693(84)90422-2 }{,
  351--356}.

\bibitem{Barr:1985hk}
S.~M. Barr, {\it {Harmless Axions in Superstring Theories}},  {\em Phys.Lett.},
  {\bf B158} (1985)\href{http://dx.doi.org/10.1016/0370-2693(85)90440-X }{,
  397}.

\bibitem{Choi:1985je}
K.~Choi and J.~E. Kim, {\it {Harmful Axions in Superstring Models}},  {\em
  Phys.Lett.}, {\bf B154}
  (1985)\href{http://dx.doi.org/10.1016/0370-2693(85)90416-2 }{, 393}.

\bibitem{Polchinski:1998rr}
J.~Polchinski, {\it {String theory. Vol. 2: Superstring theory and beyond}}, .

\bibitem{Bergshoeff:2001pv}
E.~Bergshoeff, R.~Kallosh, T.~Ortin, D.~Roest, and A.~Van~Proeyen, {\it {New
  formulations of D = 10 supersymmetry and D8 - O8 domain walls}},  {\em
  Class.Quant.Grav.}, {\bf 18}
  (2001)\href{http://dx.doi.org/10.1088/0264-9381/18/17/303 }{, 3359--3382},
  [\href{http://www.arxiv.org/abs/hep-th/0103233}{{\tt hep-th/0103233}}].

\bibitem{Grimm:2004uq}
T.~W. Grimm and J.~Louis, {\it {The Effective action of N = 1 Calabi-Yau
  orientifolds}},  {\em Nucl.Phys.}, {\bf B699}
  (2004)\href{http://dx.doi.org/10.1016/j.nuclphysb.2004.08.005 }{, 387--426},
  [\href{http://www.arxiv.org/abs/hep-th/0403067}{{\tt hep-th/0403067}}].

\bibitem{Grimm:2004ua}
T.~W. Grimm and J.~Louis, {\it {The Effective action of type IIA Calabi-Yau
  orientifolds}},  {\em Nucl.Phys.}, {\bf B718}
  (2005)\href{http://dx.doi.org/10.1016/j.nuclphysb.2005.04.007 }{, 153--202},
  [\href{http://www.arxiv.org/abs/hep-th/0412277}{{\tt hep-th/0412277}}].

\bibitem{Jockers:2004yj}
H.~Jockers and J.~Louis, {\it {The Effective action of D7-branes in N = 1
  Calabi-Yau orientifolds}},  {\em Nucl.Phys.}, {\bf B705}
  (2005)\href{http://dx.doi.org/10.1016/j.nuclphysb.2004.11.009 }{, 167--211},
  [\href{http://www.arxiv.org/abs/hep-th/0409098}{{\tt hep-th/0409098}}].

\bibitem{Haack:2006cy}
M.~Haack, D.~Krefl, D.~Lust, A.~Van~Proeyen, and M.~Zagermann, {\it {Gaugino
  Condensates and D-terms from D7-branes}},  {\em JHEP}, {\bf 0701}
  (2007)\href{http://dx.doi.org/10.1088/1126-6708/2007/01/078 }{, 078},
  [\href{http://www.arxiv.org/abs/hep-th/0609211}{{\tt hep-th/0609211}}].

\bibitem{Cicoli:2011yh}
M.~Cicoli, M.~Goodsell, J.~Jaeckel, and A.~Ringwald, {\it {Testing String Vacua
  in the Lab: From a Hidden CMB to Dark Forces in Flux Compactifications}},
  {\em JHEP}, {\bf 1107} (2011)\href{http://dx.doi.org/10.1007/JHEP07(2011)114
  }{, 114}, [\href{http://www.arxiv.org/abs/1103.3705}{{\tt arXiv:1103.3705}}].

\bibitem{Grimm:2011dx}
T.~W. Grimm and D.~V. Lopes, {\it {The N=1 effective actions of D-branes in
  Type IIA and IIB orientifolds}},  {\em Nucl.Phys.}, {\bf B855}
  (2012)\href{http://dx.doi.org/10.1016/j.nuclphysb.2011.10.019 }{, 639--694},
  [\href{http://www.arxiv.org/abs/1104.2328}{{\tt arXiv:1104.2328}}].

\bibitem{Kerstan:2011dy}
M.~Kerstan and T.~Weigand, {\it {The Effective action of D6-branes in N=1 type
  IIA orientifolds}},  {\em JHEP}, {\bf 1106}
  (2011)\href{http://dx.doi.org/10.1007/JHEP06(2011)105 }{, 105},
  [\href{http://www.arxiv.org/abs/1104.2329}{{\tt arXiv:1104.2329}}].

\bibitem{Camara:2011jg}
P.~G. Camara, L.~E. Ibanez, and F.~Marchesano, {\it {RR photons}},  {\em JHEP},
  {\bf 1109} (2011)\href{http://dx.doi.org/10.1007/JHEP09(2011)110 }{, 110},
  [\href{http://www.arxiv.org/abs/1106.0060}{{\tt arXiv:1106.0060}}].

\bibitem{Gmeiner:2009fb}
F.~Gmeiner and G.~Honecker, {\it {Complete Gauge Threshold Corrections for
  Intersecting Fractional D6-Branes: The Z6 and Z6' Standard Models}},  {\em
  Nucl.Phys.}, {\bf B829}
  (2010)\href{http://dx.doi.org/10.1016/j.nuclphysb.2009.12.011 }{, 225--297},
  [\href{http://www.arxiv.org/abs/0910.0843}{{\tt arXiv:0910.0843}}].

\bibitem{Honecker:2012qr}
G.~Honecker, M.~Ripka, and W.~Staessens, {\it {The Importance of Being Rigid:
  D6-Brane Model Building on $T^6/Z_2 x Z_6'$ with Discrete Torsion}},  {\em
  Nucl.Phys.}, {\bf B868}
  (2013)\href{http://dx.doi.org/10.1016/j.nuclphysb.2012.11.011 }{, 156--222},
  [\href{http://www.arxiv.org/abs/1209.3010}{{\tt arXiv:1209.3010}}].

\bibitem{Honecker:2013mya}
G.~Honecker and W.~Staessens, {\it {On axionic dark matter in Type IIA string
  theory}},  {\em Fortsch.Phys.}, {\bf 62}
  (2014)\href{http://dx.doi.org/10.1002/prop.201300036 }{, 115--151},
  [\href{http://www.arxiv.org/abs/1312.4517}{{\tt arXiv:1312.4517}}].

\bibitem{Abe:2014pwa}
H.~Abe, T.~Kobayashi, and H.~Otsuka, {\it {Towards natural inflation from
  weakly coupled heterotic string theory}},
  \href{http://www.arxiv.org/abs/1409.8436}{{\tt arXiv:1409.8436}}.

\bibitem{Abe:2014xja}
H.~Abe, T.~Kobayashi, and H.~Otsuka, {\it {Natural inflation with and without
  modulations in type IIB string theory}},
  \href{http://www.arxiv.org/abs/1411.4768}{{\tt arXiv:1411.4768}}.

\bibitem{Koerber:2010bx}
P.~Koerber, {\it {Lectures on Generalized Complex Geometry for Physicists}},
  {\em Fortsch.Phys.}, {\bf 59}
  (2011)\href{http://dx.doi.org/10.1002/prop.201000083 }{, 169--242},
  [\href{http://www.arxiv.org/abs/1006.1536}{{\tt arXiv:1006.1536}}].

\bibitem{Candelas:1990pi}
P.~Candelas and X.~de~la Ossa, {\it {Moduli Space of {Calabi-Yau} Manifolds}},
  {\em Nucl.Phys.}, {\bf B355}
  (1991)\href{http://dx.doi.org/10.1016/0550-3213(91)90122-E }{, 455--481}.

\bibitem{Blumenhagen:2009qh}
R.~Blumenhagen, M.~Cveti\v{c}, S.~Kachru, and T.~Weigand, {\it {D-Brane
  Instantons in Type II Orientifolds}},  {\em Ann.Rev.Nucl.Part.Sci.}, {\bf 59}
  (2009)\href{http://dx.doi.org/10.1146/annurev.nucl.010909.083113 }{,
  269--296}, [\href{http://www.arxiv.org/abs/0902.3251}{{\tt
  arXiv:0902.3251}}].

\bibitem{Ibanez:2012zz}
L.~E. Ibanez and A.~M. Uranga, {\it {String theory and particle physics: An
  introduction to string phenomenology}}, .

\bibitem{Blumenhagen:2005tn}
R.~Blumenhagen, M.~Cvetic, F.~Marchesano, and G.~Shiu, {\it {Chiral D-brane
  models with frozen open string moduli}},  {\em JHEP}, {\bf 0503}
  (2005)\href{http://dx.doi.org/10.1088/1126-6708/2005/03/050 }{, 050},
  [\href{http://www.arxiv.org/abs/hep-th/0502095}{{\tt hep-th/0502095}}].

\bibitem{Cvetic:2007ku}
M.~Cvetic, R.~Richter, and T.~Weigand, {\it {Computation of D-brane instanton
  induced superpotential couplings: Majorana masses from string theory}},  {\em
  Phys.Rev.}, {\bf D76}
  (2007)\href{http://dx.doi.org/10.1103/PhysRevD.76.086002 }{, 086002},
  [\href{http://www.arxiv.org/abs/hep-th/0703028}{{\tt hep-th/0703028}}].

\bibitem{Blumenhagen:2006xt}
R.~Blumenhagen, M.~Cvetic, and T.~Weigand, {\it {Spacetime instanton
  corrections in 4D string vacua: The Seesaw mechanism for D-Brane models}},
  {\em Nucl.Phys.}, {\bf B771}
  (2007)\href{http://dx.doi.org/10.1016/j.nuclphysb.2007.02.016 }{, 113--142},
  [\href{http://www.arxiv.org/abs/hep-th/0609191}{{\tt hep-th/0609191}}].

\bibitem{Ibanez:2006da}
L.~Ibanez and A.~Uranga, {\it {Neutrino Majorana Masses from String Theory
  Instanton Effects}},  {\em JHEP}, {\bf 0703}
  (2007)\href{http://dx.doi.org/10.1088/1126-6708/2007/03/052 }{, 052},
  [\href{http://www.arxiv.org/abs/hep-th/0609213}{{\tt hep-th/0609213}}].

\bibitem{Forste:2010gw}
S.~F{\"o}rste and G.~Honecker, {\it {Rigid D6-branes on $T^6/(Z_2 x Z_{2M} x
  \Omega R)$ with discrete torsion}},  {\em JHEP}, {\bf 1101}
  (2011)\href{http://dx.doi.org/10.1007/JHEP01(2011)091 }{, 091},
  [\href{http://www.arxiv.org/abs/1010.6070}{{\tt arXiv:1010.6070}}].

\bibitem{Honecker:2011sm}
G.~Honecker, {\it {K{\"a}hler metrics and gauge kinetic functions for
  intersecting D6-branes on toroidal orbifolds - The complete perturbative
  story}},  {\em Fortsch.Phys.}, {\bf 60}
  (2012)\href{http://dx.doi.org/10.1002/prop.201100087 }{, 243--326},
  [\href{http://www.arxiv.org/abs/1109.3192}{{\tt arXiv:1109.3192}}].

\bibitem{Rabadan:2001mt}
R.~Rabadan, {\it {Branes at angles, torons, stability and supersymmetry}},
  {\em Nucl.Phys.}, {\bf B620}
  (2002)\href{http://dx.doi.org/10.1016/S0550-3213(01)00560-0 }{, 152--180},
  [\href{http://www.arxiv.org/abs/hep-th/0107036}{{\tt hep-th/0107036}}].

\bibitem{Cremades:2002cs}
D.~Cremades, L.~Ibanez, and F.~Marchesano, {\it {Intersecting brane models of
  particle physics and the Higgs mechanism}},  {\em JHEP}, {\bf 0207}
  (2002)\href{http://dx.doi.org/10.1088/1126-6708/2002/07/022 }{, 022},
  [\href{http://www.arxiv.org/abs/hep-th/0203160}{{\tt hep-th/0203160}}].

\bibitem{Strominger:1985ks}
A.~Strominger, {\it {Yukawa Couplings in Superstring Compactification}},  {\em
  Phys.Rev.Lett.}, {\bf 55}
  (1985)\href{http://dx.doi.org/10.1103/PhysRevLett.55.2547 }{, 2547}.

\bibitem{short}
G.~Shiu, W.~Staessens, and F.~Ye, {\it {Widening the Axion Window via Kinetic
  and St\"uckelberg Mixings}},  \href{http://www.arxiv.org/abs/1503.0101}{{\tt
  arXiv:1503.0101}}.

\bibitem{Gray:2012jy}
J.~Gray, Y.-H. He, V.~Jejjala, B.~Jurke, B.~D. Nelson, {\em et.~al.}, {\it
  {Calabi-Yau Manifolds with Large Volume Vacua}},  {\em Phys.Rev.}, {\bf D86}
  (2012)\href{http://dx.doi.org/10.1103/PhysRevD.86.101901 }{, 101901},
  [\href{http://www.arxiv.org/abs/1207.5801}{{\tt arXiv:1207.5801}}].

\bibitem{Altman:2014bfa}
R.~Altman, J.~Gray, Y.-H. He, V.~Jejjala, and B.~D. Nelson, {\it {A Calabi-Yau
  Database: Threefolds Constructed from the Kreuzer-Skarke List}},  {\em JHEP},
  {\bf 1502} (2015)\href{http://dx.doi.org/10.1007/JHEP02(2015)158 }{, 158},
  [\href{http://www.arxiv.org/abs/1411.1418}{{\tt arXiv:1411.1418}}].

\bibitem{Blumenhagen:2007sm}
R.~Blumenhagen, S.~Moster, and E.~Plauschinn, {\it {Moduli Stabilisation versus
  Chirality for MSSM like Type IIB Orientifolds}},  {\em JHEP}, {\bf 0801}
  (2008)\href{http://dx.doi.org/10.1088/1126-6708/2008/01/058 }{, 058},
  [\href{http://www.arxiv.org/abs/0711.3389}{{\tt arXiv:0711.3389}}].

\bibitem{Collinucci:2008sq}
A.~Collinucci, M.~Kreuzer, C.~Mayrhofer, and N.-O. Walliser, {\it {Four-modulus
  'Swiss Cheese' chiral models}},  {\em JHEP}, {\bf 0907}
  (2009)\href{http://dx.doi.org/10.1088/1126-6708/2009/07/074 }{, 074},
  [\href{http://www.arxiv.org/abs/0811.4599}{{\tt arXiv:0811.4599}}].

\bibitem{Cicoli:2011qg}
M.~Cicoli, C.~Mayrhofer, and R.~Valandro, {\it {Moduli Stabilisation for Chiral
  Global Models}},  {\em JHEP}, {\bf 1202}
  (2012)\href{http://dx.doi.org/10.1007/JHEP02(2012)062 }{, 062},
  [\href{http://www.arxiv.org/abs/1110.3333}{{\tt arXiv:1110.3333}}].

\bibitem{Plauschinn:2008yd}
E.~Plauschinn, {\it {The Generalized Green-Schwarz Mechanism for Type IIB
  Orientifolds with D3- and D7-Branes}},  {\em JHEP}, {\bf 0905}
  (2009)\href{http://dx.doi.org/10.1088/1126-6708/2009/05/062 }{, 062},
  [\href{http://www.arxiv.org/abs/0811.2804}{{\tt arXiv:0811.2804}}].

\bibitem{Blumenhagen:2008zz}
R.~Blumenhagen, V.~Braun, T.~W. Grimm, and T.~Weigand, {\it {GUTs in Type IIB
  Orientifold Compactifications}},  {\em Nucl.Phys.}, {\bf B815}
  (2009)\href{http://dx.doi.org/10.1016/j.nuclphysb.2009.02.011 }{, 1--94},
  [\href{http://www.arxiv.org/abs/0811.2936}{{\tt arXiv:0811.2936}}].

\bibitem{Freed:1999vc}
D.~S. Freed and E.~Witten, {\it {Anomalies in string theory with D-branes}},
  {\em Asian J.Math}, {\bf 3} (1999) 819,
  [\href{http://www.arxiv.org/abs/hep-th/9907189}{{\tt hep-th/9907189}}].

\bibitem{Blumenhagen:2006ci}
R.~Blumenhagen, B.~K{\"o}rs, D.~L{\"u}st, and S.~Stieberger, {\it
  {Four-dimensional String Compactifications with D-Branes, Orientifolds and
  Fluxes}},  {\em Phys.Rept.}, {\bf 445}
  (2007)\href{http://dx.doi.org/10.1016/j.physrep.2007.04.003 }{, 1--193},
  [\href{http://www.arxiv.org/abs/hep-th/0610327}{{\tt hep-th/0610327}}].

\bibitem{Peskin:1995ev}
M.~E. Peskin and D.~V. Schroeder, {\it {An Introduction to quantum field
  theory}}, .

\bibitem{Harvey:2005it}
J.~A. Harvey, {\it {TASI 2003 lectures on anomalies}},
  \href{http://www.arxiv.org/abs/hep-th/0509097}{{\tt hep-th/0509097}}.

\bibitem{Fujikawa:1979ay}
K.~Fujikawa, {\it {Path Integral Measure for Gauge Invariant Fermion
  Theories}},  {\em Phys.Rev.Lett.}, {\bf 42}
  (1979)\href{http://dx.doi.org/10.1103/PhysRevLett.42.1195 }{, 1195--1198}.

\end{thebibliography}\endgroup

\end{document}